\documentclass[preprint,12pt,3p]{elsarticle}
\usepackage{framed,multirow}
\usepackage{float}
\usepackage{amssymb}
\usepackage{latexsym}
\usepackage{url}
\usepackage{color}
\usepackage{mathrsfs} 
\definecolor{inchworm}{rgb}{0.7, 0.93, 0.36} 
\definecolor{orng}{rgb}{1.0, 0.31, 0.0} 
\usepackage{algorithm}
\usepackage{algorithmic}
\usepackage{amsmath,amssymb,amsfonts}
\usepackage{graphicx}
\usepackage{textcomp}
\usepackage{cuted}
\usepackage{dblfloatfix} 
\usepackage{hyperref}

\hypersetup{colorlinks = true,
    linkcolor=blue,
    filecolor=cyan,
    citecolor = blue,
    urlcolor=blue}
\definecolor{ib}{rgb}{0.0, 0.18, 0.65}
\definecolor{fg}{rgb}{0.13, 0.55, 0.13}
\usepackage[english]{babel}
\addto\extrasenglish{

} 
\addto\captionsenglish{} 
\usepackage[labelfont=bf]{caption} 
\usepackage{relsize} 
\newcommand\eatpunct[1]{} 
\usepackage{booktabs} 
\usepackage[bb=boondox]{mathalfa} 
\usepackage{float} 
\usepackage{makecell} 
\usepackage{flushend} 
\usepackage[nameinlink,capitalize]{cleveref} 
\usepackage{setspace}

\usepackage{MnSymbol} 
\usepackage{mathrsfs} 
\usepackage{tabularx,booktabs} 
\newcolumntype{C}{>{\centering\arraybackslash}X} 
\usepackage{array}
\newcolumntype{P}[1]{>{\centering\arraybackslash}p{#1}} 
\newcolumntype{L}[1]{>{\arraybackslash}p{#1}} 
\usepackage{enumerate} 
\usepackage[labelformat=simple]{subcaption}
\usepackage{arydshln} 
\usepackage{tabularray}
\usepackage{color, colortbl}

\journal{Information Fusion}

\definecolor{orange}{rgb}{1.0, 0.31, 0.0} 
\definecolor{LightCyan}{rgb}{0.88,1,1}
\begin{document}

\begin{titlepage}
\begin{center}
\vspace*{10pt}
\doublespacing
{\Large A Comprehensive Review on Hashtag Recommendation: From
Traditional to Deep Learning and Beyond}
\vspace*{25pt}

Shubhi Bansal$^{a}$ (phd2001201007@iiti.ac.in), Kushaan Gowda$^b$ (kg3081@columbia.edu),
Anupama Sureshbabu K$^{a}$ (anupama.kodothsbk@gmail.com), Chirag Kothari$^{a}$
(chiragkothari2503@gmail.com), Nagendra Kumar$^a$ (nagendra@iiti.ac.in) \\

\hspace{10pt}

\begin{flushleft}
\small  
$^a$Department of Computer Science and Engineering, Indian Institute of Technology Indore, Indore 453552, India \\
$^b$Department of Computer Science, Columbia University, New York, New York, USA\\

\vspace{1cm}
\normalsize
\textbf{Corresponding Author:} \\
Dr. Nagendra Kumar \\
Department of Computer Science and Engineering, \\
Indian Institute of Technology Indore, Indore 453552, India \\
Tel: +91-7316603225 \\
Email: nagendra@iiti.ac.in\\

This review article has been submitted to Elseiver.
\end{flushleft}        
\end{center}
\end{titlepage}
\begin{frontmatter}
\title{A Comprehensive Review on Hashtag Recommendation: From
Traditional to Deep Learning and Beyond}

\author[1]{Shubhi Bansal}
\ead{phd2001201007@iiti.ac.in}

\author[2]{Kushaan Gowda}
\ead{kg3081@columbia.edu}

\author[1]{Anupama Sureshbabu K}
\ead{anupama.kodothsbk@gmail.com}

\author[1]{Chirag Kothari}
\ead{chiragkothari2503@gmail.com}

\author[1]{Nagendra Kumar\corref{cor1}}
\ead{nagendra@iiti.ac.in}
\cortext[cor1]{Corresponding author}

\address[1]{Department of Computer Science and Engineering, Indian Institute of Technology (IIT) Indore, Indore 453552, India}

\address[2]{Department of Computer Science, Columbia University, New York, New York, USA}

\begin{abstract} 
The exponential growth of user-generated content on social media platforms has precipitated significant challenges in information management, particularly in content organization, retrieval, and discovery. Hashtags, as a fundamental categorization mechanism, play a pivotal role in enhancing content visibility and user engagement. However, the development of accurate and robust hashtag recommendation systems remains a complex and evolving research challenge. Existing surveys in this domain are limited in scope and recency, focusing narrowly on specific platforms (X and Sina Weibo), methodologies, or timeframes. 
To address this gap, this review article conducts a systematic analysis of hashtag recommendation systems, comprehensively examining recent advancements across several dimensions. We investigate unimodal versus multimodal methodologies, diverse problem formulations (encompassing ranking, classification, and generation), filtering strategies (content-based, collaborative, personalised, and hybrid), methodological evolution from traditional frequency-based models to advanced deep learning architectures. Furthermore, we critically evaluate performance assessment paradigms, including quantitative metrics, qualitative analyses, and hybrid evaluation frameworks. Our analysis underscores a paradigm shift toward transformer-based deep learning models, which harness contextual and semantic features to achieve superior recommendation accuracy. Key challenges such as data sparsity, cold-start scenarios, polysemy, and model explainability are rigorously discussed, alongside practical applications in tweet classification, sentiment analysis, and content popularity prediction. By synthesizing insights from diverse methodological and platform-specific perspectives, this survey provides a structured taxonomy of current research, identifies unresolved gaps, and proposes future directions for developing adaptive, user-centric recommendation systems. Serving as a foundational resource for researchers and practitioners, this work aims to catalyze innovation in social media content organization, thereby advancing user experience and driving innovation in social media content management and discovery. A comprehensive compilation of research papers published from 2015 onward in the domain of hashtag recommendation, as reviewed in this study, is accessible on GitHub\footnote{https://github.com/ankh77sb/A-Comprehensive-Review-on-Hashtag-Recommendation}. 
\end{abstract}

\end{frontmatter}

\tableofcontents
\section{Introduction} 
\label{sec:introduction}
Social Network Services (SNS) have revolutionized global communication, enabling 5.2 billion users\footnote{https://datareportal.com/social-media-users} worldwide generating and consuming vast amounts of content daily to exchange ideas, experiences, and opinions in real time. Platforms such as X\footnote{https://x.com/} (formerly known as Twitter), Instagram\footnote{https://www.instagram.com/}, and TikTok\footnote{https://www.tiktok.com/about} host an overwhelming volume of user-generated content (UGC)—over 500 million tweets\footnote{https://www.finalroundai.com/interview-questions/twitter-daily-tweets-volume} and 95 million Instagram posts daily\footnote{https://bernardmarr.com/how-much-data-do-we-create-every-day-the-mind-blowing-stats-everyone-should-read/}—creating a deluge that challenges effective discovery, retrieval, and organization. To combat this information overload, hashtags—metadata labels prefixed with a ``\#” symbol—have emerged as indispensable tools for navigating the digital landscape, enabling users to categorize posts, amplify reach, and foster community engagement. Since their introduction on X in 2007 by Chris Messina, hashtags have evolved beyond simple indexing mechanisms to shape cultural movements, drive trends, improve marketing strategies, enhance content visibility, and foster online communities. Research indicates that the inclusion of hashtags in tweets significantly enhances user engagement, with studies demonstrating that tweets containing hashtags achieve approximately double the level of interaction compared to those without \cite{zhang2019hashtag}.

\textbf{The Dual Role of Hashtags}\\
Hashtags play a pivotal role in shaping social media ecosystems by serving as both organizational metadata and social engagement facilitators. As organizational tools, hashtags aggregate related posts, enhancing searchability and retrieval efficiency by serving as navigational anchors within vast repositories of UGC. For instance, \#AIResearch helps researchers and practitioners track advancements in artificial intelligence across platforms such as X and LinkedIn. Simultaneously, hashtags foster social interactions by connecting users with interest-based communities, trending discussions and digital movements, as exemplified by \#BlackLivesMatter, which has mobilized global activism through UGC-driven awareness campaigns. 

\textbf{The Imperative for Hashtag Recommendation}\\
Despite their utility, the unregulated and free-form adoption of hashtags introduces significant challenges for both users and recommendation systems. Semantic noise arises from inconsistent usage and nonstandard linguistic practices, including slang (\#GOAT for ``greatest of all time"), abbreviations (\#TBT for ``Throwback Thursday"), and misspellings. Redundancy further dilutes the coherence of the topic, as seen in the proliferation of hashtags related to major events with more than 1,200 variations of \#COVID19 \cite{smith2021covid}, such as \#COVID19, \#CoronavirusPandemic, and \#COVID19Updates. Furthermore, ambiguity poses a major obstacle, as hashtags such as \#Amazon may refer to the e-commerce platform or the rainforest, making it difficult for recommendation models to infer the intended context. Despite their critical role in enhancing content visibility and user engagement, the adoption of hashtags remains notably low. Empirical evidence highlights this trend across various social media platforms and content types. For instance, in a dataset of 3,107,866 multilingual tweets, only 24.16\% of posts contained two or more hashtags \cite{bansal2024multilingual}. Similarly, 63\% of multimodal microblogs were found to include fewer than two hashtags \cite{bansal2024hybrid}. Furthermore, a significant proportion of micro-videos—nearly 78\% are uploaded without any hashtags \cite{cao2020hashtag}. These findings underscore the persistent challenge of low hashtag adoption despite their demonstrated benefits. Users struggle in selecting appropriate hashtags due to the dynamic nature of trends, evolving platform norms, semantic complexity, and cognitive overload. Trending hashtags evolve rapidly, necessitating users to stay updated to maintain visibility. Furthermore, semantic challenges, including ambiguity, redundancy, slang prevalence, and context misinterpretation, further complicate the selection. For instance, a content creator posting about film reviews might use \#Joker, but the hashtag could refer to either the 2019 movie or the playing card, leading to irrelevant audience engagement. Similarly, businesses promoting sustainability initiatives may struggle to differentiate between \#GreenEnergy, \#SustainableLiving, and \#EcoFriendly, potentially reducing reach due to improper hashtag selection. Moreover, the cognitive load associated with manually selecting hashtags results in suboptimal choices or omission altogether. 
These challenges underscore the need for automated hashtag recommendation systems that leverage artificial intelligence to generate contextually relevant suggestions in real time. By addressing the limitations of manual selection, such systems can enhance content discoverability, improve user engagement, and optimize UGC-driven strategies across diverse social media platforms.

\begin{figure}
\centering
\includegraphics[width=0.72\textwidth,keepaspectratio=True]{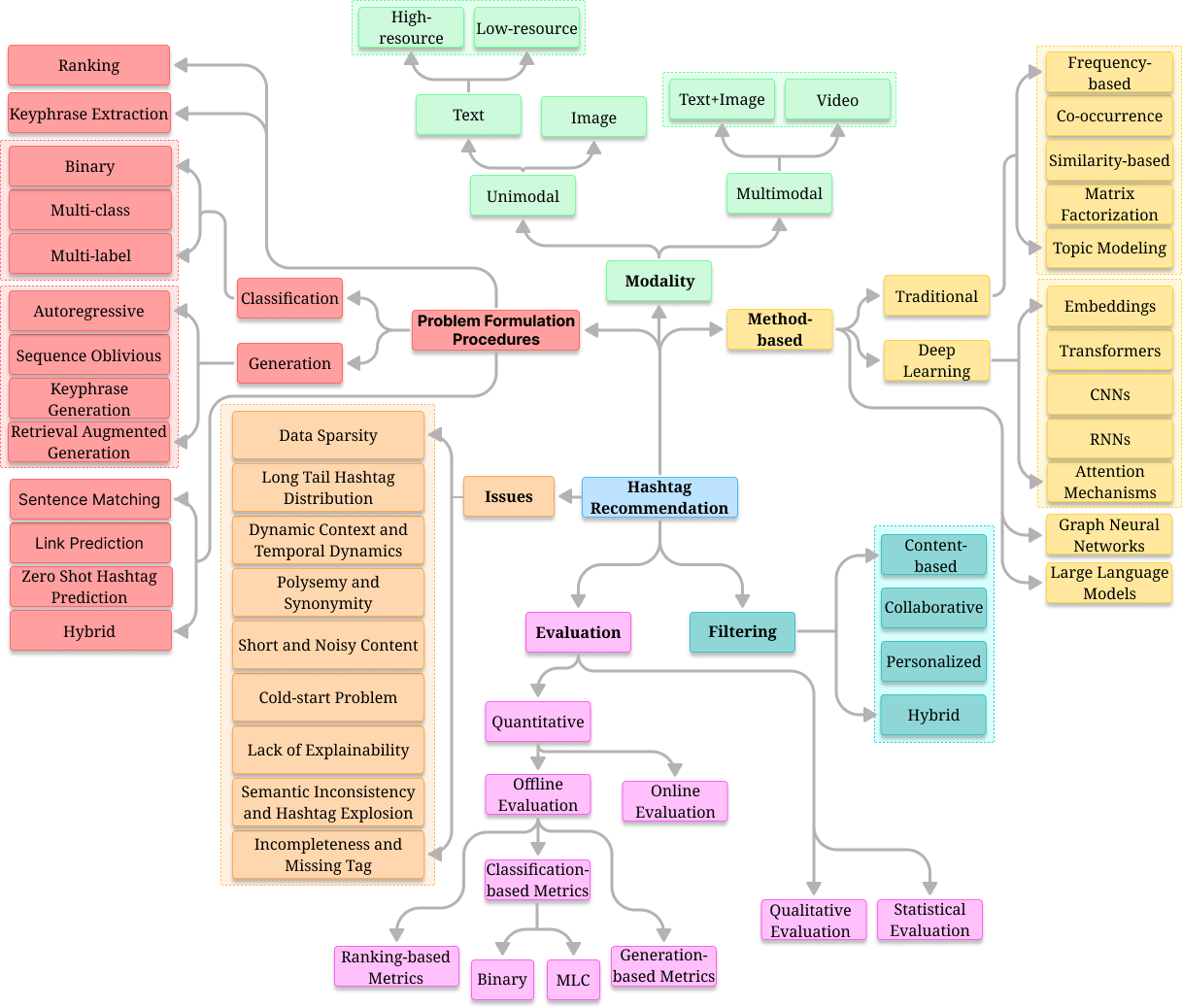}
\caption{Taxonomy of Hashtag Recommendation}
\label{fig:taxonomy}
\end{figure}

\textbf{Gaps with Existing surveys: The Need for a Holistic Survey}\\
Over the past decade, methodologies have evolved from simple frequency-based techniques to sophisticated deep learning models leveraging transformers, Graph Neural Networks (GNNs), and multimodal fusion. Despite these advancements, prior surveys remain limited in scope, either by focusing on specific platforms (e.g., X), outdated timeframes (pre-2020), or isolated technical methodologies (e.g., keyword extraction), lacking a unified perspective on recent multimodal and AI-driven innovations. A comprehensive review that integrates advancements across modalities, problem formulations, and evaluation frameworks is needed to unify the field and guide future research. This survey addresses this gap by providing a systematic, multidisciplinary analysis of hashtag recommendation systems, encompassing research from 2015. We introduce a hierarchical taxonomy organized around six pivotal dimensions, as depicted in \autoref{fig:taxonomy}.
\begin{enumerate}
\item Modality: Unimodal (text, image) vs. multimodal (image+text, video) approaches.
\item Problem Formulation: Tasks such as ranking, classification (binary, multi-label), keyphrase extraction, sentence matching, generation, and link prediction.
\item Filtering Approaches: Content-based, collaborative filtering, personalized, and hybrid approaches.
\item Methods: Traditional techniques (co-occurrence analysis, matrix factorization) vs. deep learning architectures (CNNs, RNNs, transformers).
\item Evaluation: Quantitative metrics (precision, BLEU), qualitative assessments, and hybrid frameworks.
\item Applications: Engagement analytics, trend prediction, and ethical AI.
\item Challenges: Several issues such as long tail hashtag distribution, noisy nature of content, data sparsity, among many others.
\item Future Research Directions: Several fundamental issues, methodological advancements, and evaluation procedures.
\end{enumerate}

\textbf{Contributions of the Review Article}\\
This review article makes three core contributions:
\begin{itemize}
\item This review represents the first extensive and comprehensive analysis of hashtag recommendation, addressing the task from a wide range of perspectives. Covering research from 2015 onwards, it provides a thorough and contemporary synthesis of advancements in the field, offering a holistic understanding of its evolution.
\item We introduce a hierarchical framework that systematically categorizes nearly 150 studies based on modality, problem formulation, filtering approaches, methodology, datasets, evaluation, challenges, and applications. This taxonomy provides a structured and comprehensive overview of the field, enabling researchers to navigate its complexities with clarity.
\item We conducted a systematic evaluation of emerging trends and identified unresolved challenges, including explainability, scalability, and adaptability to low-resource languages. Our analysis includes a comparative synthesis of prior methodologies, highlighting their strengths and limitations in real-world hashtag recommendation scenarios.
\item We offer actionable best practices for dataset selection, ethical deployment, and evaluation, providing valuable insights for researchers and practitioners aiming to implement hashtag recommendation systems in real-world settings.
\end{itemize}
\begin{figure}
\centering
\includegraphics[keepaspectratio=true,width=0.9\textwidth]{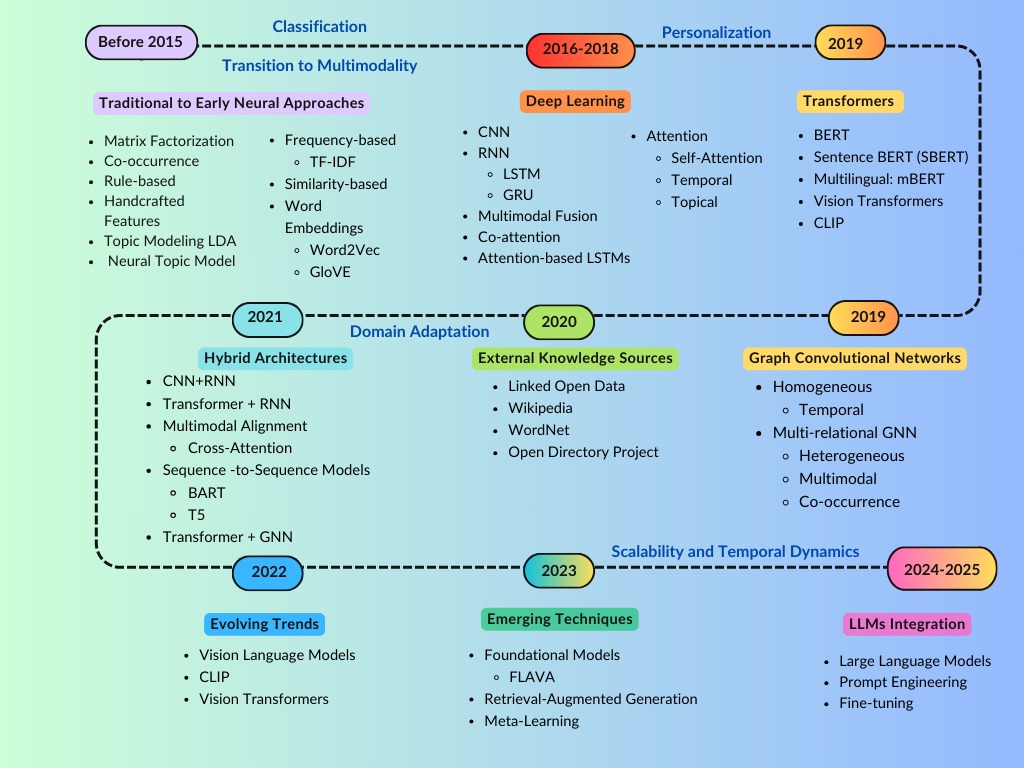}
\caption{Timeline for Evolution in the Domain of Hashtag Recommendation}
\label{fig:timeline}
\end{figure}



\begin{figure}
\centering
\includegraphics[width=0.55\textwidth,keepaspectratio=True]{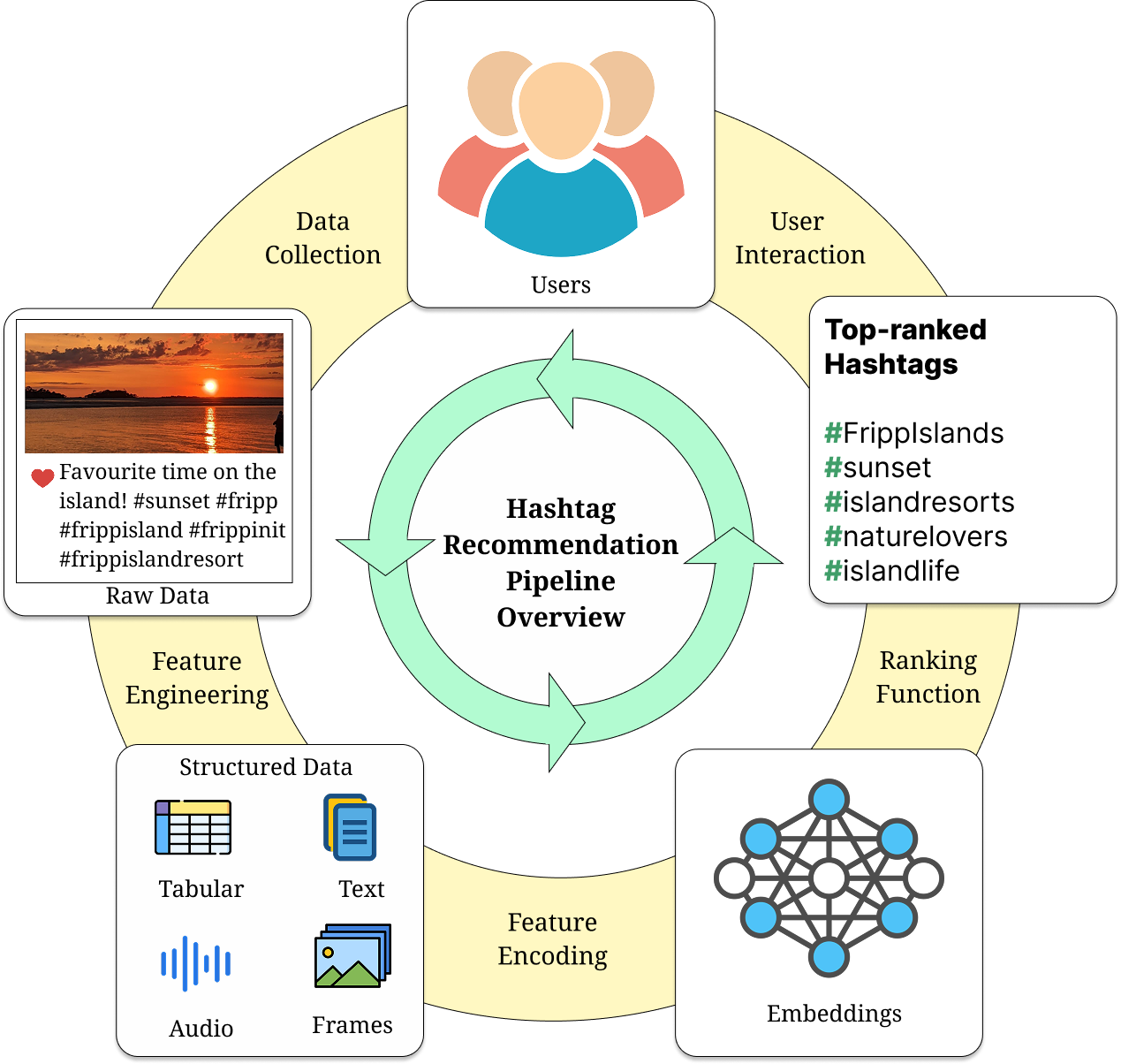}
\caption{Illustration of Hashtag Recommendation System Pipeline}
\label{fig:overveiw_htrc}
\end{figure}

\begin{figure}
\begin{minipage}{0.6\textwidth} 
\centering
\includegraphics[width=\textwidth]{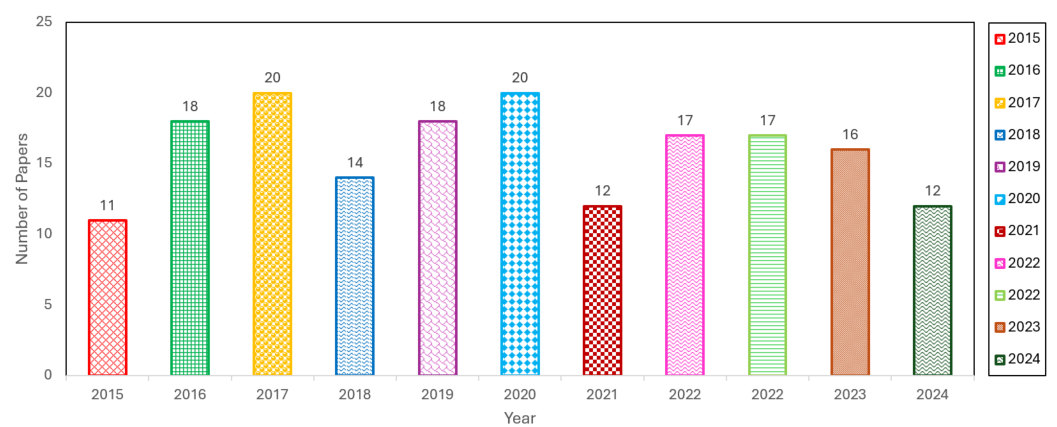} 
\caption{Statistics of Publications Related to Hashtag Recommendation per Year (2015 onwards)}
\label{fig:lit_bar}
\end{minipage}\hfill
\begin{minipage}{0.39\textwidth} 
\centering
\includegraphics[width=\textwidth]{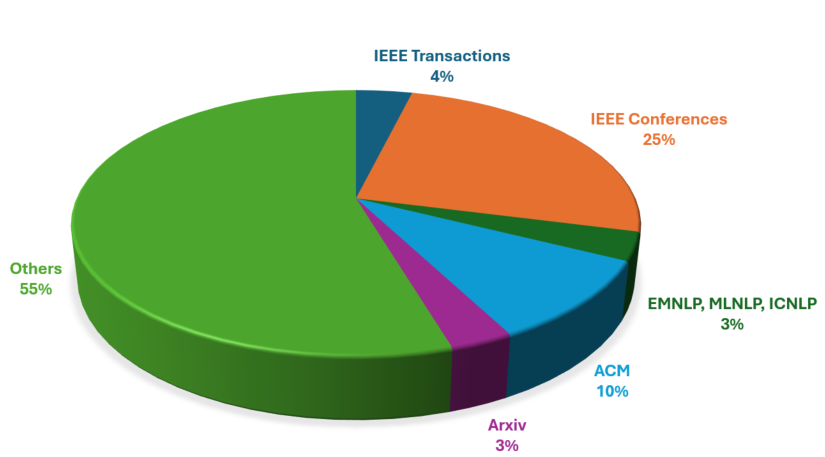} 
\caption{Venue Distribution of Publications Related to Hashtag Recommendation}
\label{fig:lit_pie}
\end{minipage}
\end{figure}

The evolution of the field of hashtag recommendation, encompassing methods, challenges, and modalities from 2015 onwards, is illustrated in \autoref{fig:timeline}. \autoref{fig:overveiw_htrc} presents the overall pipeline for hashtag recommendation systems in SNS, which comprises several  stages: data collection (including user interactions), feature engineering, feature encoding, and ranking function. The system takes as input a user-uploaded piece of content and generates as output a ranked list of top-k hashtags, tailored to the content of the post and the user’s historical behavior and preferences. Furthermore, the distribution of analyzed papers over a decade (2015–2024) is examined from two perspectives: (1) the annual distribution of published papers in this domain, starting from 2015 as depicted in \autoref{fig:lit_bar}, and (2) the distribution of publications across various venues as shown in \autoref{fig:lit_pie}.

\textbf{Flow of the review article}\\
The remainder of this paper is structured as follows. In the subsequent sections, we conduct a comprehensive analysis of a decade of literature in the domain of hashtag recommendation, examining it through multiple dimensions. \autoref{sec:modalities} explores modalities, focusing on unimodal and multimodal approaches. \autoref{sec:problem_formulation} delves into problem formulations, including ranking, classification, keyphrase extraction, sentence matching, generation, and link prediction. \autoref{sec:filtering} discusses filtering approaches, while \autoref{sec:methods} reviews methodological advancements. \autoref{sec:datasets} provides an overview of datasets curated by various researchers in this field, and \autoref{sec:evaluation} examines evaluation strategies and metrics. \autoref{sec:discussion} addresses challenges and real-world applications, highlighting practical considerations in a wide range of sectors and downstream tasks. Finally, in \autoref{sec: conclusion_and_fw}, we conclude the survey by summarizing key findings and outlining prospective future research directions.
\section{Modality-based Hashtag Recommendation}
\label{sec:modalities}
The recommendation of suitable hashtags for microposts encompassing multimodal content stands as a pivotal challenge for numerous SNS. The accuracy of
multimodal hashtag recommendation algorithms relies heavily
on the comprehension of multimodal information, user historical
information, and the reasoning ability based on such information. Multimodal hashtag recommendation aims to
assist users in categorizing content creation while facilitating
the discovery of communities with similar interests or experiences. Effective multimodal hashtag recommendation can enhance the quality of SNS platforms, increase user engagement, and improve the browsing experience.

Hashtag recommendation systems leverage various modalities to understand content and suggest relevant hashtags. These systems can be broadly categorized based on modalities they utilize: unimodal (using a single modality) and multimodal (combining multiple modalities). Given that roughly 85\% of social media data is unstructured\footnote{https://www.business-standard.com/article/technology/-85-of-world-s-data-is-unstructured-106100301029\_1.html}, effectively tagging this content is crucial for information retrieval and classification. \autoref{table:modality} presents a categorization of research papers based on the modalities they address, along with the corresponding methods employed within each category.
\begin{table}\footnotesize
\centering
  \caption{Categorization of Methods for Hashtag Recommendation Across Modalities}
  \label{table:modality}
  \begin{tabular}{cccc}
    \toprule
    \textbf{Modality} & \textbf{Methods} & \textbf{Papers }\\
    \midrule
    Text & Transformers & \cite{cantini2021learning,kaviani2020emhash,jain2024nlp}\\
     & LSTM & \cite{li2016hashtag,gong2016hashtag,li2016tweet}\\
    & Semantic and Topic Models & \cite{cantini2021learning,dey2017semtagger,gong2015hashtag,kumar2019fully}\\
    & Attention Mechanisms & \cite{shen2019hashtag,ma2018temporal,huang2016hashtag,zhang2021howyoutagtweets}\\
    & Temporal and Dynamic Solutions & \cite{mireshghallah2022non,mireshghallah2023simple,kowald2017temporal}\\
    \midrule
    Image & CNNs & \cite{park2016harrison,hachaj2020image}\\ 
    & Graph-based Modeling & \cite{kolyszko2024dynamic,chen2021tagnet} \\
    & Multi-Label Optimization &  \cite{hachaj2020image,jocic2017image}\\
    & Dynamic Learning &  \cite{kolyszko2024dynamic}\\
    & Efficiency-Focused Techniques & \cite{kurunkar2022image}\\
    & Attention Mechanisms &  \cite{wu2018hashtag}\\
    \midrule
    Image + Text & Co-Attention Mechanisms & \cite{zhang2019hashtag,zhang2017hashtag,bansal2022hybrid}\\
    & Multimodal Transformers &  \cite{khalil2023cross,feng2023tnod}\\
    & Graph-based Models &  \cite{khalil2024mrlkg}\\
    \midrule
    Video & Attention Mechanisms & \cite{yang2020sentiment,yu2023generating}\\
    & Generative Models & \cite{yu2023generating}\\
    & Graph-based Models & \cite{bansal2024hybrid,wei2019personalized,li2019long}\\
    & Hybrid Models & \cite{bansal2024hybrid,bansal2022hybrid}\\
    \bottomrule
  \end{tabular}
\end{table}
\subsection{Unimodal Hashtag Recommendation}
Unimodal hashtag recommendation refers to the process of generating relevant hashtags by analyzing a single type of data modality, such as text or images, independently. This approach is particularly effective when the input content is dominated by one modality, enabling specialized models to extract meaningful patterns and features from that specific data type. In this section, we review literature from 2015 onwards, first covering textual modality and then visual modality. Text-based methods are widely studied due to the prevalence of textual content on social media platforms, while image-based approaches address the growing use of visual content, such as photos and graphics, in user-generated posts.
\subsubsection{Text-based Hashtag Recommendation}
Textual content dominates social media, making text-based hashtag recommendation a widely studied area. While manual tagging is cumbersome, automated hashtag recommendation systems offer assistance by suggesting relevant hashtags. However, traditional tagging approaches focus on domain-specific, long-form text, posing challenges for open-domain tagging of short and informal nature of social media posts.
Short text, common on social media, presents unique difficulties. Its brevity, informal language, and poor composition \cite{guo2013linking,garcia2010towards} hinder effective feature extraction using conventional statistical methods designed for long-form text \cite{dovgopol2015twitter,belem2017survey}.  Furthermore, open-domain hashtag recommendation requires broader knowledge and computational resources than domain-specific approaches \cite{jayaratne2017content}.

Text-based hashtag recommendation models analyze the textual content of social media posts to predict relevant hashtags. These models must address the informal nature of social media language, adapt to domain-specific terminology, and remain current with the evolving trends in hashtag usage. A key factor differentiating text-based approaches is the availability of linguistic resources, which impacts the choice of techniques employed. 
The ultimate objective is to predict relevant hashtags that enhance the discoverability, categorization, and searchability of social media content, while adapting to real-world constraints, such as noisy data, evolving hashtags, and user intent mismatches. Unlike multimodal systems, text-only approaches seek to leverage semantic depth and contextual representation as their primary tools. We now examine research papers focused on hashtag recommendation for text in high-resource languages, followed by those addressing low-resource languages. This distinction allows us to explore the unique challenges and advancements in each category, highlighting the disparities in resource availability and the tailored approaches developed to address them.
\paragraph{High-Resource Languages}
Research on text-based hashtag recommendation increasingly focuses on high-resource languages such as English \cite{,zhang2019hashtag,wang2019microblog} and Chinese \cite{zhang2019hashtag,ma2018temporal,javari2020weakly,kou2018hashtag,mao2022attend}, leveraging the availability of extensive datasets and powerful pre-trained language models. English, for example, dominates platforms such as X, encompassing almost 53\% of the total volume of tweets\footnote{https://semiocast.com/top-languages-on-twitter-stats/}. This focus has likely been facilitated by progress in processing these languages, as well as the availability of extensive English and Chinese language resources. 
These works benefited from the advancement in the processing of these languages and the presence of
rich English and Chinese-based knowledge resources.

In the context of high-resource languages on social media, text-based hashtag recommendation involves analyzing the semantic and contextual information within textual content. This modality sets itself apart from multimedia data (images, audio, or video) due to several defining characteristics and challenges:

\textbf{Characteristics of Text:}
Text data often carries explicit semantic information, making it more aligned to the linguistic nature of hashtags. Descriptive captions, keywords, and hashtags inherently rely on textual semantics for meaning representation.

\textbf{Challenges of Text-based Hashtag Recommendation:}
\begin{itemize}
\item Polysemy and Ambiguity: A word in the text (e.g., fall) might have multiple interpretations depending on its context (e.g., seasonal, action-based).
\item Length and Noise: Social media posts are often short, informal, and noisy (e.g., abbreviations, spelling errors, emojis), complicating semantic understanding.
\item Domain-Specificity: Hashtags often reflect cultural or temporal trends (e.g., \#OOTD or \#MetGala), requiring models to understand evolving linguistic patterns.
\end{itemize}
\begin{enumerate}
\item \textbf{Short, Noisy, and Informal Textual Data:} Social media posts are typically short (tweets capped at 280 characters) and often include informal elements such as slang, abbreviations, emojis, and misspellings \cite{kumar2019fully,dovgopol2015twitter}. The unstructured and noisy nature of this data makes traditional approaches such as rule-based systems or simple statistical models ineffective.
\item \textbf{Contextual Representation:} Unlike images or videos, where spatial or temporal features define context, text requires understanding word sequences, semantic relationships, and syntax to capture the meaning of posts \cite{li2016hashtag,shen2019hashtag}. This can include subtle linguistic nuances, idiomatic expressions, and implicit user intent.
\item \textbf{Dynamic Trends:} Hashtag usage evolves rapidly, with specific hashtags trending for only brief periods. Text-modality systems must adapt to these shifts in real time, accounting for changes in context and hashtag meaning \cite{ma2018temporal,mireshghallah2023simple}.
 \item \textbf{Cold Start and Long-Tail Issues:} Many hashtags are infrequent or newly introduced, leading to a sparse distribution.  Text-only data can exacerbate this due to the lack of complementary information from other modalities \cite{kumar2019fully,shen2019hashtag,yang2019self}.
\item \textbf{Multilingualism:} Social media often features multilingual content, posing challenges for semantic modeling and hashtag prediction, including tokenization of mixed-language text and code-switching \cite{dovgopol2015twitter}.
\end{enumerate}

\textbf{Suitable Methods for Text-based Hashtag Recommendation}
The unique characteristics of text data make certain methods particularly effective:
\begin{enumerate}
\item Contextual Understanding with Transformers: Transformer-based models such as Bidirectional Encoder Representations from Transformers (BERT) outperform simpler models due to their ability to capture intricate semantic relationships and contextual nuances in text \cite{cantini2021learning,kaviani2020emhash}. Their use of self-attention mechanisms enables the system to weigh word importance relative to the overall context, making them highly suitable for tasks such as hashtag prediction, where capturing the entire post’s meaning is crucial.
\item Sequence Management with LSTMs: Long Short Term Memory (LSTMs) and their variants are effective in modeling sequential dependencies and capturing long-range contextual information in social media posts \cite{li2016hashtag,gong2016hashtag,shen2019hashtag}. This differs from image-based or audio data, where temporal coherence or object recognition might play a larger role.
\item Limited Relevance of Graph-based Models: Although GNNs are effective for image or multimodal recommendation, where they handle hashtag co-occurrence and similarity between hashtags, their role is less prominent for purely text-modality recommendation. This is because textual content rarely benefits from graph-based interactions unless integrated with network-level features or user interactions \cite{huang2016hashtag,zhang2021howyoutagtweets}.
\item Topic Models for Simplicity: Topic models such as Latent Dirichlet Allocation (LDA) provide an interpretable, resource-efficient way to extract semantic topics from text and align them with hashtags \cite{dey2017semtagger,gong2015hashtag}, but they do not perform well with noisy, dynamic data, making them less suitable compared to neural models such as transformers.
\item Temporal Adaptation with Lightweight Models: Non-parametric dense retrieval and cognitive decay models address dynamic hashtag trends, enabling real-time adaptation to hashtag distribution shifts without the computational overhead of retraining \cite{kowald2017temporal,mireshghallah2023simple,mireshghallah2022non}.
\item Semantic Mapping for Low-Resource Contexts: Few-shot or zero-shot learning maps text embeddings to hashtag spaces, enabling use of pre-trained embeddings for recommending unseen or infrequent hashtags \cite{kumar2019fully,yang2019self}. These methods are uniquely suited to text-based recommendation, where embeddings form the backbone of semantic understanding.
\end{enumerate}

\textbf{Specific Techniques and Examples}
Several techniques have been employed for text-based hashtag recommendation:
\begin{enumerate}
\item Deep Learning:
\begin{itemize}
\item LSTM Variants: Effective for modeling sequence-level dependencies and capturing word correlations in short posts \cite{li2016hashtag,gong2016hashtag,li2016tweet}.
\item Transformer-based Models: Widely adopted for their superior capacity to handle contextual, semantic, and syntactic relationships in text \cite{cantini2021learning,kaviani2020emhash,jain2024nlp}.
\end{itemize}
\item Attention Mechanisms:
\begin{itemize}
\item Self-Attention: Enhances understanding of long-range word relationships, especially in transformer models for text \cite{shen2019hashtag,ma2018temporal}.
\item Hierarchical Attention: Integrates tweet content and external memory (user history) for personalized hashtag prediction \cite{huang2016hashtag,zhang2021howyoutagtweets}.
\end{itemize}
\item Semantic and Topic Modeling:
\begin{itemize}
\item Topic Models: Lightweight approaches such as LDA or its neural extensions for thematic tagging from noisy text \cite{dey2017semtagger,gong2015hashtag}.
\item Semantic Mapping: Embedding-based methods that map sentences to hashtags \cite{cantini2021learning,kumar2019fully}.
\end{itemize}
\item Temporal and Dynamic Solutions:
\begin{itemize}
\item Non-Parametric Dense Retrieval: Adapts hashtag recommendations to recent trends without retraining models \cite{mireshghallah2023simple,mireshghallah2022non}.
\item Temporal Decay Models: Incorporates time-sensitive hashtag trends through cognitive-inspired learning frameworks \cite{kowald2017temporal}.
\end{itemize}
\end{enumerate}
Text-based hashtag recommendation for high-resource languages on social media presents unique challenges, due to noisy, short-text dynamics, and semantic complexity. Unlike other modalities, text requires models capable of capturing intricate linguistic structures and evolving trends. Transformer-based models, LSTMs, and semantic embedding techniques dominate the space, effectively addressing challenges such as contextual representation, temporal drift, and cold start issues. Future research directions include integrating user personalization and handling multilingual content. For a detailed discussion of specific methods, refer to Section \ref{sec:methods}. 
\paragraph{Low-resource Languages}
These methods face challenges due to limited labeled data and the scarcity of effective pre-trained language models. They often rely on techniques such as transfer learning, cross-lingual transfer, or exploiting available resources in related high-resource languages to address the data scarcity issue. The scarcity of written texts, audio recordings, and other digital resources for low-resource languages, coupled with noisy or incomplete data, makes direct application of high-resource language hashtag recommendation methods infeasible. Developing linguistic knowledge for these languages often requires specialized expertise or native speaker proficiency. 

Recommending hashtags for textual content can be framed as a text categorization problem \cite{dogra2022complete,lei2020tag,li2022unified,li2023integration}. However, while text categorization in low-resource Indic languages has seen some attention \cite{pathak2022muboost,rehman2023user,sanghvi2023fine}, hashtag recommendation for such languages remains relatively unexplored \cite{zhang2022twhin}. This is a significant gap, given the increasing popularity of social media platforms such as X in regions where low-resource languages are prevalent. For example, India represents a substantial X user base, with a significant portion of users tweeting in languages other than English. This trend highlights the growing need for hashtag recommendation systems that cater to diverse linguistic communities. Local content creators, brands, language learners, and researchers studying multilingualism all face challenges in finding relevant hashtags in low-resource languages. Moreover, the challenges associated with low-resource languages—such as limited annotated data, morphologically rich structures, and sparse training datasets—remain underexplored.
\begin{figure}[h]
\includegraphics[width=\linewidth]{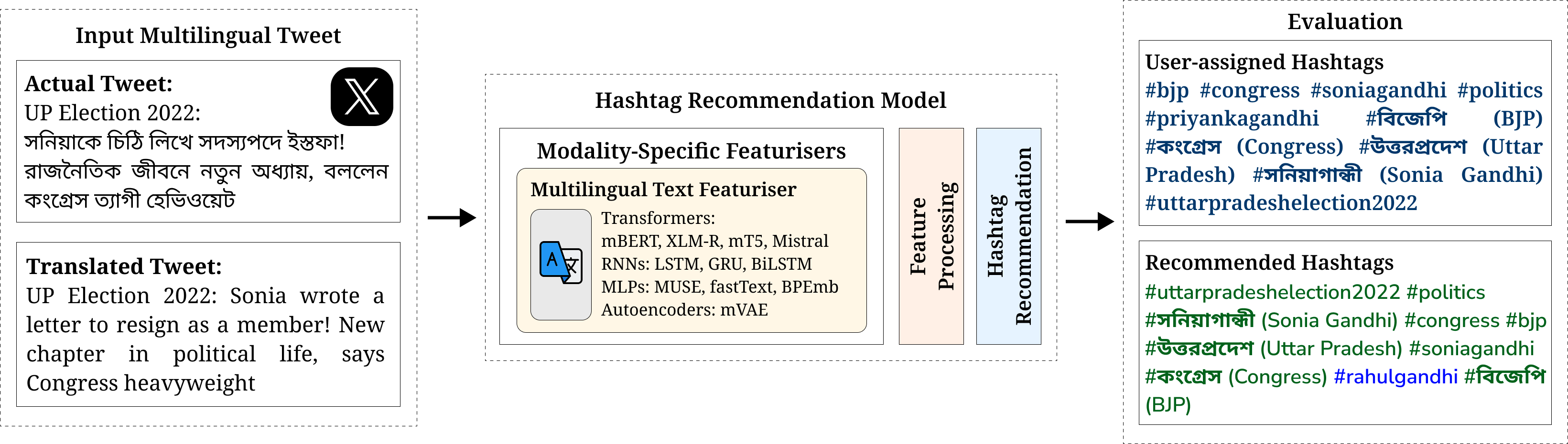}
\caption{Overview of Hashtag Recommendation System for Multilingual Microblogs}
\label{fig:multilingual_htrc}
\end{figure}
\autoref{fig:multilingual_htrc} presents an automated system for multilingual hashtag recommendation. It processes user-generated tweets in their native language, utilizing a ``Multilingual Text Featurizer" to extract features through Transformers, RNNs, MLPs, and Autoencoders. These features inform a ``Hashtag Recommendation Model", which suggests relevant hashtags. The system's performance is evaluated by comparing these suggestions to user-assigned hashtags, employing metrics such as precision, recall, and F1-score. This approach facilitates content organization and discoverability for low-resource languages on social media. 

\begin{table}[h]\footnotesize
\centering
  \caption{Categorization of Papers on Hashtag Recommendation by Language Coverage}
\label{table:multilingual_htrc}
  \begin{tabular}{cccc}
    \toprule
    Paper & No. of Languages & Languages\\
    \midrule
    Zhang \textit{et al.} \cite{zhang2022twhin} & 50 &Indic and Non-Indic\\
    Bansal \textit{et al.} \cite{bansal2024multilingual} & 8 & Indo-Aryan, Dravidian Language, English\\
    Walunj \textit{et al.} \cite{walunjtag} & 1 & Marathi \\
    Alagha \textit{et al.} \cite{alagha2020tag}  & 1 & Arabic\\
    \bottomrule
  \end{tabular}
\end{table}
\autoref{table:multilingual_htrc} summarizes existing work in this area.
Zhang \textit{et al.} \cite{zhang2022twhin} developed TwHIN-BERT, a socially enriched multilingual transformer trained on 7 billion multilingual tweets, incorporating X engagement data (e.g., retweets, likes) as a social objective alongside textual inputs. This model represents a significant advancement in tailoring transformers to informal, crowd-sourced, and noisy datasets while enabling multilingual generalization. By framing hashtag prediction as a multi-class classification problem, TwHIN-BERT leverages a large corpus of tweets and social interactions, utilizing Approximate Nearest Neighbor search to identify socially similar tweet pairs. However, it does not explicitly address individual user interests, language usage styles, topical interests, or linguistic relationships within language families. Despite these limitations, TwHIN-BERT enhances text representation for noisy user-generated content through in-domain training and social engagement data. Its relevance lies in its focus on informal text and hashtag prediction benchmarks, although it lacks domain specialization and emphasis on high-resource languages. This work aligns with the broader trend of developing multilingual models enriched with social network features, pushing the boundaries of transformer applications in socially contextualized environments. Walunj \textit{et al.} \cite{walunjtag} designed a tag recommendation system for Marathi news articles using multi-label classification with Binary Relevance (One vs. Rest). This approach trains a separate binary classifier such as logistic regression or SVM for each tag, predicting its relevance to a given article. Each tag is treated as an independent binary classification problem. Specifically, after pre-processing the text data, numerical feature vectors are generated using techniques such as Term Frequency-Inverse Document Frequency (TF-IDF) or word embeddings. For model training, One vs. Rest method is employed. This involves creating a binary label vector for each tag, indicating its presence or absence in the training documents. To tag new documents, the presence or absence of each tag is predicted using these trained classifiers, and their outputs are combined using thresholding or ranking to select the most relevant tags. While the One vs. Rest technique can handle numerous tags and offers flexibility in classifier choice, it may be less efficient for very large datasets or high-dimensional feature spaces. Tagging Arabic text presents challenges due to the complexities of the Arabic language and a scarcity of Arabic knowledge resources. To this end, Alagha \textit{et al.} \cite{alagha2020tag} utilized Latent Semantic Analysis (LSA) to identify underlying concepts within Wikipedia. The authors generated similarity scores between documents, concepts, and terms, retrieving relevant Wikipedia articles that best correspond to a given text, and deriving tags from titles and categories that provide both specific and general topic coverage. These selected tags were then ranked based on factors such as title-text overlap, article ranking, and category frequency within news articles.
Bansal \textit{et al.} \cite{bansal2024multilingual} developed an automated system to recommend hashtags for tweets in low-resource Indic languages by employing a graph-based deep neural network. \autoref{table:multilingual_htrc} presents a categorization of research papers based on number and linguistic diversity of languages covered when recommending hashtags for user-uploaded content. 
\subsubsection{Image-based Hashtag Recommendation}
Image-based hashtag recommendation systems help users tag their posts using techniques to extract visual characteristics and associate them with relevant hashtags. This modality relies exclusively on semantic information derived from visual content, presenting distinct challenges compared to text-based or multimodal approaches. Research in this area has evolved from CNN-based frameworks to advanced transformer architectures and graph-based methods, with many methods leveraging attention mechanisms. Unlike text, which directly encodes semantic meaning through structured words and phrases, images provide unstructured pixel-level data that must be interpreted by computational models to infer meanings, contexts, and abstract associations. 
The visual representation of the real world is universal (think of an image depicting a cat or apple) and such a
representation can be understood by every human; as a further advantage, images are often
available for free on the Internet and they are of good quality. Finally, it is relatively easy to
verify an image and to uniquely assign a meaning to each image.

\textbf{Characteristics of Images:}
Images primarily convey visual semantics, which are rich in contextual and aesthetic cues (e.g., style, content) but lack explicit categorical information. Visual elements such as colors, objects, or scenes correlate to specific hashtags, often subtly tied to sociocultural contexts.

\textbf{Challenges of Image-based Hashtag Recommendation}
Image-based hashtag recommendation faces unique challenges stemming from the nature of visual data:
\begin{enumerate}
\item Semantic Gap: Images inherently encode low-level, raw data (pixel values, colors, edges), whereas hashtags require high-level semantic understanding (abstract themes, emotions, or social trends). Bridging this semantic gap is a central issue \cite{park2016harrison,hachaj2020image}. It is difficult to map low-level pixel information to high-level semantic concepts (e.g., linking a sunset photo to \#Travel or \#Wanderlust).
\item Polysemy and Ambiguity: A single image can evoke multiple interpretations that vary depending on the viewer's perspective or the social platform's context. For example, a photo of a sunset could be relevant to hashtags such as \#nature, \#travel, or \#photography. This multi-label nature adds complexity not commonly faced in text-based hashtag recommendation \cite{hachaj2020image,kolyszko2024dynamic}.
\item Noisy and Imbalanced Data: In social media, hashtags are user-generated and often skewed towards popular trends, creating imbalanced datasets where dominant hashtags may not accurately reflect image content \cite{park2016harrison,jocic2017image}.
\item Evolving Hashtag Trends: Hashtags change frequently due to trends and cultural events, requiring dynamic and adaptive models \cite{kolyszko2024dynamic}.
\item Visual Abstractions and Context: Hashtags often depend on subjective and cultural interpretations of image context (e.g., identifying trends or symbolic imagery).
\item Heterogeneity in Image Quality: Low-resolution, blurry, or noisy images make visual representation extraction challenging.
These challenges distinguish image-based hashtag recommendation from text-based approaches, which benefit from structured linguistic patterns, and from video/audio, which incorporate temporal dimensions.
\end{enumerate}

\textbf{Suitable Methods for Image-based Hashtag Recommendation}

The specific characteristics of image data necessitate specialized methods:\\
\begin{enumerate}
\item Feature Extraction with Deep Learning
\begin{itemize}
  \item Convolutional Neural Networks (CNNs): CNNs effectively extract hierarchical image features (textures, edges, patterns). Pre-trained CNNs (ResNet, VGG, Inception) are often fine-tuned on social media datasets \cite{park2016harrison,hachaj2020image}. 
CNNs excel in classifying fixed categories of objects within images but may struggle with abstract or ambiguous contexts that go beyond visible objects.
\item Graph-based Modeling:
Hashtag co-occurrence is an important characteristic in social media. Graph-based approaches, such as those using Graph Convolutional Networks (GCNs), are instrumental in modeling relationships between hashtags and enabling incremental adaptation to newly emerging trends or rarely seen hashtags. For example, Incremental GCNs have been shown to handle evolving trends by incorporating dynamic hashtag correlations effectively \cite{kolyszko2024dynamic}.
\item Vision Transformers and Transfer Learning:
Vision Transformers (ViTs) offer powerful long-range dependency modeling. Self-supervised models such as CLIP can bridge the data gap by projecting image embeddings into semantic spaces, though their adoption is still emerging.
\item Attention Mechanisms:
These mechanisms are used to capture finer semantic dependencies within an image. Attention layers can focus on local regions of an image and their global relationships (object-to-scene relationships), improving hashtag prediction in contextually ambiguous cases \cite{wu2018hashtag}.
\end{itemize}
\item Multi-Label Optimization:
Hashtags are inherently multi-label, as a single image may correspond to multiple appropriate hashtags. Optimization techniques such as binary cross-entropy loss, multi-label attention mechanisms, and class rebalancing methods (e.g., focal loss) help address class imbalance and improve prediction accuracy \cite{hachaj2020image,jocic2017image}.

\item Dynamic Learning:
The rapidly evolving nature of hashtags makes models trained on static datasets fragile over time. Incremental learning paradigms, such as those employing GCNs or continual learning, are critical for adapting dynamically to emerging hashtags and evolving platform trends \cite{kolyszko2024dynamic}.
\item Efficiency-Focused Techniques:
The deployment of hashtag recommendation systems on social media platforms demands speed and scalability. Techniques such as model pruning, quantization, and knowledge distillation are used to balance accuracy and inference efficiency \cite{kurunkar2022image}.
\end{enumerate}
Image-only hashtag recommendation for social media is distinguished by its reliance on visual feature extraction, posing challenges such as semantic gap, hashtag ambiguity, and dataset noise. Effective methods for this modality seek to derive rich semantic embeddings from images, address multi-label classification problems, and dynamically adapt to changing social media trends. CNN-based feature extraction, graph-based representation learning, and multi-label optimization are the cornerstones of current methodologies. However, emerging methods such as ViT, contrastive learning, and incremental learning offer promising directions for addressing limitations of static CNNs and evolving hashtag trends. Achieving robust hashtag recommendation at scale will depend on optimizing algorithmic performance while addressing the unique data characteristics and constraints of social media applications.
For detailed discussions of specific methods (CNNs, GCNs, Transformers), please refer to Section \ref{sec:methods}.

While unimodal approaches focus on a single data type, such as text or images, they fail to capture semantics and contextual richness embedded within social media posts, which frequently combine multiple modalities. 
\subsection{Multimodal Hashtag Recommendation}
Multimodality, in the context of hashtag recommendation, refers to the synergistic combination of information derived from multiple distinct data modalities. This fusion is predicated on the principle that integrating information from diverse data sources, such as text, images, and videos, provides a more complete and accurate reflection of the underlying content to generate relevant hashtags. For instance, a social media post may include an image accompanied by a caption. While the image provides visual context, the text offers semantic details that are not immediately apparent from the visual content alone. Multimodal methods aim to bridge this gap by combining features extracted from constituent modalities, enabling a more comprehensive understanding of the post. 
In this section, we review papers that employ multimodal approaches, beginning with methods that simultaneously utilize text and image data. We then explore advancements in micro-video-based hashtag recommendation, which represents a more complex and emerging area of research.
\subsubsection{Image and Text-based Hashtag Recommendation}
Social media platforms are predominantly characterized by the prevalence of multimedia content. The vast volume of data generated on these platforms encompasses diverse modalities, including both visual and textual elements, which collectively contribute to the richness of the information ecosystem. Multimodal hashtag recommendation systems leverage the complementary information present in text and image data to generate more relevant hashtags. These methods usually involve jointly embedding image and text representations and then using this combined representation for hashtag recommendation. Combining these modalities presents unique characteristics and challenges compared to unimodal approaches. When incorporating multimodal data, such as text and images, distinct challenges and characteristics emerge owing to the differences in data modalities, their semantic representations, and the ways these modalities interact. 
\begin{figure}
\includegraphics[width=\textwidth]{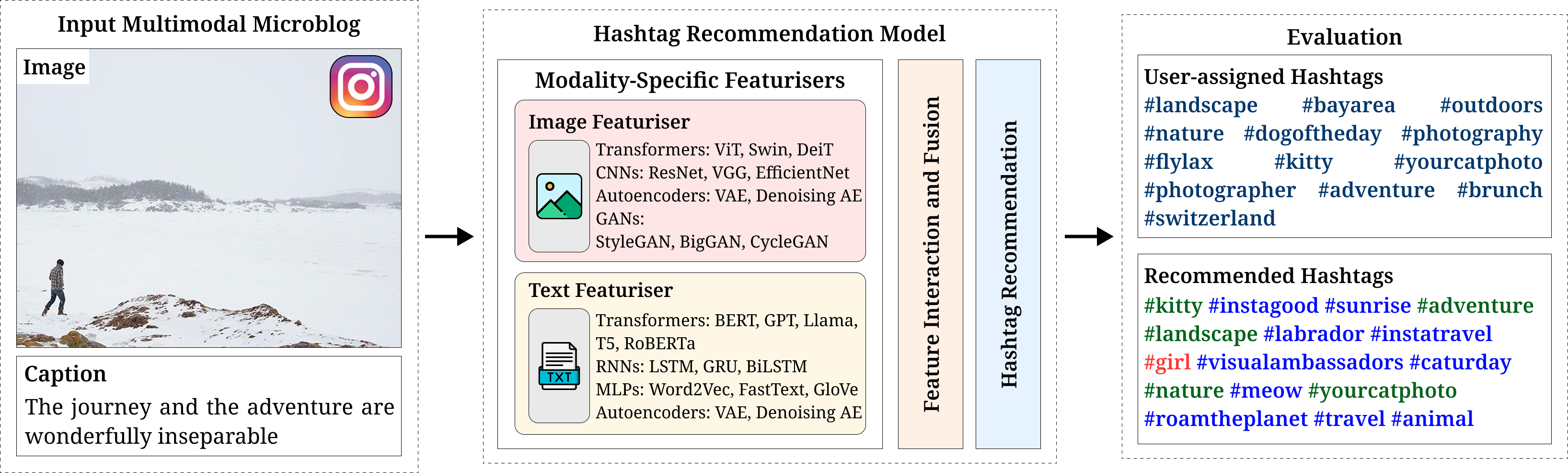}
\caption{Overview of Hashtag Recommendation for Multimodal Microblogs}
\label{fig:multimodal_htrc}
\end{figure}
\autoref{fig:multimodal_htrc}  presents a schematic overview of a sophisticated hashtag recommendation system designed for multimodal microblogs composed of three primary modules: input processing, hashtag recommendation, and evaluation. The features extracted from the different modalities are then subjected to ``Feature Interaction and Fusion". This process aims to combine the multimodal information effectively, potentially using techniques such as attention mechanisms or graph neural networks. The fused feature representation is then fed into the ``Hashtag Recommendation" module to predict relevant hashtags which are then compared against a set of user-assigned hashtags for the given microblog to discern the effectiveness of the devised system.

\textbf{Challenges and Characteristics of Text+Image-based Hashtag Recommendation}

Integrating text and images introduces complexities related to semantic representation, contextual ambiguities, and cross-modal interactions:
\begin{itemize}
\item Interdependence: Text and images often complement each other in a post. Captions provide context that imagery lacks, while visual elements may reinforce or expand the scope of textual descriptions.
\item Alignment: Aligning textual and visual modalities into a unified representation is a core obstacle. For example:
\begin{itemize}
\item Heterogeneous Representations: Textual data is sequential and embedded in high-dimensional linguistic spaces, while image data is spatial and represented via pixel-level or feature-level embeddings. Aligning these heterogeneous representations is a key challenge. 
\item Modal Imbalance: Posts may have richer textual descriptions than visual data, or vice versa, resulting in unbalanced contributions from each modality to hashtag prediction.
\item Context Modeling: Effective hashtag recommendation requires understanding the interaction between text and images, ensuring semantic coherence across modalities.

Effective hashtag recommendation requires models to capture interdependencies and semantic coherence across modalities (how captions relate to the objects depicted in images).
\end{itemize}
\end{itemize}
\textbf{Suitable Methods for Text+Image-based Hashtag Recommendation}
The suitability of certain methods for hashtag recommendation is inherently tied to how well the techniques leverage the characteristics of text and image data to address the aforementioned challenges:
\begin{enumerate}
\item Attention Mechanisms: Co-attention mechanisms \cite{zhang2019hashtag,zhang2017hashtag,ma2019co,bansal2022hybrid} aligns textual and visual data by modeling their mutual dependencies, capturing how objects or scenes relate to hashtags in captions.
\item Multimodal Transformers: Shared latent spaces (e.g., LXMERT \cite{khalil2023cross} or multimodal CLIP architectures) embed text and images into unified representations, directly addressing cross-modal alignment. TNOD \cite{feng2023tnod}) unify representations at granular and contextualized levels effectively. Joint embeddings using cross-modal pretraining incorporate vision-language alignment for hashtag recommendation.
\item Graph-based Learning: Multimodal graphs align text-image pairs and contextualize hashtags through global relational reasoning (keyword-guided GCNs in \cite{khalil2024mrlkg}).
\end{enumerate}
\textbf{Specific Techniques and Examples}
Aligning two disparate modalities requires robust techniques for fusing sequential text and spatial image data while preserving their unique characteristics and interactions. Common techniques include:
\begin{itemize}
    \item \textbf{Joint Embeddings:}  Using transformers (e.g., LXMERT) or cross-attention.
    \item \textbf{Co-attention Networks:} For interactive learning between text and images \cite{zhang2019hashtag,zhang2017hashtag,bansal2022hybrid}.
    \item \textbf{Multimodal Transformers:} Unifying representations at granular and contextualized levels \cite{khalil2023cross,feng2023tnod}.
    \item \textbf{GCNs:} Combining multimodal inputs for context-aware recommendations \cite{khalil2024mrlkg}.
\end{itemize}
Multimodal hashtag recommendation systems on social media uniquely benefit from the complementary nature of text and image data. Text offers explicit semantic cues, while images provide implicit contextual richness.
These systems face unique challenges—particularly in cross-modal alignment, hashtag polysemy, and contextual coherence. Cross-modal techniques, such as co-attention and transformer-based joint embedding models, integrate text and image data effectively which excel in pairing sequential and spatial data for robust predictions, while graph reasoning handles relational and contextual challenges.
While these methods perform well on static data, open challenges such as evolving hashtags, dynamic social trends, domain drift, dataset variability and the subjective nature of visual semantics remain areas for future exploration. For detailed discussions of specific methods (Transformers, CNNs, GCNs), refer to Section \ref{sec:methods}.
\subsubsection{Video-based Hashtag Recommendation}
\begin{figure}
\includegraphics[width=\textwidth]{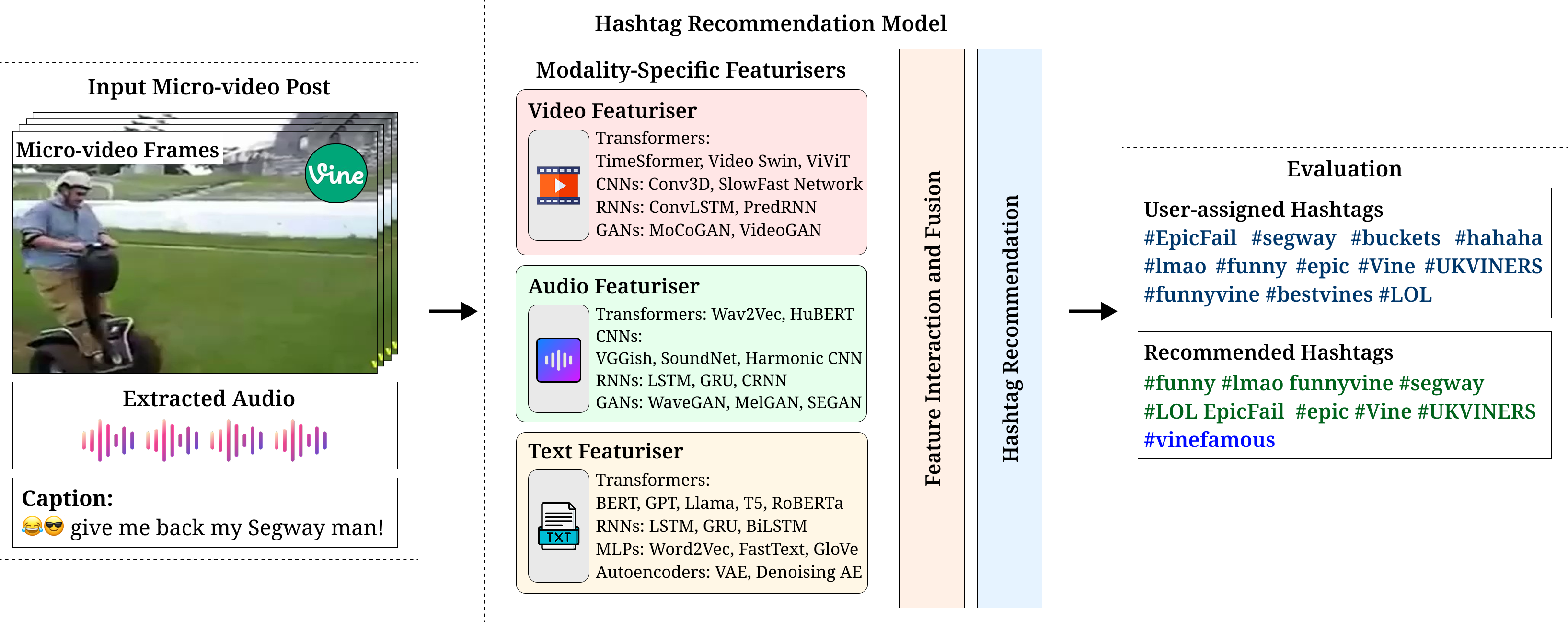}
\caption{Overview of Hashtag Recommendation for Micro-videos}
\label{fig:microvideo_htrc}
\end{figure}

Hashtag recommendation systems for video-based content have gained increasing importance due to the explosion of video-sharing platforms and their reliance on hashtags for discoverability, engagement, and content categorization. Videos are inherently multimodal, involving visual and auditory components alongside textual metadata (e.g., video descriptions, user-generated captions) and user-related factors (e.g., historical preferences, engagement patterns). This multimodality sets video-based hashtag recommendation apart, necessitating approaches that effectively fuse and align heterogeneous data streams. 
\autoref{fig:microvideo_htrc} depicts a broad overview of automated hashtag recommendation system for micro-videos. It analyzes the micro-video's frames, audio, and caption using modality-specific featurizers including Transformers, CNNs, and RNNs besides many others. These features are then processed and fused to recommend relevant hashtags, which are evaluated against user-assigned hashtags for performance analysis.
 This section examines the characteristics of multimodal video data, suitable methods, and specific techniques employed in this domain.
 
\textbf{Characteristics of Multimodal Data in Videos}
\begin{enumerate}
\item Visual Data (Frames, Motion Features):
Videos provide rich visual information extracted from individual frames or sequences of frames. These features capture objects, actions, and scene context relevant for hashtag prediction. Unlike static images, videos offer temporal dynamics such as motion and progression, which add complexity but also enable richer understanding.

\textbf{Unique Challenges for Visuals:}
\begin{itemize}
\item Computational demand grows with the need to process sequences instead of static images.
\item Temporal consistency between frames must be maintained to identify relevant patterns over time.
\item Misalignment between low-level features (e.g., pixel data) and high-level semantics (e.g., actions, abstract concepts) remains a significant obstacle, often termed the semantic gap.
\end{itemize}
\item Audio Data (Speech, Sound Effects, Music):
Audio signals convey additional contextual information, such as dialogue, soundtracks, or even environmental sounds that align with video hashtags.
Unique Challenges for Audio:
\begin{itemize}
\item Audio signals are continuous and non-linguistic, requiring specialized techniques such as spectrogram analysis to transform them into features compatible with text-based models.
\item Background noise and overlapping sounds often obscure important auditory signals, making accurate analysis difficult.
\item Speech data within audio streams requires natural language processing for transcription, creating dependencies between audio and text modalities.
\end{itemize}
\item Textual Metadata:
Videos include user-written captions, platform-generated subtitles, or hashtags. Text data is often more structured and interpretable than visual or audio data, making it a valuable modality for hashtag recommendation.
However, text is often sparse or incomplete, and captions or descriptions may not fully capture the video's content. Metadata may heavily reflect user bias or platform trends, leading to skewed recommendations.
\item User Information and Behavioral Data:
User preferences (historical hashtags, watched videos) and engagement patterns (likes, shares, comments) provide critical context for personalizing hashtag recommendations.
Unique Challenges for User Data:
\begin{itemize}
\item Cold-start issues for new users with little behavioral data.
\item Privacy concerns when leveraging detailed user information.
\end{itemize}
\end{enumerate}
Goals of Video-based Hashtag Recommendation
\begin{itemize}
\item Semantic Alignment: Bridging the semantic gap between raw video/audio features and meaningful hashtags is a core goal. High-level representations must relate visual and audio inputs to abstract concepts.
\item Multimodal Fusion: Effectively combining visual, audio, textual, and user data while preserving complementary information and minimizing redundancies is critical.
\item Real-Time Scalability: Systems must scale to millions of videos on dynamic platforms such as Instagram or TikTok, where trends evolve rapidly, and response times must be instantaneous.
\item Personalization: Customizing hashtags based on individual user profiles and community trends while ensuring fairness and reduced bias across demographics.
\end{itemize}
\textbf{Suitable Methods for Video-based Hashtag Recommendation}
\begin{enumerate}
\item  \textbf{Graph-based Models:} GNNs and GCNs are highly suitable for video-visual data because they can model interactions between videos, hashtags, and users as nodes in a graph. These methods leverage co-occurrence patterns (e.g., hashtags commonly used together for similar video content) to improve hashtag prediction accuracy \cite{wei2019personalized,li2019long,bansal2024hybrid}.
\item \textbf{Attention Mechanisms:} 
Attention-based models, especially Transformers and ViTs, allow the system to focus on the most contextually relevant frames or regions in a video \cite{yang2020sentiment,yu2023generating}. They account for temporal dynamics by prioritizing meaningful sequences or objects across time, addressing the high dimensionality of video frame data.
\item \textbf{Sequential Models:}
Sequential models such as Recurrent Neural Networks (RNNs), LSTMs and Gated Recurrent Units (GRUs), process continuous audio streams by modeling sequential dependencies. Spectrograms (visual representations of audio frequencies) are often used as input features, aligning audio data with visual or textual embeddings.
\item \textbf{Hybrid Architectures:} Combine collaborative filtering, content fusion, and multimodal embeddings for personalization and cold-start handling \cite{bansal2022hybrid,bansal2024hybrid}.
\item \textbf{Generative Models:} Modern sequence-to-sequence (Seq2seq) frameworks (transformer-based generators) are well-suited for creating novel and dynamic hashtags, especially with pretrained Vision-Language Models (VLMs) \cite{yu2023generating}.
\end{enumerate}
Video-based hashtag recommendation presents unique challenges and opportunities across its multimodal components. Emerging innovations, such as generative models and pretrained VLMs, are pushing the state-of-the-art, enabling dynamic and context-aware hashtag recommendations. For detailed discussions of specific methods (GNNs, Transformers, RNNs), please refer to Section \ref{sec:methods}.
The analysis of audiovisual features in videos presents a viable approach for generating recommendations; however, the process becomes computationally intensive when conducted on a frame-by-frame basis. In the domain of audio classification, audio files of specific durations are utilized as input data to categorize sounds into distinct classes. This task, however, poses greater challenges in the context of deep learning compared to image classification, as it lacks the visual component that facilitates feature extraction. Yang \textit{et al.} \cite{yang2022interpretable} investigated the extraction of audio features from an online audio clip-sharing platform, focusing on identifying spoken words and topics to derive relevant tags. While audiovisual features can enhance recommendation systems, the frame-level analysis required for such tasks often renders the process inefficient and resource-intensive.
In contrast to the work of Yu \textit{et al.} \cite{yu2023generating}, who introduced a guided generation model for hashtag creation using multimodal inputs and visual language model-based retrieval signals, Cao\textit{ et al.} \cite{cao2020hashtag} adopted a multiview representation learning framework to extract and integrate feature representations from micro-video modalities. Their approach utilized regularized projections and hashtag embeddings within a neural collaborative filtering framework to generate hashtag recommendations. Building on this foundation, subsequent studies such as Yang \textit{et al.} \cite{yang2020sentiment} and Gupta \textit{et al.} \cite{gupta2023shrse} incorporated sentiment analysis and semantic embeddings of hashtags through weighted concatenation. While these methods demonstrated the ability to recommend sentiment-aware hashtags by leveraging all three micro-video modalities and employing sequential modeling, they predominantly relied on concatenating modality-specific features before projecting them into a latent space. Mehta \textit{et al.} \cite{mehta2021open} proposed a heterogeneous graph structure that connected hashtags based on semantic co-occurrence, videos through shared hashtags, and direct links between videos and their assigned hashtags. They employed a graph convolutional network (GCN)-based node update mechanism to generate micro-video embeddings for hashtag recommendation. Similarly, Li \textit{et al.} \cite{li2019long} addressed the challenge of hashtag long-tail distribution by constructing a hashtag graph enriched with external knowledge and utilizing a pairwise interactive embedding network to model interactions between hashtags, micro-videos, and users. However, their approach represented users by averaging historical hashtags and micro-videos, which neglected the nuanced, modality-specific preferences of users that could significantly enhance the personalization of hashtag recommendations. Liu \textit{et al.} \cite{liu2020user} derived user embeddings from historical hashtags and demographic metadata such as age, gender, location, and country, which were used to guide attention mechanisms at both image and video levels. While this method improved micro-video representations, its reliance on demographic data may lead to inaccuracies in cases where individual user interests deviate from broader demographic trends. Wei \textit{et al.} \cite{wei2019personalized} employed GCNs on a heterogeneous graph comprising users, micro-videos, and hashtags to model their complex relationships for personalized hashtag recommendations. Their approach refined node representations through message passing, enabling the model to learn micro-video features and hashtag embeddings conditioned on user preferences. However, this method did not account for the intricate dynamics of user-user interactions, which could further enrich the recommendation process.
\section{Hashtag Recommendation based on Problem Formulation}
\label{sec:problem_formulation}
The formulation of hashtag recommendation as a computational problem has evolved significantly over the years, and researchers have approached it from various perspectives. In this section, we review papers from 2015 onward that frame hashtag recommendation through various problem formulations, including ranking, classification, keyphrase extraction, sentence matching, generation, and link prediction. A summary of the same has been provided in \autoref{table:problem_formulation}.
\begin{table}\scriptsize
\centering
\caption{Problem Formulation Types}
\label{table:problem_formulation}
\resizebox{\textwidth}{!}{
\begin{tabular}{@{\extracolsep{\fill}} lccccccccc}
\toprule
Formulation Type & Loss &	Papers\\
\midrule
Ranking & Hinge loss, WARP loss, triplet loss, NDCG loss, MAP loss  & \cite{denton2015user,wu2018starspace,park2016harrison,gong2016hashtag,wu2018hashtag,wang2019microblog,zhang2019hashtag,yang2020sentiment,kaviani2020emhash} \\
Classification & Cross Entropy loss, Pairwise loss, Ranking loss & \cite{li2016tweet,li2019topical,peng2019modeling,won2023extra,liu2018fasttagrec,he2022ptm4tag,bansal2024multilingual} \\
Keyphrase Extraction  & BCE,  Sequence labelling loss & \cite{gong2015hashtag,zhang2016keyphrase,zhang2018encoding}\\
Sentence Matching & BCE, Contrastive loss & \cite{zheng2020attentive,li2023code,li2023dual}  \\
Generation & Cross entropy & \cite{wang2019microblog,zheng2021news,mao2022attend, yang2020amnn,chen2017reading,zhang2023relevance,he2023merging}  \\
Link Prediction & BCE, Pairwise Ranking loss & \cite{wei2019personalized,wang2022micro,mehta2021open}\\
\bottomrule
\end{tabular}
}
\end{table}
\subsection{Keyphrase Extraction}
Keyphrase extraction approaches for hashtag recommendation formulate the task as identifying relevant phrases within the source post that can serve as effective hashtags \cite{gong2015hashtag,zhang2016keyphrase,zhang2018encoding}. These approaches aim to pinpoint existing phrases within the post, rather than generating entirely new ones.

\textit{Output:}
$P = {p_1, p_2,..., p_l}$: A set of $l$ extracted keyphrases from the input $x$.

\textit{Mathematical Definition:}
\textit{The goal is to learn a function $f(x)$ that maps $x$ to a set of relevant keyphrases $P$.}
Various techniques can be applied to model keyphrase extraction function $f(x)$
\begin{itemize}
\item N-grams and Part-of-Speech Tagging: These methods can be used to identify candidate phrases based on sequences of words and their grammatical roles. For example, noun phrases are often good candidates for keyphrases.
\item Graph-based Ranking: Similar to TextRank, graph-based methods can represent the text as a graph of words or phrases and use ranking algorithms to identify the most important ones.
\item Supervised Machine Learning: Machine learning models (e.g., sequence labeling models \cite{zhang2016keyphrase} or models incorporating conversational context \cite{zhang2018encoding} can be trained to identify keyphrases. These models learn patterns from labeled data and can be used to predict which phrases are most relevant. Word embeddings \cite{marujo2015automatic} can be incorporated into these models to improve performance.
\end{itemize}
The loss function depends on the specific technique used.  For supervised learning, common choices include:
\begin{itemize}
\item Binary Cross-Entropy: If the task is framed as classifying whether a phrase is a keyphrase or not, binary cross-entropy can be used.
\item Sequence Labeling Loss: For sequence labeling approaches, a loss function appropriate for sequence prediction (Conditional Random Field loss) would be used.
\end{itemize}

As noted by Zhang \textit{et al.} \cite{zhang2016keyphrase}, hashtags themselves can be valuable keywords for keyphrase extraction, but simply extracting them from the source text can be suboptimal. Marujo \textit{et al.} \cite{marujo2015automatic} demonstrated the effectiveness of word embeddings over TF-IDF for keyphrase extraction on tweets. The authors formulated the problem as binary classification and showed that word embeddings in a system such as MAUI \cite{medelyan2009human} perform better than TF-IDF \cite{sparck1972statistical} for keyphrase extraction on general tweets. Zhang \textit{et al.} \cite{zhang2016keyphrase} formulated the problem as a sequence-labeling task which allows extracting keyphrases of arbitrary lengths, without being constrained by some fixed number of classes. Zhang \textit{et al.} \cite{zhang2018encoding} extends the work of \cite{zhang2016keyphrase} by encoding conversational context. 
A key limitation of keyphrase extraction for hashtag recommendation is that it restricts extracted hashtags to those already present in the source text. This approach fails to capture the creative and dynamic nature of hashtag usage, where users often invent new hashtags or combine existing words in novel ways, owing to their background, proficiency level and linguistic style, resulting in suboptimal performance. This motivates the need for generative approaches that can create hashtags beyond the confines of the input text.
\subsection{Sentence Matching}
Sentence matching approaches frame hashtag recommendation as a task of assessing the semantic similarity between a post (or software object) and a candidate hashtag or set of hashtags.  This aims to capture relationships between posts and hashtags, addressing limitations of methods that predict individual tags in isolation.

\textit{Input:}
$x$: Input query (social media post), $h$: A single hashtag.\\
\textit{Output:}
$y$: A binary label indicating whether the hashtag $h$ is relevant to the query $x$.

\textit{Mathematical Definition:}
\textit{The goal is to learn a sentence matching model $f(x, h)$ that predicts the similarity between the input query $x$ and the hashtag $h$.}
$f(x, h)$: Can be modeled using:
\begin{itemize}
\item Siamese Networks: Two encoders (one for $x$ and one for $h$) that map the inputs to a shared embedding space, and the similarity is computed in this space.
\item BERT-based Models: Fine-tune a pre-trained BERT model to predict the similarity.
\end{itemize}
Loss Function:
Binary cross-entropy or contrastive loss can be used.

While sentence matching methods \cite{zheng2020attentive,li2023code,li2023dual} offer improvements over MLC by considering post-hashtag relationships, they struggle to capture complex interdependencies between multiple hashtags and nuanced textual meaning. These approaches are limited by a focus on lexical similarity, failing to capture the core focus  of the content. This motivates exploring alternative approaches such as sequence generation.
\subsection{Ranking}
Hashtag recommendation, similar to search and traditional recommendation tasks, can be formulated as a ranking problem. This involves receiving an input query (post) and returning a ranked list of relevant items (hashtags) based on a scoring or ranking function. These functions aim to maximize the likelihood that relevant hashtags appear at the top of the list, sometimes by directly optimizing for the logit values predicted by the top-k \cite{covington2016deep,wu2018starspace}. 

\textit{Mathematical Definition}:
\textit{The core idea is to learn a scoring function $s(x, h)$ that measures the relevance of hashtag $h$ to a query $x$. The ranking is then obtained by sorting hashtags according to their scores}. This scoring function can be modeled as $s(x,h)=f(x,h; \theta)$ where $f(.)$ is a function parameterized by $\theta$. The function $f(.)$ can take various forms, such as:
\begin{itemize}
\item Linear Model: $f(x,h; \theta) = w^T  \phi(x, h)$ where $w$ is a weight vector and $\phi(x, h)$ is a feature vector representing the interaction between $x$ and $h$.
\item Neural Network: $f(x,h; \theta)$ can be a neural network with parameters $\theta$. This could include architectures such as siamese networks, point-wise networks, or pair-wise networks.
\item Similarity Function: $f(x,h; \theta)$  can compute the similarity between embeddings of $x$ and $h$, such as cosine similarity, dot product, or approximate nearest neighbor. Here, $embedding(x)$ and $embedding(h)$ are vector representations of the query post and hashtag, respectively, which could be generated from a model trained on a related task.
\end{itemize}
The hashtags are then sorted in descending order of $s(x, h)$, and top-k hashtags are recommended.  The loss functions used to train these models include pairwise ranking loss (for example, hinge loss, WARP loss \cite{weston2011wsabie}), triplet loss, and listwise ranking loss (e.g., NDCG loss, MAP loss).
\begin{itemize}
\item Pairwise Ranking Loss: Aims to make the score of a relevant hashtag higher than the score of an irrelevant hashtag. Examples include:
$L = max(0, 1 - s(x, h^+) + s(x, h^-))$ where $h^+$ is a relevant hashtag and $h^-$ is an irrelevant hashtag.
\item Triplet Loss: Aims to make the score of a relevant hashtag higher than the score of an irrelevant hashtag by a certain margin.
$L = max(0, margin - s(x, h^+) + s(x, h^-))$
\item Listwise Ranking Loss: Considers the ranking of the entire list of hashtags. Examples include NDCG loss, MAP loss.
\end{itemize}

Several studies have explored hashtag recommendation using a ranking framework \cite{zhang2019hashtag,kaviani2020emhash,gong2016hashtag,park2016harrison,wu2018hashtag,wang2019microblog,yang2020sentiment,wu2018starspace,denton2015user}. Early work focused on learning scoring functions to rank hashtags based on input features. Park \textit{et al.} \cite{park2016harrison} used visual feature extractors from images combined with multi-label classifiers to calculate the score of each hashtag and provide top-k hashtag recommendations. Other approaches handle multimodal input (image and text) by projecting features into a common representation space and optimizing a pairwise ranking loss \cite{,wu2018starspace,denton2015user}, such as the weighted approximate-rank pairwise (WARP) loss \cite{weston2011wsabie}, as the training objective.

Conventional tag recommendation models \cite{yang2020sentiment,denton2015user,wu2018starspace} project input and tag embeddings into a shared space and learn with pairwise ranking losses. Building on this, more generalized context-tag mapping models merge encoded features to represent the context. At inference time, these models select top-k hashtags nearest to the context embedding in the shared space.  

However, a significant limitation of many standard ranking approaches is their treatment of recommended items in isolation, neglecting the inter-dependencies that often exist between them. This is particularly relevant in hashtag recommendation, where hashtags within a relevant set are often semantically related or co-occur frequently. Despite their effectiveness, these ranking-based approaches fail to explicitly consider these inter-dependencies among generated hashtags. Just as term dependencies are crucial in information retrieval ranking, accounting for hashtag relationships can significantly enhance hashtag recommendation. When tags are interdependent, especially with a given query, it is desirable to incorporate these dependencies into the tag selection process. Choosing a hashtag should not be an independent event but rather a decision influenced by the set of already-recommended hashtags, leading to a more cohesive and semantically meaningful set of recommendations.  This contrasts with conventional information retrieval ranking techniques, which also often neglect such inter-dependencies.
\subsection{Classification}
Classification, a supervised learning approach, trains a classifier to predict a class label for a given instance. In hashtag recommendation, it translates to predicting relevant hashtags for a given post or user. Classification-based hashtag recommendation offers some advantages. The abundance of posts and hashtags provides a vast amount of labeled data for training classifiers, enabling them to learn robust representations \cite{weston2014tagspace}. Furthermore, classification methods require less task-specific engineering compared to approaches such as topic-based recommendation \cite{weston2014tagspace}.

Framing hashtag recommendation as a classification problem presents several challenges. Traditional classification-based methods rely on a predefined set of candidate hashtags \cite{li2016tweet,wei2019personalized,yang2020sentiment,cao2020hashtag}. While this approach works well for static environments, it struggles in dynamic settings where new and unforeseen hashtags emerge constantly. This limitation is not unique to classification; most machine learning techniques face similar challenges, as models are typically constrained to recommending items (or hashtags) they have encountered during training (i.e., those present in their vocabulary). However, classification-based methods are particularly affected because they require the candidate hashtag set to be explicitly defined and fixed during training. Continuously updating and retraining models to accommodate an ever-expanding hashtag vocabulary is computationally expensive and impractical for real-time adaptation, such as in event-oriented hashtag recommendation. To address this limitation, some approaches extend the model's output vocabulary beyond the predefined hashtag set, for example, by incorporating words from the text or leveraging external knowledge sources. However, such extensions often introduce additional complexity and may not fully resolve the challenges of handling emerging hashtags in dynamic environments. Another challenge stems from the nature of hashtags themselves. Hashtags can be noisy labels for classification due to variations in spelling and formatting. The long-tail distribution of hashtag popularity, where a few hashtags are very popular while most have low usage, can bias classifiers towards more frequent hashtags, leading to inaccurate recommendations. Consequently, many classification-based methods restrict the number of hashtag labels, which can limit their ability to capture the full spectrum of hashtag usage in social media.

Classification-oriented approaches can be further categorized into binary, multi-class, and multi-label classification, depending on the nature of the prediction task.
\subsubsection{Binary Classification}
Binary classification predicts whether a single hashtag is relevant to a given post or user.  

\textit{Mathematical Definition}: \textit{The goal is to learn a classifier function $f(x, h)$ that predicts a binary label $y$, indicating relevance (1) or irrelevance (0) of hashtag $h$ to query $x$}. 
The classifier function $f(x, h)$ which predicts the probability of $h$ being relevant to $x$ can be modeled using various techniques such as:
\begin{itemize}
\item Logistic Regression: $f(x, h) = sigmoid(w^T * \phi(x, h))$ where $w$ is a weight vector and $\phi(x, h)$ is a feature vector.
\item Support Vector Machines (SVM): Finds a hyperplane that separates relevant and irrelevant hashtags.
\item Neural Networks: $f(x, h) = g(x, h; \theta)$ where $g$ is a neural network with parameters $\theta$.
\end{itemize}
Binary cross-entropy is a common loss function for this task.

Several studies have framed hashtag recommendation as a binary classification problem. Wei \textit{et al.} \cite{wei2019personalized} used pairwise loss between user-specific micro-video representations and user-specific hashtag representations, constructing triplets of micro-videos, positive hashtags, and negative hashtags for pairwise ranking. Cao \textit{et al.} \cite{cao2020hashtag} framed hashtag recommendation for micro-videos as binary classification problem. The authors optimized pointwise log loss \cite{he2017neural} to predict interaction scores between micro-videos and hashtags. Observed interactions are assigned a target value of 1, and non-interactions are assigned 0. Negative interactions are sampled to pair with each observed interaction. Li \textit{et al.} \cite{li2019long} proposed an interactive model incorporating hashtags, micro-videos, and users simultaneously, using attention mechanisms to filter noise and identify relevant information. The authors aimed to predict a score for each triplet (user, micro-video, hashtag), where a triplet represents a valid interaction if user $u_k$ added hashtag $h_j$ to their posted micro-video. Six negative hashtags were randomly sampled per positive instance. Yang \textit{et al.} \cite{yang2020sentiment} also framed hashtag recommendation as binary classification, optimizing cross-entropy loss to predict video-hashtag interaction scores. The authors paired each positive instance with 100 randomly sampled negative hashtags during testing. Then each method outputs prediction scores for these 101 hashtags.

These prior works, treating video hashtag recommendation as a binary classification problem, select hashtags from a limited candidate set (101 \cite{yang2020sentiment,cao2020hashtag} or 1001 \cite{wei2019personalized}) by computing recommendation scores individually. This approach is generally time-consuming and impractical for real-world applications. 
\subsubsection{Multi-class Classification}
Multi-class classification extends binary classification to handle multiple hashtag categories \cite{li2016hashtag,li2016tweet,peng2019modeling}. Each hashtag category is treated as a distinct class label. 

\textit{Mathematical Definition}:
\textit{The aim is to learn a classifier function $f(x)$ that predicts the probability distribution over the set of hashtag categories. The output is a single class label $y$ representing the most relevant hashtag category for the query $x$ (chosen from a predefined set of categories).}

The classifier function $f(x)$ predicts the probability distribution over the hashtag categories, can be modeled using:
\begin{itemize}
\item Multinomial Logistic Regression: Extends logistic regression to multiple classes.
\item Neural Networks: A neural network with a softmax output layer to produce a probability distribution.
\end{itemize}
Li \textit{et al.} \cite{li2016tweet} employed a co-attention network to integrate textual and visual information, framing the task as a multi-class classification problem. Their approach utilized a Long Short-Term Memory (LSTM) network for processing textual data and a pre-trained VGG network for extracting image features. In a distinct methodological contribution, Zhang \textit{et al.} \cite{zhang2022twhin} introduced the Twitter Heterogeneous Information Network (TwHIN), a polyglot language model trained on an extensive corpus of tweets, designed to address the task of multi-class hashtag prediction. Similarly, Li \textit{et al.} \cite{li2019topical} proposed a Topical Co-Attention Network, which incorporates content and topic attention mechanisms to enhance the accuracy of hashtag recommendations. These studies collectively highlight the significance of leveraging multimodal data and advanced neural architectures for improving hashtag prediction tasks.

Multi-class approaches often use a small, predefined number of candidate hashtags (e.g., 20 as in \cite{li2016tweet}). While it allows recommending hashtags not present in the tweet, preselecting a fixed number of candidates makes these approaches impractical for dynamic scenarios.
\subsubsection{Multi-Label Classification}
In Multi-label Classification (MLC), an instance can be associated with multiple labels simultaneously, unlike traditional classification where only a single label is assigned. MLC finds applications in various domains, including image classification (e.g., an image tagged with ``ocean," ``sand," and ``sun"), text classification (e.g., a news article categorized as ``politics", ``sports", and ``entertainment"), recommendation systems (e.g., suggesting movies that are both ``dramatic" and ``romantic"), and bioinformatics (e.g., identifying multiple functional properties of a protein). 
\begin{figure}
\begin{minipage}{0.31\textwidth} 
\centering
\includegraphics[width=\textwidth]{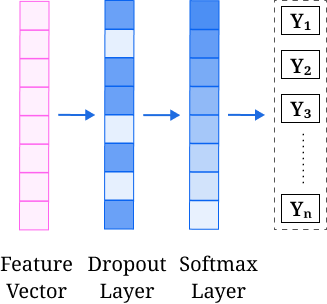} 
\caption{Multi-Label Classification}
\label{fig:mlc}
\end{minipage}\hfill
\begin{minipage}{0.64\textwidth} 
\centering
\includegraphics[width=\textwidth]{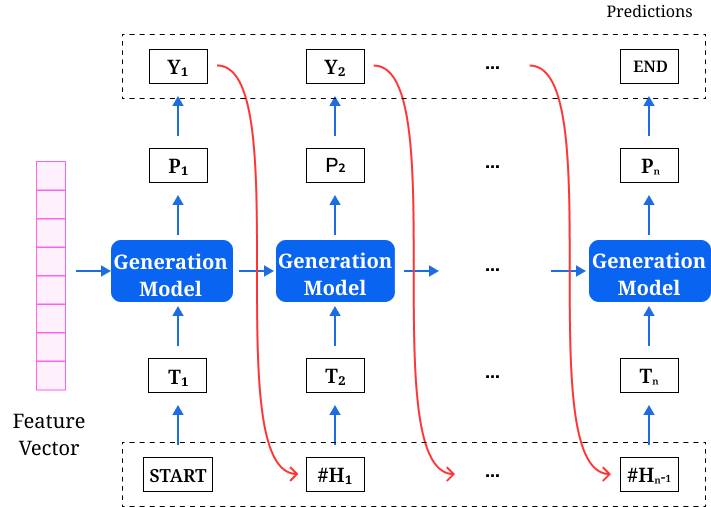} 
\caption{Generation}
\label{fig:sg_full}
\end{minipage}
\end{figure}
A common approach to multi-label recognition involves training independent binary classifiers for each label. However, this method overlooks relationships between labels \cite{li2019long} . Prior work in the domain of hashtag recommendation has framed the problem as MLC \cite{bansal2024multilingual,bansal2024hybrid,zhou2017scalable,li2019tagdeeprec,liu2018fasttagrec,he2022ptm4tag}. In multi-label classification (MLC), the training dataset comprises instances, each of which is associated with a subset of labels. The objective of MLC is to predict the set of relevant labels for unseen instances by using patterns and relationships derived from the training data, where set of labels for each instance are explicitly known (see \autoref{fig:mlc}). MLC-based approaches represent hashtags as sparse one-hot vectors, treating them as independent labels \cite{yu2023generating}. This representation fails to capture semantic information and co-occurrence patterns among hashtags. Furthermore, these methods are limited to a predefined set of candidate hashtags, hindering their adaptability to the evolving nature of online content.

\textit{Input:}
\begin{itemize}
\item $x$: Input query (e.g., a post represented as a vector of features).
\item $H = \{h_1, h_2,..., h_n\}$: Set of candidate hashtags.
\end{itemize}

\textit{Output:}
$Y=\{y_1, y_2,..., y_n\}$: A set of binary labels, where $y_i$ indicates whether hashtag $h_i$ is relevant to the query x (1 for relevant, 0 for irrelevant).\\

\textit{Mathematical Definition:}
\textit{The goal is to learn a classifier function $f(x, h_i)$ for each hashtag $h_i$ that predicts the binary label $y_i$.}
Classifier Function:
$f(x, h_i)$: Predicts the probability of $h_i$ being relevant to $x$. This can be modeled similarly to binary classification, but with a separate classifier for each hashtag.

\textbf{Loss Function}
\begin{itemize}
\item Binary Cross-Entropy: Applied to each hashtag independently.
\item Ranking Loss: Can be used to encourage correct ranking of relevant hashtags (though this moves closer to a ranking-based formulation).
\end{itemize}
Jain \textit{et al.} \cite{jain2024nlp} proposed a BERT-embedding-based LSTM (BELHASH) model for hashtag recommendation, formulated as MLC. BELHASH leveraged BERT for global semantic understanding with LSTM for capturing sequential information, aiming to leverage both short-term and long-term dependencies. 
Won \textit{et al.} \cite{won2023extra} employed FLAVA, a language and vision alignment Transformer model with three encoders based on the Vision Transformer (ViT) architecture. Classification logits are computed by passing the final hidden representation corresponding to the [CLS M] position through two MLP layers. External knowledge is integrated using the Open Directory Project (ODP). Binary cross-entropy serves as the loss function.
Bansal \textit{et al.} \cite{bansal2024multilingual} addressed hashtag recommendation for low-resource Indic language tweets and micro-videos \cite{bansal2024hybrid} as MLC problem. The authors predicted the relevance probability of each hashtag from a preconfigured pool for a given post, using these probabilities for ranking and hashtag selection.

While MLC-based approaches can outperform topic and extractive models, they often underperform Sequence Generation (SG)-based approaches \cite{bansal2022hybrid}. A key limitation of MLC is their reliance on a predefined candidate list, which can lead to suboptimal recommendations due to the imbalanced and dynamic nature of hashtag usage. These models often struggle to capture correlations between hashtags, where the presence of one hashtag can significantly influence the likelihood of others. This is further complicated by the dynamic nature of hashtags. User-assigned hashtags frequently appear neither in target posts nor in the candidate list. The vast and ever-changing vocabulary of hashtags, driven by user freedom on platforms such as social media and software information sites, coupled with the rapid creation of new hashtags due to diverse and evolving topics, makes it challenging for classification models to effectively capture the complex relationships in hashtag usage. The constant emergence of new trends in hashtag usage further exacerbates this challenge. These inherent limitations of classification methods motivate the exploration of alternative approaches better suited to the dynamic and evolving nature of hashtags.  While SG-based approaches also operate within the constraints of a predefined vocabulary, they offer greater flexibility by generating hashtags token-by-token. This allows SG models to combine tokens in novel ways, potentially creating new hashtags that were not explicitly present in the training data. 
\subsection{Generation}
Generative approaches offer a flexible paradigm for hashtag recommendation, moving beyond simple extraction to create novel hashtags. These methods learn to generate hashtags from scratch, eliminating reliance on predefined candidate lists and enabling the creation of hashtags that may be rare or unseen in the training data. The necessity of this approach is underscored by the sparse nature of hashtag occurrences in social media data. An analysis of 36,000 tweets revealed that over 80\% of hashtags occur at most five times \cite{zheng2021news}.

\textit{Mathematical Definition}
\textit{The core objective is to learn a generative model $p(H|x)$ that predicts the probability of generating a set of hashtags $H$ given the input query $x$.}
Generative Model $p(H|x)$ can be modeled using various techniques:
\begin{itemize}
\item Seq2seq Models: Encode the input $x$ and decode it into a sequence of hashtags.
\item Language Models: Generate hashtags conditioned on the input $x$ using a language model.
\item Conditional Variational Autoencoders (CVAEs): Learn a latent representation of hashtags and generate hashtags from this representation conditioned on $x$.
\end{itemize}
A typical loss function for these models is cross-entropy, aiming to maximize the likelihood of generating the correct hashtags.

Several studies have framed hashtag recommendation as a generation task \cite{wang2019microblog,mao2022attend,zheng2021news}, recognizing that hashtags are often generated in a sequence. SG-based methods capture deeper semantic relationships between texts and tags by framing tag recommendation as a conditional language modeling task. Considering tags as a sequence allows them to capture dependencies, ensuring each subsequent tag is contextually relevant to previously generated ones, as shown in \autoref{fig:sg_full}. This approach allows for the innovative combination of words, yielding tags that did not appear as complete sequences in the training data. This flexibility arises from SG’s ability to learn a probability distribution over word sequences. However, prior research has often given little consideration to popular or trending hashtags. While these approaches produce semantically relevant hashtags, recommended hashtags might not be widely used, hindering the discoverability of content.  In this subsection, we categorize generation-based methods into four key paradigms: autoregressive sequence generation, sequence-oblivious generation, keyphrase generation, and retrieval-augmented generation. Each paradigm addresses the task from a unique perspective, utilizing distinct methodologies to optimize the relevance, diversity, and creativity of recommended hashtags. 
\subsubsection{Autoregressive Sequence Generation}

\begin{figure}
\begin{minipage}{0.3\textwidth} 
\centering
\includegraphics[width=\textwidth]{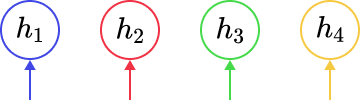} 
\caption{Ranking}
\label{fig:ranking}
\end{minipage}\hfill
\begin{minipage}{0.3\textwidth} 
\centering
\includegraphics[width=\textwidth]{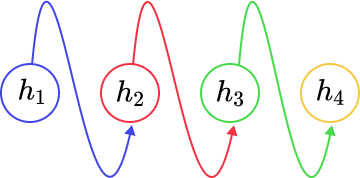} 
\caption{Autoregressive Sequence Generation}
\label{fig:sg}
\end{minipage}\hfill
\begin{minipage}{0.3\textwidth} 
\centering
\includegraphics[width=\textwidth]{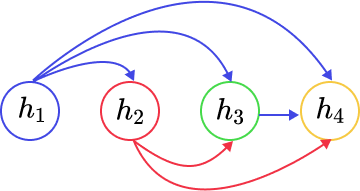} 
\caption{Sequence Oblivious Generation}
\label{fig:sog}
\end{minipage}
\end{figure}

\textit{Mathematical Definition:}
\textit{The goal is to learn a model $p(h_t | x, h_{<t})$ that predicts the probability of generating hashtag $h_t$ at time step $t$, given the input $x$ and the previously generated hashtags $h_{<t} = \{h_1, h_2,..., h_{t-1}\}$.}
This conditional probability distribution is modeled using RNNs, such as GRUs, or transformer architectures as shown in \autoref{fig:sg}. A cross-entropy loss function is commonly used to maximize the likelihood of the generated hashtag sequence.

Several studies have explored autoregressive sequence generation for hashtag recommendation. Yang \textit{et al.} \cite{yang2020amnn} developed a Seq2seq encoder-decoder framework, employing attention mechanisms to combine visual and textual embeddings before feeding them into a GRU for sequential hashtag generation. Diao \textit{et al.} \cite{diao2023hashtag} introduced a two-stage framework designed for low-resource tweet classification, utilizing hashtag guidance to enhance performance. Their approach incorporates a transformer-based hashtag generator equipped with attention mechanisms to capture both topical and entity-level information. Similarly, Mao \textit{et al.} \cite{mao2022attend} developed a transformer-based model that integrates an encoder to filter irrelevant data and a segment selector to restructure text segments prior to sequential hashtag prediction. Their methodology employs a sequential decoding algorithm to generate hashtags, demonstrating the effectiveness of transformer architectures in handling complex text data for hashtag recommendation tasks. Tang \textit{et al.} \cite{tang2019integral} introduced an encoder-decoder framework with RNNs and attention for autoregressive tag generation. The decoder employs a prediction path to model dependencies between sequentially generated tags. An indicator function handles content-tag overlap by determining the probability of using existing text words as tags.
 
However, existing sequence generation methods, especially those relying on encoder-decoder frameworks with RNNs \cite{wang2019microblog, yang2020amnn} or Transformers \cite{mao2022attend, diao2023hashtag}, face challenges. RNNs, while adept at capturing sequential information, struggle with long-range dependencies, which can be critical even within the limited character count of tweets. Transformers, while generally robust, can produce generic or repetitive hashtags when confronted with the noisy, informal language and grammatical errors common in certain contexts, such as disaster situations. This can hinder their effectiveness in generating hashtags that accurately reflect the rapidly evolving needs and information during such events. Furthermore, while these generative models consider the dependency between tags in the generated sequence, they often overlook the fundamental orderlessness of hashtag sets. The fact that hashtags are generated sequentially does not reflect the reality that the order of hashtags in a set is typically irrelevant. As noted earlier, even methods using GRUs \cite{wang2019microblog, yang2020amnn} to capture inter-tag dependencies still treat hashtags as ordered sequences, neglecting the crucial aspect of orderlessness. This tension between interdependency and orderlessness remains a key challenge in hashtag recommendation.
\subsubsection{Sequence Oblivious Generation}
Traditional hashtag recommendation methods rely on either ranking or autoregressive (AR) generation. However, these methods struggle to fully capture the inherent characteristics of hashtag sets: interdependency and orderlessness. Ranking-based approaches disregard relationships between hashtags (see \autoref{fig:ranking}), while AR methods, including those using powerful architectures such as transformers \cite{vaswani2017attention}, impose a sequential order on hashtag generation (see \autoref{fig:sg}), despite hashtag order being inconsequential. This sequential dependency, particularly within the decoder of AR models (which relies heavily on the immediately preceding token), unnecessarily constrains the generation process. Furthermore, maximizing sequence likelihood, common practice in text generation, is overly restrictive for hashtag recommendation, as shuffling generated hashtags does not affect their overall relevance. Prior generative approaches that utilize GRUs \cite{wang2019microblog, yang2020amnn,tang2019integral}, to model inter-tag dependencies, still treat hashtags as ordered sequences. This highlights a central challenge: simultaneously modeling both the interdependency and orderlessness of hashtag sets, a crucial aspect often overlooked by conventional encoder-decoder frameworks used in text generation \cite{cho2014properties, bahdanau2014neural, NIPS2014_a14ac55a, radford2018improving, yang2019xlnet, chi2020cross}. To address these limitations, hashtag recommendation can be formulated as Sequence Oblivious Generation (SOG). SOG generates each hashtag independently of the order of previously generated hashtags and the order of ground-truth hashtags in the training data (see \autoref{fig:sog}). 

\textit{Mathematical Definition:}
\textit{The goal is to learn a model $p(H|x)$ that predicts the probability of generating a set of hashtags ($H = {h_1, h_2,..., h_k}$), without considering their order.}
This is a challenging modeling problem that necessitates permutation-invariant models capable of operating on sets rather than sequences. Set-based loss functions are employed to evaluate the similarity between the generated hashtag set and the ground-truth hashtag set.

Kang \textit{et al.} \cite{kang2020leveraging} framed hashtag generation as SOG by leveraging the transformer encoder and its self-attention mechanism to process all input features simultaneously, eliminating the sequence dependency of decoder-based generation. Unlike AR training that maximizes sequence likelihood (1-to-1), SOG maximizes the probability of the entire hashtag set (1-to-M) at each generation step, training the model to be insensitive to ground-truth hashtag order.
\subsubsection{KeyPhrase Generation}
Keyphrase generation (KPG) focuses on producing phrases that capture the most important information within a given text. KPG for hashtag recommendation aims to create new, relevant hashtags, rather than simply extracting existing phrases from the source text. This allows for greater creativity and the capture of nuances not explicitly present in the input.

\textit{Mathematical Definition:}
\textit{The goal of KPG for hashtag recommendation is to learn a function $g(x)$ that maps an input text $x$ to a set of generated hashtags $H$.}

As outlined by Meng \textit{et al.} \cite{meng2021empirical}, current KPG approaches fall into two primary categories: One2One \cite{meng2017deep}, which generates individual keyphrases sequentially, and One2Seq \cite{yuan2020one}, which produces a sequence of keyphrases simultaneously. One2One approach can be simpler to implement but might miss relationships between hashtags. One2Seq allows to capture dependencies between hashtags and generate more coherent sets. Yu \textit{et al.} \cite{yu2023generating} further refined One2Seq approach by randomly shuffling target hashtags during training to mitigate the influence of presentation order. Wang \textit{et al.} \cite{wang2019microblog} generated hashtags for microblogs as a specialized adaptation of KPG for social media content.

Framing hashtag recommendation as KPG task offers several advantages. Critically, it enables the generation of novel hashtags, reflecting the creative practice of hashtag invention by users. Moreover, KPG can capture subtle meanings, emotions, or contextual information not explicitly present in the input text. By leveraging a deeper understanding of the text, KPG can generate more relevant and informative hashtags.  However, this framing also presents challenges. KPG is inherently more complex than keyphrase extraction, demanding more sophisticated models and training strategies. Evaluating the quality of generated hashtags poses a significant challenge, as traditional metrics such as precision and recall are inadequate, necessitating human evaluation. Finally, maintaining control over the generated output to ensure appropriateness and relevance remains a key area of research, as models may occasionally produce nonsensical or offensive hashtags.
\subsubsection{Retrieval Augmented Generation}
Leveraging Retrieval Augmented Generation (RAG) offers a promising solution for hashtag recommendation in the dynamic and noisy landscape of social media. The challenge arises from the rapid evolution of online language, user vocabulary, and the diverse and often idiosyncratic ways hashtags are employed as forms of self-expression, combining words, using varied spellings and short forms. This dynamic environment, particularly during events and topical discussions, creates a constant influx of new information and evolving user needs. Existing retrieval-based methods \cite{gong2016hashtag,zhang2017hashtag}, relying on static resources such as fixed hashtag lists or databases, struggle to keep pace with the rapid evolution of language and topics online. While generation-based methods \cite{wang2019microblog,mao2022attend,zheng2021news} are better equipped to understand new information, they often struggle to generate accurate and relevant hashtags without additional guidance. RAG bridges this gap by combining strengths of pretrained generative models with information retrieval techniques \cite{asai2023retrieval,kocon2023chatgpt}. 
This approach allows RAG models to leverage existing knowledge while adapting to new information, crucial for effective hashtag recommendation. By capitalizing on both retrieval and generation, RAG models can capture the evolving needs of online communities, effectively filter and process the informal language and misspellings prevalent in social media data, and generate hashtags that not only reflect the current situation but also anticipate future trends, thereby resulting in hashtag recommendation systems that can navigate the complexities of social media discourse. 

Previous research has explored RAG to address information-driven tasks \cite{chen2017reading,zhang2023relevance,he2023merging,li2023generative,wang2022micro,wang2021discover,mysore2023editable,chen2022corpusbrain,chen2023continual}. It has been applied in Natural Language Processing (NLP) tasks, such as generating image captions \cite{ramos2023retrieval}, producing keyphrases \cite{kim2021structure,gao2022retrieval}, neural machine translation \cite{gu2018search,hossain2020simple}, answering open-ended questions \cite{lee2019latent,guu2020retrieval,lewis2020retrieval} and knowledge-based dialogue generation \cite{lian2019learning}. Furthermore, RAG has proven valuable in mitigating shortcomings of Large Language Models (LLMs), such as mitigating factual inaccuracies \cite{cao2020factual,raunak2021curious}, compensating for outdated knowledge \cite{he2022rethinking}, and in improving performance in specialized domains \cite{li2023chatgpt}.
Furthermore, its applicability extends into multimodal domains, with implementations for generating captions in open-book video contexts \cite{zhang2021open} and question answering by integrating visual and textual modalities \cite{gao2022transform}. 

However, applicability of RAG to hashtag recommendation is relatively nascent. A few recent studies have begun to explore RAG for hashtag recommendation. Lu \textit{et al.} \cite{lu2024retrieval} retrieved relevant information from external knowledge sources to enrich post representations. This retrieved knowledge is then integrated with the post's multimodal content using a cross-modal context-aware attention mechanism, which focuses on extracting targeted features from different modalities such as title and code, guided by the main modality (description). 
Zheng \textit{et al.} \cite{zheng2021news} introduced HashNews, a retriever-generator framework. The retriever component goes beyond traditional information retrieval by incorporating a time-aware entity-focused ranking function that considers the temporal popularity of entities. The system constructs a series of candidate corpora based on timestamps and calculates the temporal popularity of entities by comparing their inverse document frequency values in recent news corpora with a time-independent corpus. This ranking function emphasizes emerging entities, enabling the retrieval of news articles that are both relevant and timely. The generator component then employs a novel hybrid bi-attention mechanism to jointly model the post and the retrieved news, with a particular focus on important entities extracted from the news.  This approach highlights the importance of considering the temporal dynamics of information when generating relevant hashtags. 

Building on these advancements, current research is actively exploring the extension of RAG to multimodal hashtag generation, aiming to capture the dynamic and nuanced nature of online conversations.
\subsection{Link Prediction}
Framing hashtag recommendation as a link prediction problem offers a powerful approach to leveraging the relationships between posts (or users) and hashtags. This approach represents these relationships as a graph, where nodes correspond to posts (or users) and hashtags, and edges represent their association.  Specifically, an edge exists if a hashtag is relevant to or used in a particular post (or by a user). The task of hashtag recommendation is then translated into a link prediction problem: \textit{given a post (or user) node, predict which hashtag nodes it is most likely to connect with, effectively identifying the most relevant hashtags.} This graph-based representation allows hashtag recommendation systems to exploit the rich structural information inherent in how hashtags are used within the network. Instead of analyzing post content in isolation, link prediction considers the broader context of hashtag usage, capturing relationships between posts and hashtags.  
For instance, if posts P1 and P3 share similar vocabulary and P1 uses hashtags \#news and \#sports, a link prediction algorithm might infer that P3 is also likely relevant to those hashtags, even if P3's content doesn't explicitly mention them. This is because the algorithm learns from the graph structure that posts similar to P1 tend to be associated with those hashtags.

\textit{Mathematical Definition:} \textit{The goal is to predict the probability or likelihood of a link existing between a post (or user) $x$ or $u$ and a hashtag $h$. This is often represented as a link prediction function}:
$p(u, h)$: probability of a link between node $u$ and node $h$. 
This function can be modeled using various techniques:
\begin{itemize}
\item GNNs: GNNs learn representations for nodes and edges, and $p(u, h)$ can be derived from these representations (e.g., $p(u,h) = sigmoid(z_u^T z_h)$ where $z_u$ and $z_h$ are the GNN embeddings of $u$ and $h$).
\item Matrix Factorization: The adjacency matrix of the graph is factorized into lower-dimensional matrices, and the link probability is calculated from these factors.
\item Neighborhood-based methods: $p(u, h)$ is based on the number of common neighbors between $x$ or $u$ and $h$. It may also consider hashtags used by similar posts/users.
\end{itemize}
For a given node $u$, the hashtags are then ranked in descending order of $p(u, h)$, and the top-k hashtags are recommended. Common loss functions used in this context include binary cross-entropy and pairwise ranking loss, similar to the ranking-based formulation but applied to link probabilities.

Several studies have explored this link prediction paradigm for hashtag recommendation. Wei \textit{et al.} \cite{wei2019personalized} approached the task of personalized hashtag recommendation for micro-videos by modeling it as a link prediction problem and employing graph-based methodologies. However, their framework does not account for the challenges associated with incorporating unseen trending hashtags or extending the approach to long-form video content. Similarly, Wang \textit{et al.} \cite{wang2022micro} conceptualized micro-video tagging as a link prediction problem, constructing a heterogeneous network that integrates tag ontology, video tag annotations, and user follow relationships. While these studies highlight the utility of graph-based techniques in hashtag recommendation, they underscore the need for further exploration into handling dynamic and evolving hashtag trends, as well as scalability to diverse video formats. Mehta \textit{et al.} \cite{mehta2021open} further advanced this approach by applying GCNs to a heterogeneous graph. This graph consisted of videos and hashtags as nodes, with edges representing hashtag-to-hashtag, video-to-hashtag, and video-to-video relationships, learning joint representations for more effective hashtag recommendations.
\section{Filtering Approaches for Hashtag Recommendation}
\label{sec:filtering}
\begin{figure}
\includegraphics[width=\textwidth]{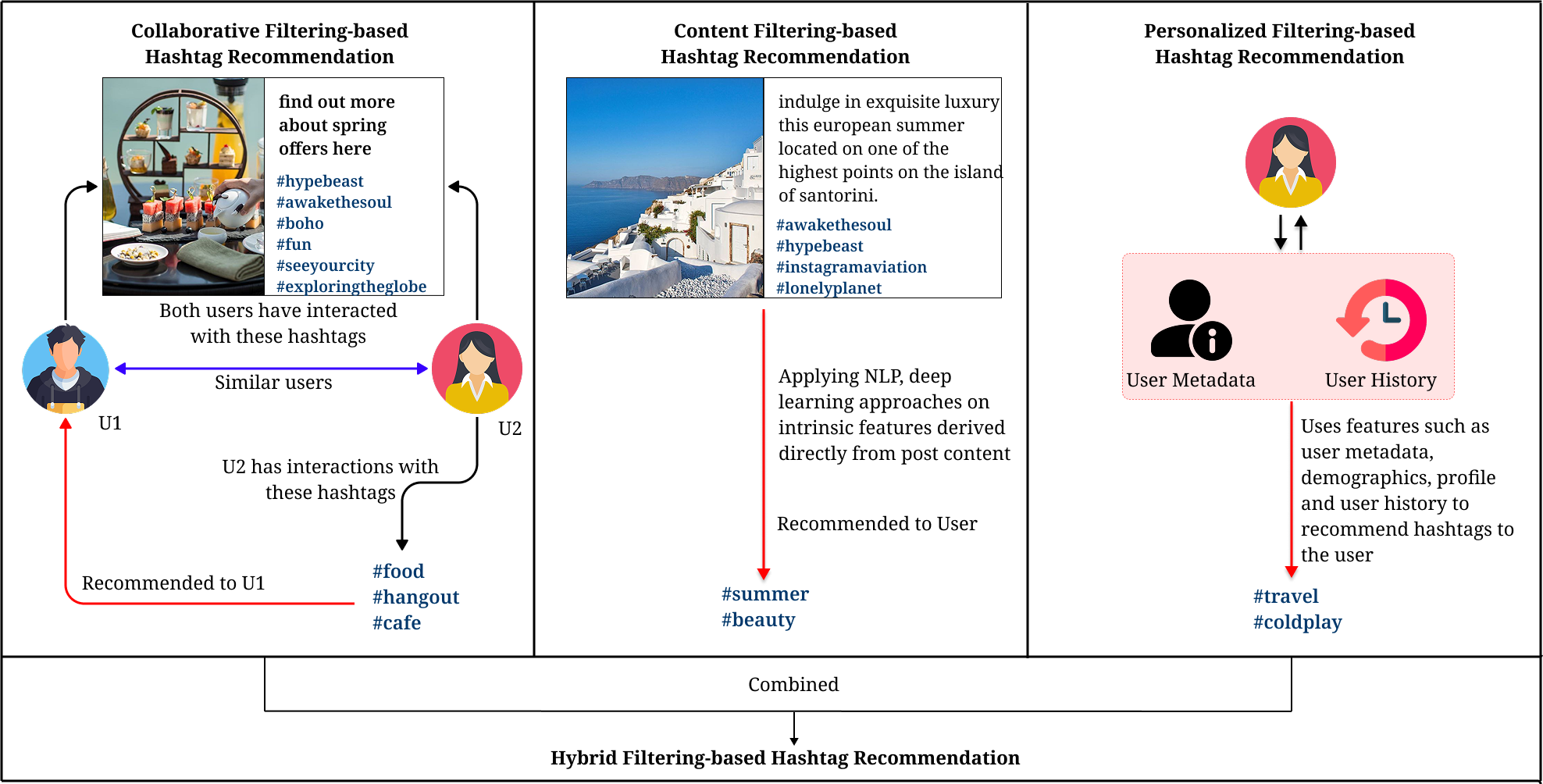}
\caption{Comparative Overview of Filtering Approaches for Hashtag Recommendation}
\label{fig: filtering}
\end{figure}
Effective hashtag recommendation relies on filtering approaches to enhance the relevance and utility of suggested hashtags. These approaches are designed to address the challenges of information overload and relevance, ensuring that recommended hashtags align with the content, user preferences, and contextual dynamics of social media posts. In this section, we explore four primary filtering approaches: content-based filtering, collaborative filtering, personalized filtering, and hybrid filtering. Each approach offers unique mechanisms for identifying and ranking hashtags, catering to different aspects of the recommendation task.
\autoref{fig: filtering} provides a comparative overview of these filtering approaches.
\subsection{Content-based Filtering}
Content-based filtering for hashtag recommendation relies on analyzing the semantic and contextual attributes of social media posts to predict relevant hashtags using Natural Language Processing (NLP) and deep learning. Unlike collaborative or hybrid approaches, it focuses entirely on the content of a post, such as text embeddings, semantic structures, or statistical properties, without requiring user history or social network features. Over the last decade (2015–2024), content-based hashtag recommendation methods have evolved significantly, leveraging advancements in NLP, deep learning, and semantic knowledge integration. Content-based filtering approaches for hashtag recommendation focus on leveraging intrinsic features derived directly from post content—such as textual, visual, audio, or video features—without relying on user interaction data or social graphs. This line of research is particularly valuable for addressing cold-start scenarios and platforms with limited user-behavioral data, where collaborative filtering or hybrid approaches are not applicable. In the context of multimodal social media posts, significant progress has been made in developing methods that integrate multiple modalities, utilize deep learning architectures, and improve semantic understanding for hashtag prediction during 2015–2024.

The field initially focused on text-only models that addressed hashtag recommendation using text-based representations or topic modeling techniques. Early works \cite{gong2015hashtag} explored Dirichlet Process Mixture Models (DPMMs) to recommend hashtags by modeling hashtag types as latent variables, while Dey \textit{et al.} \cite{dey2017emtagger} introduced EmTaggeR, which leveraged word embeddings for textual content. Later, advanced text representation techniques such as Tree-Structured LSTMs \cite{zhu2019learning} and contextual BERT embeddings \cite{cantini2021learning} enhanced the semantic richness of textual features, improving performance in purely text-based settings.

Beyond textual content, multimodal approaches emerged to integrate visual, textual, and, in some cases, audio or video features into unified architectures. Research studies \cite{zhang2017hashtag,gong2018hashtag} pioneered the use of co-attention mechanisms to jointly model text and image features, while subsequent works \cite{feng2023tnod,yang2020amnn} utilized transformer architectures to align and fuse multimodal content through attention-based frameworks. A notable example is LXMERT4Hashtag \cite{khalil2023cross}, which employs a cross-attention transformer model to align text and image embeddings for multimodal hashtag prediction. Other methods have incorporated additional modalities, such as audio and sentiment information for micro-video posts in TOAST \cite{yang2020sentiment}, indicating initial explorations into richer multimodal datasets.

More recent innovations have focused on integrating graph-based reasoning and secondary features into recommendation systems. Khalil \textit{et al.} \cite{khalil2024mrlkg} introduced MRLKG, a model that combines transformer-driven multimodal representations with keyword-guided GCNs to incorporate relational context within hashtags. Similarly, sentiment-aware learning \cite{yang2020sentiment} highlighted the potential of incorporating emotional context into multimodal hashtag prediction. However, while most multimodal methods focus on visual and textual fusion, the inclusion of audio and video features remains underexplored. Temporal modeling has also been sparsely addressed, despite the inherently dynamic nature of hashtag usage over time. Few works \cite{gruetze2015learning} integrated time-decay or temporal embeddings to capture evolving hashtag trends, highlighting an important gap in adapting content-based methods for real-time social media dynamics.

Content oriented hashtag recommenders excel at contextual representation and scalability while maintaining independence from user data. These approaches lack novelty, because they
generate tags that are already part of the target content \cite{belem2017survey}. 
However, challenges surrounding sparse data, hashtag evolution, and multimodal integration persist (including underutilized modalities such as audio and video). Generative models such as Generative Pre-trained Transformer (GPT) or Text-to-Text Transfer Transformer (T5) have seen limited exploration in this domain. Future research could combine semantic enrichment, temporal learning, and multimodal strategies while preserving the scalability and user-independence of content-based approaches.
\subsection{Collaborative Filtering}
Collaborative Filtering (CF) is a prevalent technique in recommendation systems, including those designed for hashtag recommendation. CF leverages the collective behavior of users to identify similar preferences and recommend items that like-minded users have positively interacted with. The underlying assumption is that users exhibiting similar past behavior are likely to share future preferences. In hashtag recommendation, CF systems analyze user-hashtag interactions to discern patterns and generate relevant recommendations. These methods involve constructing a user-hashtag matrix, where rows correspond to users, columns to hashtags, and matrix entries reflect user interactions with hashtags such as frequency of hashtag usage or binary indicator of use. Two primary approaches within CF are commonly employed:
\begin{itemize}
\item User-based CF: Identifies users with comparable hashtag usage patterns. For a target user, the system computes similarity scores with other users based on their past hashtag interactions. Hashtags frequently used by these similar users, but not yet used by the target user, are then recommended. \autoref{table:colloborative_filtering} shows the categorization of user-based collaborative filtering  approaches for hashtag recommendation.
\begin{table}\footnotesize
\centering
  \caption{Categorization of Collaborative Filtering  Approaches in Hashtag Recommendation}
\label{table:colloborative_filtering}
  \begin{tabular}{cccc}
    \toprule
    \textbf{Paper} & \textbf{Type of User-based CF}\\
    \midrule
    Torres \textit{et al.} \cite{torres2020hashtags} & Behavioral\\
    Kowald \textit{et al.} \cite{kowald2017temporal}& Behavioral\\
    Kou \textit{et al.} \cite{kou2018hashtag} &	Behavioral\\
    Wen \textit{et al.} \cite{wen2022implicit}&	Behavioral\\
    \midrule
    Alsini \textit{et al.} \cite{alsini2020utilizing} & Social\\
    Wang \textit{et al.} \cite{wang2022micro} & Social\\
    \bottomrule
  \end{tabular}
\end{table}
\begin{itemize}
\item Behavioral: A prominent subcategory of user-based CF, focuses on identifying users with similar past hashtag usage and topical interests.
For instance, Torres \textit{et al.} \cite{torres2020hashtags} utilized a user-item matrix and a memory-driven KNN algorithm, leveraging both user and hashtag interactions to predict missing valuess. Temporal information was incorporated by Kowald \textit{et al.} \cite{kowald2017temporal} to identify similar users based on past activity. Kou \textit{et al.} \cite{kou2018hashtag} proposed a hybrid approach, combining content similarity, collaborative filtering, and topical representations. The CF component of their system focuses on user similarity derived from hashtag usage, a clearly behavioral aspect. While the topical representations could incorporate social information, representations are primarily based on usage patterns and content analysis, making the overall approach lean towards behavioral rather than social. Wen \textit{et al.} \cite{wen2022implicit} proposed a method to connect users with shared interests by extracting implicit tags from item text using LDA and calculating feature weights for item-tag associations. Their algorithm integrates an item-tag matrix with a user scoring matrix to infer user tagging preferences based on historical behavior. As this approach relies on observed tagging patterns and item-tag relationships derived from user actions, it is categorized as a behavioral method. However, its dependence on historical user data, such as past hashtags and shared topics, renders it susceptible to the cold-start problem. In scenarios involving new users, the lack of sufficient historical data impedes the identification of similar users or items, thereby limiting the effectiveness of hashtag recommendations. Alsini \textit{et al.} \cite{alsini2020utilizing} explicitly examined the influence of social relationships, including followers and mentions, on hashtag suggestion, in addition to hashtag usage and topical relevance. By incorporating social network structures such as followers and mentions, their approach is classified as social. Similarly, Wang \textit{et al.} \cite{wang2022micro} addressed micro-video tagging as a link prediction problem within a heterogeneous network that integrates tag ontology and user follow relationships. The inclusion of user follow relationships further categorizes this as a social approach.

Nevertheless, social network information, such as followers, friends, and mentions, may be unavailable for cold-start users who are new to the platform. Additionally, cold-start users are typically unfamiliar with prevailing trends and often seek recognition and visibility for their posted micro-videos. This lack of social and behavioral data poses significant challenges for personalized hashtag recommendation systems, particularly in addressing the needs of new users.
\end{itemize}
\item Item-based CF: It focuses on relationships between hashtags. It calculates similarity scores between hashtags based on their co-occurrence in user posts. To recommend hashtags to a user, the system identifies hashtags similar to those the user has previously used and suggests these similar hashtags. In essence, item-based CF in this context leverages the idea that if two hashtags are frequently used together by many users, they are likely related and thus, if a user has used one, they might be interested in the other.

\end{itemize}
\subsection{Personalized Filtering}
Personalized hashtag recommendation systems aim to go beyond generic or popularity-based recommendations, tailoring hashtag suggestions to individual users based on their specific preferences and behaviors. These systems typically leverage two primary data types: user behavior (e.g., historical post activity, interactions, and usage patterns) and metadata (e.g., content of posts, hashtags, images, location, and temporal information). Personalized hashtag recommendation, specifically, tailors these suggestions to individual user preferences by leveraging two primary data sources: (1) User behavior, such as historical hashtag usage, interactions, and engagement patterns, and (2) metadata, which includes post-level attributes such as text, images, timestamps, location, and hashtags. The integration of these inputs poses a unique challenge: systems must model complex relationships between users, hashtags, and content while addressing issues such as cold-start problems, hashtag sparsity, and the dynamic nature of social media platforms. \autoref{table:personalised_filtering} shows a fine-grained categorization of personalized filtering approaches for hashtag recommendation showcasing the corresponding type of metadata, user behavior and methods employed.
\begin{table}\scriptsize
\centering
\caption{Categorization of Personalized Filtering Approaches for Hashtag Recommendation}
\label{table:personalised_filtering}
\resizebox{\textwidth}{!}{
\begin{tabular}{@{\extracolsep{\fill}} lccccccccc}
\toprule
\textbf{Paper}& \textbf{Metadata} &\textbf{User Behavior}&\textbf{Model Type} &	\textbf{Attention}\\
\midrule
Zhang \textit{et al.} \cite{zhang2021howyoutagtweets}& Hashtags emb & User Hashtag History &NTM+BERT, topic-based emb & Topic Attention \\
Tan \textit{et al. }\cite{tan2024llm} & Text+Image & User Posting History & LLM with multimodal alignment & No\\
Padungkiatwattana \textit{et al.} \cite{padungkiatwattana2022pac} & word-level and post attributes & user-user, user-hashtag interactions & High-order Multi-Relation GNN &Attention for multiple relations\\
Liu \textit{et al.} \cite{liu2020user} & Video metadata, age, country,& historical likelihood of hashtags & Hierarchical Multi-head Attention & Hierarchical Multi-head Attention\\
Bansal \textit{et al.} \cite{bansal2022hybrid} & Text+Image & user posting history & Co-Attention + DNN & Word-level and parallel co-attention\\
Li \textit{et al.} \cite{li2019long} & Video content & User-post interactions &GCN & Graph Attention\\
Denton \textit{et al.} \cite{denton2015user} & Images, user & User conditioning on emb.&CNN + Metadata Fusion& No\\
Tran \textit{et al.} \cite{tran2018hashtag}& Trending social metadata & Hashtag co-occurrence & Personalized ranking & No\\
Tao \textit{et al.} \cite{tao2022personalized} & Content emb & 
Episodic user learning& Meta-learning Few-shot & No\\
Chen \textit{et al.} \cite{chen2021tagnet} & Images, text & user-post-image relativity & Aggregated Graph Convolution & Triplet Attention\\
Huang \textit{et al.} \cite{huang2016hashtag} & post-level hashtag trends & Hierarchical user history & End-to-End Memory Net & Hierarchical attention\\
Durand \textit{et al.} \cite{durand2020learning} & image-hashtag alignment & user-history embeddings & Open vocabulary cross-modal learning & Cross-modal attention\\
Bansal \textit{et al.} \cite{bansal2024hybrid} & text, images, audio & Interaction modeling & GNN + collaborative filtering\\
\bottomrule
\end{tabular}
}
\end{table}
Zhang \textit{et al.} \cite{zhang2021howyoutagtweets} combined user-specific hashtagging histories (user behavior) and hashtag metadata (e.g., semantic contexts). The authors used personalized topic attention to align user embeddings with hashtag relevance, adapting to user-specific interests, thereby fully integrating both user behavior and metadata for personalized hashtag recommendations. Wei \textit{et al.} \cite{wei2019personalized} explicitly modeled user-hashtag-micro-video relationships through a graph structure. User preferences are learned from historical hashtag interactions.
GCN propagates personalized features using user-hashtag-content embeddings. Padungkiatwattana \textit{et al.} \cite{padungkiatwattana2022pac} captured high-order user behavior through social relationships (user-user) and shared hashtag patterns (hashtag-hashtag).
The authors employed graph-based multi-relational model creates personalized user and hashtag embeddings. Liu \textit{et al.} \cite{liu2020user} modeled user-guided preferences by integrating historical hashtag usage and user profiles with micro-video post content.
They used hierarchical attention to fuse user preferences with multimodal post data. Bansal \textit{et al.} \cite{bansal2022hybrid} incorporated user-specific preferences using multimodal inputs (text, images) and historical user behavior. Li \textit{et al.} \cite{li2019long} used user-video-hashtag interactions to capture and model user-specific behavior for personalized recommendations. The authors propagated personalized embeddings via GCN-based models focusing on long-tail hashtags. Tao \textit{et al.} \cite{tao2022personalized} addressed cold-start personalization for users by learning user-specific embeddings that generalize across users and hashtags.
The authors proposed a meta-learning framework trains a personalized adaptation model using user-specific historical posts. Huang \textit{et al.} \cite{huang2016hashtag} incorporated user history as external memory, with attention mechanisms prioritizing past behavior relevant to current recommendations. Hierarchical attention extracts personalized patterns across historical and post-level features.
Durand \textit{et al.} \cite{durand2020learning} leveraged user behavior (historical image hashtags and posts) to build personalized user representations for hashtag prediction. Open-vocabulary embedding aligns user-specific visual data with hashtag diversity.
Bansal \textit{et al.} \cite{bansal2024hybrid} developed a personalized hashtag recommendation system for micro-videos by modeling users' modality-specific tagging preferences. This was achieved through the construction of a heterogeneous graph that links users to the constituent modalities of their previously posted micro-videos, enabling the capture of individual tagging behaviors.

While personalized hashtag recommendation systems focus on aligning recommendations with the preferences of content creators, they often overlook the significance of incorporating trending topics and community interests. This limitation restricts the potential exposure of content to a broader yet relevant audience. As noted by Mehta \textit{et al.} \cite{mehta2021open}, such systems treat all hashtags uniformly, prioritizing the personal preferences of creators while neglecting the strategic use of trending hashtags to enhance content visibility and engagement. Consequently, this approach may hinder the ability of content to reach a wider audience that aligns with both individual and collective interests.

Despite significant progress, challenges remain in incorporating richer contextual metadata (e.g., temporal or spatial data \cite{padungkiatwattana2022pac,tran2018hashtag} and mitigating fairness or bias issues in personalized recommendations. High-quality, extensive historical data is often required to train personalized models effectively. For new users or those with limited activity (cold-start users), the lack of sufficient data can impact performance.
Incomplete or noisy metadata (e.g., missing geolocation, timestamps, or hashtag tags) can reduce the accuracy of predictions. Excessive reliance on historical behavior data may result in over-personalized recommendations, leading to a feedback loop that reinforces the user’s existing preferences while suppressing diversity and exploration.
For example, users may not be exposed to trending or novel hashtags that fall outside their historical patterns.
Personalized models are prone to replicate or even amplify biases inherent in the training data. For example, demographics, social disparities, or stereotypical associations in user behavior can affect hashtag recommendations.
Fairness-aware or debiasing techniques are not explored in sufficient detail in the current literature, and issues such as underrepresentation of minority hashtags or diversity trade-offs remain problematic.

\subsection{Hybrid Filtering}
Traditional approaches to hashtag recommendation include content-based filtering, collaborative filtering, and personalized filtering, each with distinct advantages and limitations. Content-based filtering analyzes textual, visual, or multimodal features of a post to generate semantically relevant hashtag recommendations \cite{li2016hashtag,huang2016hashtag}. Collaborative filtering exploits co-occurrence patterns in hashtag usage and user-item interactions but struggles with challenges such as data sparsity and cold-start problem \cite{javari2020weakly,torres2020hashtags}. Personalized filtering emphasizes user-specific preferences, such as historical hashtag usage or social connections, to deliver tailored suggestions \cite{zhang2021howyoutagtweets,tao2022personalized}. However, standalone methods often fail to address the nuanced and multidimensional nature of social media data, leading to the rise of hybrid filtering techniques that combine two or more approaches.

Recent research has explored a variety of hybrid filtering architectures to improve the robustness and relevance of hashtag recommendations as shown in \autoref{table:hybrid_filtering}. These hybrid systems integrate content, collaborative, and personalized data signals into unified frameworks to mitigate cold-start issues, improve scalability, and dynamically adapt to evolving hashtag trends. Specifically, hybrid approaches address various limitations of standalone methods. The fusion of all three signals—content, collaborative, and personalized is particularly promising, with recent advances leveraging graph-based models and attention mechanisms for comprehensive and dynamic recommendation systems
\begin{table}\scriptsize
\centering
\caption{Categorization of Hybrid Filtering Approaches for Hashtag Recommendation}
\label{table:hybrid_filtering}
\resizebox{\textwidth}{!}{%
\begin{tabular}{@{\extracolsep{\fill}} lccc}
\toprule
\textbf{Combination Type}&	\textbf{Key Strengths}&	\textbf{Key Weaknesses}&	\textbf{Papers}\\
\midrule
CBF + CF &	Good hashtag relevance via CF and CBF	& Lacks user-specific adaptation&	\cite{wei2019personalized,zhang2019hashtag}\\
CF + Personalization &	Deep personalization and cold-start handling&	No new content adaptability&	\cite{zhang2021howyoutagtweets,padungkiatwattana2022pac,tao2022personalized}\\
CBF + Personalization &	Learns hashtag meanings well, new hashtag adaptability&No user interactions included&\cite{badami2018cross}	\\
CBF + CF + Personalization &Best overall performance, all benefits&High computational complexity&\cite{bansal2024hybrid}\\
\bottomrule
\end{tabular}%
}
\end{table}
Recent studies have explored various hybrid filtering architectures. Graph-based methods, particularly those employing GNNs, have emerged as a dominant framework for hashtag recommendation. GNNs have been used to model complex relationships between users, posts, and hashtags, enabling seamless integration of multimodal content (text, images) with collaborative signals (user-hashtag co-usage) and personalized preferences \cite{kolyszko2024dynamic,javari2020weakly,wei2019personalized,bansal2024hybrid,padungkiatwattana2022pac}. Bansal \textit{et al.} \cite{bansal2024hybrid} propose hybrid GNN models that combine content, collaborative, and personalized features to address cold-start problems and enhance recommendation accuracy across multimodal microblogs. Similarly, research studies \cite{wei2019personalized,padungkiatwattana2022pac} employed GNNs to incorporate hashtag-user interactions and high-order relations in user communities, demonstrating improvements over traditional models.

Beyond GNNs, attention-based methods have been widely adopted to enhance hybrid filtering. These systems integrate personalized user behavior with content features by emphasizing critical components using mechanisms such as co-attention or topic attention \cite{zhang2021howyoutagtweets,zhang2017hashtag,ma2019co,bansal2022hybrid}. Personalized topic attention has been applied to align user hashtag preferences with evolving semantic content, as shown in works by Zhang \textit{et al.} \cite{zhang2021howyoutagtweets}. Meanwhile, multimodal hybrid systems \cite{yang2020amnn,chen2018topic}, fuse textual and visual content effectively to improve hashtag recommendations. Temporal and real-time dynamics are also increasingly addressed in hybrid systems, with incremental learning mechanisms emerging as a trend. Kolyszko \textit{et al.} \cite{kolyszko2024dynamic} introduce a class incremental learning approach for adapting to the rapidly changing nature of hashtag usage, while Peng \textit{et al.} \cite{peng2019modeling} leveraged sequential models to capture long-term user post histories for personalized hashtags. These works highlight the growing emphasis on dynamic and adaptive systems to address evolving user preferences and social media trends.

Despite these advances, challenges such as hashtag sparsity, long-tail hashtag distributions, and explainability remain underexplored. Models that explicitly address underrepresented hashtags or provide interpretable results through attention-based or explainable AI frameworks are still limited \cite{javari2020weakly,li2019long}. Additionally, while GNNs and multimodal architectures dominate the research landscape, broader integration of collaborative filtering signals with personalized and content-based features in unified systems remains an open area \cite{bansal2022hybrid,bansal2024hybrid,padungkiatwattana2022pac}.

\section{Method-based Hashtag Recommendation}
\label{sec:methods}
\begin{table}\scriptsize
    \centering
    \begin{tabular}{llp{6cm}}
        \toprule
        \textbf{Broad Category} & \textbf{Sub-Category} & \textbf{Papers} \\
        \midrule
        \multirow{5}{*}{Traditional Methods} & Frequency based & \cite{dovgopol2015twitter}, \cite{otsuka2014design}, \cite{alvari2017twitter}, \cite{nelaturu2015hashtag}, \cite{zhao2016personalized}, \cite{masood2017mfs} \\
        & Co-occurrence based & \cite{dovgopol2015twitter}, \cite{torres2020hashtags}, \cite{ben2017extended} \\
        & Matrix Factorization based & \cite{torres2020hashtags}, \cite{alsini2021hashtag} \\
        & Similarity-based & \cite{dey2017semtagger}, \cite{kowald2017temporal}, \cite{tran2018hashtag}, \cite{tajbakhsh2016microblogging} \\
        & Topic Modeling-based & \cite{li2016hashtag}, \cite{lei2020tag}, \cite{ma2019co}, \cite{tang2019integral}, \cite{gu2022time}, \cite{jafari2023unsupervised} \\
        &  & \cite{kalloubi2023listnet}, \cite{blei2003latent}, \cite{ma2013tag}, \cite{alash2020improve}, \cite{tangpong2023hashtag}, \cite{xu2020hybrid} \\
        \midrule
        \multirow{5}{*}{Deep Learning-based} & Embeddings-based & \cite{cantini2021learning}, \cite{kaviani2020emhash}, \cite{jain2024nlp}, \cite{li2016tweet}, \cite{dey2017emtagger} \\
        & CNN-based & \cite{gong2016hashtag}, \cite{li2016tweet}, \cite{zhang2017hashtag}, \cite{denton2015user} \\
        & RNN-based & \cite{li2016hashtag}, \cite{shen2019hashtag}, \cite{ma2018temporal} \\
        & Attention Mechanisms & \cite{zhang2019hashtag}, \cite{li2016hashtag}, \cite{gong2016hashtag}, \cite{shen2019hashtag}, \cite{hachaj2020image}, \cite{wu2018hashtag}, \cite{zhang2017hashtag} \\
        &  & \cite{khalil2023cross}, \cite{javari2020weakly}, \cite{yang2019self}, \cite{yang2020amnn}, \cite{fan2024right} \\
        & Transformer-based & \cite{cantini2021learning}, \cite{kaviani2020emhash}, \cite{bansal2022hybrid}, \cite{khalil2023cross}, \cite{feng2023tnod}, \cite{khalil2024mrlkg}, \cite{fan2024right} \\
        \midrule
        Graph-based & - & \cite{bansal2024hybrid}, \cite{kolyszko2024dynamic}, \cite{chen2021tagnet}, \cite{khalil2023cross}, \cite{khalil2024mrlkg}, \cite{wei2019personalized} \\
        &  & \cite{li2019long}, \cite{zhang2022twhin}, \cite{mehta2021open}, \cite{padungkiatwattana2022pac}, \cite{kipf2016semi} \\
        \midrule
        LLM-based & - & \cite{kumar2019fully}, \cite{tan2024llm}, \cite{jafari2023popular} \\
        \midrule
        External Knowledge-based & - & \cite{li2019long}, \cite{dovgopol2015twitter}, \cite{jayaratne2017content}, \cite{won2023extra}, \cite{ben2017extended}, \cite{vairavasundaram2015data} \\
        &  & \cite{al2019multi}, \cite{hong2018semantic}, \cite{kumar2021hashtag} \\
        \bottomrule
    \end{tabular}
    \caption{Categorization of Method-based Hashtag Recommendation}
    \label{tab:methods}
\end{table}
The field of hashtag recommendation has witnessed significant evolution between 2015 and 2024. Early approaches relied on traditional statistical methods such as TF-IDF and topic modeling. However, the increasing complexity of online content has driven the development of more sophisticated techniques, including deep learning, graph-based models, multimodal architectures, and, most recently, LLMs. These advancements aim to address challenges such as handling diverse data modalities (text, images, audio), adapting to real-time trends, and generating contextually relevant and personalized hashtags.The evolution of hashtag recommendation systems has been shaped by a wide range of methodologies, each designed to tackle distinct challenges and capitalize on specific data attributes. In this section, we systematically classify method-based approaches into five primary subfields: traditional methods, deep learning-based methods, graph-based methods, large language model (LLM)-based methods, and external knowledge-based methods as shown in \autoref{tab:methods}. 
\subsection{Traditional Methods for Hashtag Recommendation}
Despite the rise of modern neural approaches, traditional methods remain foundational, leveraging techniques such as frequency-based heuristics (TF-IDF), hashtag co-occurrence, matrix factorization, similarity-based methods, and topic models (LDA and LSA). These approaches exploit textual content, statistical patterns, and user interaction data to predict relevant hashtags in short, noisy social media text, as detailed below.
\subsubsection{Frequency-based
Methods for Hashtag Recommendation} These methods focus on extending TF-IDF for hashtag recommendation, addressing short-text sparsity, temporal sensitivity, or hashtag-specific adjustments.
Frequency-based methods, particularly TF-IDF, have been widely applied because of their simplicity and effectiveness in identifying semantically important terms. Variants such as Hashtag Frequency Inverse Hashtag Ubiquity (HF-IHU) \cite{otsuka2014design} and its time-sensitive extension (THF-IHU) \cite{gu2022time} adapt TF-IDF for short, dynamic social media data by incorporating hashtag-specific features, temporal weights, and document length normalization. Semantic augmentations further improve TF-IDF’s performance in handling sparsity and noise of short texts, enabling enhanced semantic relevance for hashtag prediction \cite{tajbakhsh2016microblogging}. Jafari \textit{et al.} \cite{jafari2023unsupervised} evaluated statistical keyword extraction methods (TF-IDF) for hashtag recommendation in unsupervised scenarios. Tajbaksh \textit{et al.} \cite{tajbakhsh2016microblogging} proposed semantic TF-IDF by incorporating semantic similarity metrics; improves accuracy sixfold over standard TF-IDF.
Alsini \textit{et al.} \cite{alsini2021hashtag} reviewed the use of TF-IDF for text-based hashtag recommendation, noting limitations in noisy, short text. Gu \textit{et al.} \cite{gu2022time} proposed a time-sensitive extension of TF-IDF (THF-IHU) that incorporates temporal adjustments and document length weighting; claimed $> 35\%$ improvement in recall.
Dovogpol \textit{et al.} \cite{dovgopol2015twitter} explored TF-IDF-based methods in real-time hashtag recommendation but emphasizes limitations due to sparseness in tweet-level content.
\subsubsection{Co-occurrence-based Methods for Hashtag Recommendation}  
These methods recommends hashtags via co-occurrence metrics or frequency analysis, addressing popularity bias or real-time needs. Torres \textit{et al.} \cite{torres2020hashtags} relied on statistical relationships between hashtags but face challenges in mitigating bias toward popular hashtags, addressed through weighting schemes \cite{ben2017extended}. These methods are simple and scalable but prone to over-recommending frequently used hashtags. 
Dovgopol \textit{et al.} \cite{dovgopol2015twitter} used frequency, co-occurrence graphs, and similarity measures for recommending hashtags on X.
\subsubsection{Matrix Factorization-based Methods for Hashtag Recommendation}
Matrix factorization techniques, such as weighted low-rank factorization \cite{alvari2017twitter} and collaborative filtering \cite{torres2020hashtags}, modeled latent relationships between users and hashtags, offering robust performance when tweet content is unavailable. This technique is effective where content-based features are unavailable. However, due to purely statistical relationships, these methods lack semantic grounding, necessitating hybrid solutions. 
\subsubsection{Similarity-based Methods for Hashtag Recommendation}
Textual similarity methods form the backbone of many early studies, utilizing metrics such as TF-IDF, cosine similarity, and co-occurrence frequency.
Poudel \textit{et al.} \cite{poudelcontent} further benchmarked textual methods such as Naive Bayes, KNN, and cosine similarity, demonstrating their performance on shared datasets. Beyond simple textual analysis, semantic similarity methods exploit latent topic structures and external knowledge bases to enrich content understanding. Dey \textit{et al.} \cite{dey2017semtagger} applied LDA to model topics and calculate semantic similarity for hashtag assignments. While traditional methods remain valuable, their reliance on static linguistic features and limited contextual awareness often restricts their applicability to short and noisy social media texts. Hybrid methods, which integrate content, temporal, and user-based features, represent a notable evolution. Tran \textit{et al.} \cite{tran2018hashtag} combined content similarity, user characteristics, and hashtag popularity trends for personalized recommendations. Kowald \textit{et al.} \cite{kowald2017temporal} combined TF-IDF-based content similarity with temporal dynamics in hashtag reuse. 

Despite advances, challenges persist, including the ineffective handling of rare or emerging hashtags (cold-start) and the lack of robust engagement-based evaluation metrics.
\subsubsection{Topic Modeling-based Methods for Hashtag Recommendation}
Topic modeling approaches, particularly LDA  \cite{blei2003latent}, are commonly used to uncover latent themes in tweets, aligning hashtags with extracted topics \cite{nelaturu2015hashtag,zhao2016personalized}. Extensions such as Tag-LDA \cite{ma2013tag} incorporate hashtags directly into the generative process, while multi-feature space LDA (MFS-LDA) addresses cold-start problems and integrates additional metadata \cite{masood2017mfs}. Despite their prevalence, LDA-based methods struggle with short-text sparsity, prompting solutions such as pseudo-document pooling, hashtag-informed pre-processing \cite{nelaturu2015hashtag}, and semantic knowledge integration \cite{alash2020improve}. Zhao \textit{et al.} \cite{zhao2016personalized} personalized hashtag recommendations using an LDA-based topic model adapted to individual users.

However, traditional LDA-based hashtag recommendation overlooks inter-dependencies between generated hashtags and can struggle with the limited term co-occurrence information present in short texts such as tweets \cite{li2016hashtag,tangpong2023hashtag}. While effective for long documents, LDA may not be ideal for tweets or similar short-form content. Extensions such as incorporating topical attention in LSTMs have been explored \cite{li2016hashtag}. Other topic modeling approaches include constrained LDA, which incorporates relationships between latent topics and tags \cite{ma2019co}, and hybrid models that combine tag recommendation and classification \cite{xu2020hybrid}. Association rule mining with keyword extraction has also been used in conjunction with topic modeling \cite{lei2020tag}. Neural Topic Models (NTM), such as Hashtag-NTM \cite{tang2019integral,tangpong2023hashtag}, utilize neural networks to improve topic modeling and capture word-hashtag relationships.  Specifically, Hashtag-NTM enhances topic distribution complexity and word-hashtag relationship modeling.
 
\textbf{Strengths}
\begin{itemize}
\item Computationally lightweight.
\item Effective for small datasets or when complex semantic relationships are not critical.
\item Useful in domain-specific applications (disaster response, COVID-19 tweets).
\end{itemize}

\textbf{Limitations}
\begin{itemize}
\item Limited ability to capture polysemy (multiple meanings of words) or relationships beyond co-occurrence.
\item Requires extensive preprocessing 
\item Can struggle with rare hashtags.
\end{itemize}
These shortcomings have motivated the development of more advanced techniques, including deep learning, graph-based models, multimodal architectures, and, more recently, LLMs, which will be discussed in subsequent sections.
\subsection{Deep Learning-based Methods for Hashtag Recommendation}
The emergence of deep learning has substantially advanced representation learning in the domain of hashtag recommendation. Deep learning architectures, such as Convolutional Neural Networks (CNNs), Recurrent Neural Networks (RNNs), and Transformers, have established a robust foundation for semantic representation, contextual modeling, and multi-label hashtag prediction. Neural networks have gained widespread popularity in recent years due to their capacity to model complex non-linear relationships between entities and their exceptional expressive power. Commonly employed neural network architectures in encoding models include embedding-based models, CNNs, RNNs, attention mechanisms, and Transformers. In this subsection, we classify deep learning-based approaches into five principal subfields: embedding-based methods, CNN-based methods, RNN-based methods, attention mechanisms, and Transformer-based methods. 
\subsubsection{Embeddings-based Methods for Hashtag Recommendation}
Embedding models represent a core component of deep learning-based hashtag recommendation.
\begin{itemize}
\item Word Embeddings:
Early research relied on static word embeddings such as Word2Vec and GloVe to represent words and predict hashtags.
EmTaggeR \cite{dey2017emtagger} combined global and hashtag-specific embeddings to compute semantic similarity.
Li \textit{et al.} \cite{li2016tweet} used Word2Vec to generate vector-based features for neural networks.
\item Contextual Embeddings:
Context-sensitive models such as BERT and SBERT produce embeddings that vary based on the surrounding context, addressing semantic nuances. EmHash \cite{kaviani2020emhash} utilized BERT embeddings, outperforming methods based on simpler word embeddings or LDA in hashtag recommendations.
HASHET \cite{cantini2021learning} created latent spaces for both text and hashtag embeddings, enabling semantically rich recommendations.
Jain \textit{et al.} \cite{jain2024nlp} combined BERT embeddings with LSTM to perform multi-label hashtag classification for COVID-related tweets.
\end{itemize}

\textbf{Strengths}
\begin{itemize}
\item Contextual embeddings (BERT) capture variations in meaning based on context, significantly outperforming static embeddings.
\item Facilitates semantic understanding of the text and its relation to hashtags.
\end{itemize}

\textbf{Limitations}
\begin{itemize}
\item Embedding models can struggle with domain-specific hashtags unless supplemented with specialized training.
\item High computational cost for contextual embedding models such as BERT.
\end{itemize}

Future directions can explore lightweight and efficient adaptations ( DistilBERT, MobileBERT) for embedding-based hashtag recommendation.
\subsubsection{CNN-based Methods for Hashtag Recommendation}
CNN-based methods have been widely explored for their ability to extract local patterns, such as n-grams, from text data. Gong \textit{et al.} \cite{gong2016hashtag} proposed one of the earliest deep learning models for hashtag recommendation, leveraging attention-enhanced CNNs to identify trigger words for more effective tag prediction and demonstrated improved F1-score.
Li \textit{et al.} \cite{li2016tweet} combined CNN with LSTMs to capture semantic sentence vectors from tweets.
CNNs have also been applied to extract features from images for hashtag prediction. Denton \textit{et al.} \cite{denton2015user} used CNN-based visual features combined with user metadata for hashtag prediction, framing it as a multi-label classification problem. Bharadwaj \textit{et al.} \cite{bharadwajprediction} employed CNNs for image-based hashtag classification using a custom image dataset with hashtag labels under various categories. 
Subsequent studies incorporated multimodal inputs, integrating textual and visual features using CNN architectures with co-attention mechanisms to align text and images, resulting in improved accuracy for posts on X \cite{zhang2017hashtag}.

\textbf{Strengths}

CNNs are effective for short texts such as tweets and are computationally less intensive than RNNs. They have also shown promise in extracting visual features for image-based hashtag recommendation.
However, while effective for localized feature extraction, CNNs struggle to model long-range semantic dependencies, limiting their standalone utility in more complex scenarios \cite{khalil2023cross}.

\textbf{Limitations}
\begin{itemize}
\item Limited attention was paid to hashtag relationships or domain-specific hashtag evolution
\item Lack of advanced interaction mechanisms: Simple CNN pipelines failed to capture complex patterns or relationships among regions in an image.
\item Dataset dependency: Performance heavily relied on the curation quality of image datasets.
\item Limited ability to capture long-distance dependencies in text.
\end{itemize}
\subsubsection{RNN-based Methods for Hashtag Recommendation}
RNNs, particularly LSTMs, are well-suited for sequence modeling due to their ability to capture sequential dependencies.
Li \textit{et al.} \cite{li2016hashtag} introduced a topical attention-enhanced LSTM that incorporated topic modeling to improve context alignment, while Ma \textit{et al.} \cite{ma2018temporal} developed temporal-enhanced sentence-level attention mechanisms to handle temporal hashtag trends and noisy data. Shen \textit{et al.} \cite{shen2019hashtag} used LSTMs with self-attention, reframing hashtag recommendation as a sequence labeling task, enabling the generation of new hashtags. Advances in RNN-based methods further leveraged Tree-LSTMs to incorporate syntactic structure for hashtag representation, achieving significant improvements over standard LSTMs.

\textbf{Strengths}

RNNs capture temporal and sequential patterns, which is particularly relevant for hashtags derived from context-specific phrases.

\textbf{Limitations}

Vanilla RNNs and LSTMs face challenges with scalability, inefficiencies in training, vanishing gradients over very long sequences.
\subsubsection{Attention Mechanisms for Hashtag Recommendation}
Attention mechanisms highlight the most relevant parts of constituent modalities of input posts, enabling the model to focus on key features for hashtag prediction. These mechanisms are crucial for improving interpretability and managing noisy or incomplete data \cite{ma2018temporal,zhang2017hashtag}.

\textbf{Attention in CNNs:}
First introduced in CNN models to enhance semantic understanding.
\cite{gong2016hashtag} combined CNNs with attention for improved hashtag prediction by focusing on trigger words.

\textbf{Attention in LSTMs:} Attention in LSTMs helps in capturing contextual clues and word importance dynamically.
Li \textit{et al.} \cite{li2016hashtag} employed an attention-based LSTM incorporates topical relevance into hashtag prediction for context alignment.
Shen \textit{et al.} \cite{shen2019hashtag} used self-attention in LSTMs to identify highly relevant content within posts, improving sequence tagging.
Gong \textit{et al.} \cite{gong2016hashtag} proposed attention-enhanced CNN for local pattern prioritization and trigger word selection. Zhang \textit{et al.} \cite{zhang2019hashtag} devised a parallel co-attention mechanism for multimodal inputs.
Zhang \textit{et al.} \cite{zhang2017hashtag} proposed co-attention for cross-modal (text + images) hashtag alignment in microblogs.
Javari \textit{et al.} \cite{javari2020weakly} leverages user-followee/follower links and weakly supervised attention to embed users/hashtags in a shared latent space. Yang \textit{et al.} \cite{yang2019self} utilized self-attention mechanisms to enhance hashtag recommendations by identifying relevant text features, addressing contextual and semantic relationships. Khalil \textit{et al.} \cite{khalil2023cross} devised a cross-attention transformer capturing text and image features for multimodal hashtags.
Yang \textit{et al.} \cite{yang2020amnn} proposed a multimodal attention-based model for posts containing text and images

Transformer-based architectures inherently rely on self-attention, making this method a fundamental enhancement.
Li \textit{et al.} \cite{li2016hashtag} applied BERT embeddings with contextual self-attention for hashtag predictions.
Fan \textit{et al.} \cite{fan2024right} used transformers in retrieval-augmented generation (RIGHT).

\textbf{Strengths:}
Attention mechanisms improve interpretability by emphasizing essential features.
Effective for noisy and sparse text data typical in social media.
Limitations:
Attention models often require substantial processing overhead.
Over-attention to certain words may cause bias, ignoring less obvious yet significant terms.

\textbf{Advances in Attention Mechanisms for Visual Hashtag Prediction:}

The integration of attention mechanisms into visual-only hashtag recommendation systems introduced a way to focus on the most relevant regions of an image, improving semantic alignment between image content and hashtags.
Wu \textit{et al.} \cite{wu2018hashtag} introduced an attention-based neural network exclusively for visual hashtag generation. Their attention mechanism dynamically assigned weights to different image patches to prioritize regions most relevant to the predicted hashtags.
Hachaj \textit{et al.} \cite{hachaj2020image} combined deep neural network outputs for hashtag generation. The system utilized self-attention within the convolutional layers to focus on regions of interest to generate hashtags from image content alone.

\textbf{Methods
Self-Attention Mechanisms}: Enabled better focus on semantically rich regions within input images. Self-attention fed into downstream prediction layers for better image-to-hashtag mappings \cite{,hachaj2020image,wu2018hashtag}.
Region Prioritization: Attention modules dynamically weighted features extracted from different regions of the image, aligning more closely with relevant hashtags.

These works demonstrated that introducing attention mechanisms into visual hashtag recommendation pipelines improved the depth of feature extraction, enhancing system accuracy across datasets.
\subsubsection{Transformer-based Methods for Hashtag Recommendation}
Pre-trained transformers provide robust contextual embeddings, outperforming CNNs and RNNs in tasks requiring long-range semantic understanding
Transformers, such as BERT and generative approaches such as GPT, have become dominant due to their ability to model long-range dependencies and context comprehensively \cite{kaviani2020emhash,bansal2022hybrid,khalil2023cross}. Fine-tuned transformers integrating visual and textual inputs through cross-attention mechanisms, excel in dynamic hashtag prediction \cite{khalil2023cross,khalil2024mrlkg}. Studies such as \cite{cantini2021learning,kaviani2020emhash} used BERT embeddings for hashtag prediction and semantic mapping.
Fan \textit{et al.} \cite{fan2024right} proposed a hybrid system that combines retrieval and transformer-based generative methods for recommending mainstream hashtags. Khalil \textit{et al.} \cite{khalil2023cross} integrated transformer-based contextual embeddings with a keyword-guided GCN to capture post-specific and relational trends in hashtags. 

\textbf{Strengths}: Context-aware, capable of handling diverse linguistic structures.
Limitations: Computational resource demands, particularly for large datasets, and sensitivity to overfitting on small datasets.

\textbf{Strengths of Deep Learning}:
Captures context, semantics, and relationships between words and potential hashtags.
Handles sparsity in textual data (short posts, noisy text).
Challenges:
Increased computational and training costs with deep learning models.
Neural networks demand large annotated datasets, which might not always be readily available.
Overfitting, especially on imbalanced datasets with infrequent hashtags.
Future Directions:
Improving scalability for large-scale hashtag datasets using parameter-efficient methods such as prefix-tuning and adapters.
Exploring multitask transformer-based architectures, combining hashtag prediction with related auxiliary tasks (sentiment prediction).

Transformer Architectures for Visual Hashtags
The adoption of transformer-based architectures, particularly ViT, began around 2020, replacing CNNs as the standard feature extractors for image-only tasks. Feng \textit{et al.}  \cite{feng2023tnod} leveraged a Vision Transformer (ViT)-based model to focus on object-level entities within images. The system used multi-layer attention mechanisms and object segmentation to detect semantically meaningful regions for hashtag prediction.
Although \cite{khalil2023cross} focused on broader multimodal goals, portions of this work showcase a purely vision-based transformer as a backbone for generating hashtag embeddings.
Methods
ViT: These focus on dividing an image into small patches, encoding each as a token, and using self-attention to model global dependencies across patches.
Object Detection Modules: TNOD \cite{feng2023tnod} integrated object detection layers to identify and prioritize entities related to hashtags, improving tagging accuracy on benchmarks.
Contributions
Transformers offered a significant leap forward in modeling long-range dependencies within images, capturing complex interactions that CNNs and attention mechanisms struggled to address. ViTs dramatically improved precision and recall in datasets involving diverse and challenging visual content.
\subsection{Graph-based Methods for Hashtag Recommendation}
Given the complex, graph-structured relationships inherent in social media—spanning users, posts, hashtags, and interactions—graph-based methods have gained significant traction in recent years. These approaches leverage graph embeddings to model these relationships and integrate features such as temporal dynamics, user preferences, and multimodal content, making them particularly effective for hashtag recommendation tasks.
Graph-based hashtag recommendation systems represent social media data as heterogeneous or multi-relational graphs. Nodes represent entities (e.g., users, posts, hashtags) and edges encode user interactions, co-occurrence, or content similarity. GCNs \cite{kipf2016semi} have become a popular tool for embedding these graph structures into latent vector spaces, enabling downstream recommendation models to identify relevant hashtags.
\begin{figure}
\includegraphics[width=\textwidth]{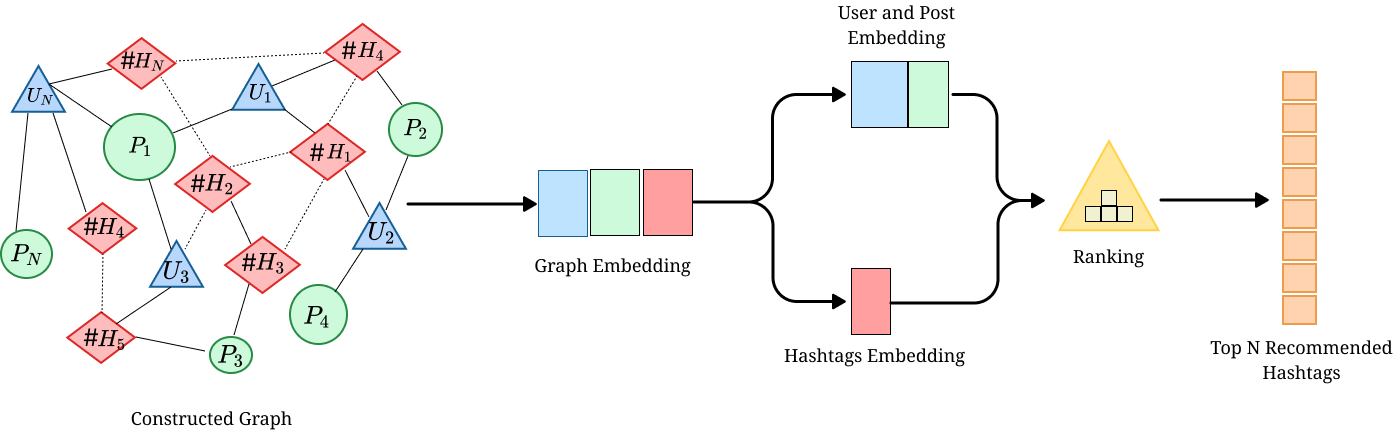}
\caption{Overview of GNN-based Method for Hashtag Recommendation}
\label{fig:gnn_htrc}
\end{figure}
\autoref{fig:gnn_htrc} presents a graph-based hashtag recommendation system utilizing GNNs. A heterogeneous graph is constructed, connecting users, posts, and hashtags. Embeddings are learned for each node type, capturing individual features and relationships. GNN processes the graph structure and node embeddings to generate a graph embedding, encoding the overall network information. This graph embedding, along with hashtag embeddings, is used to rank and recommend the most relevant hashtags for a given user or post.

Numerous studies have investigated diverse strategies for graph construction and utilization in hashtag recommendation systems. Li \textit{et al.} \cite{li2019long} addressed the challenge of long-tail hashtag distribution by constructing a hashtag graph enriched with external knowledge and employing a pairwise interactive embedding network to model interactions between hashtags, micro-videos, and users. However, this approach does not account for the modality-specific preferences of users, which, if incorporated, could significantly enhance the personalization of hashtag recommendations. Similarly, TAGNet \cite{chen2021tagnet} integrated visual similarity between images directly into a graph structure. Utilizing CNNs for initial visual embeddings, a  GCN propagated hashtag labels to semantically similar images, based on the assumption that visually analogous images often share hashtags. The model incorporated a Triplet Attention module to capture the influence of visual content, captions, and user interactions on node features, which were subsequently aggregated and propagated by the GCN for hashtag recommendation.

Other research efforts have focused on heterogeneous graph construction. Mehta \textit{et al.} \cite{mehta2021open} developed a graph that connected hashtags (based on semantic co-occurrence), videos (through shared hashtags), and direct links between videos and their assigned hashtags. GCNs were employed to generate micro-video embeddings for hashtag recommendation. Zhang \textit{et al.} \cite{zhang2022twhin} constructed a bipartite graph of tweets and users to identify socially similar tweets for multilingual hashtag prediction. Despite its innovative approach, TwHIN-BERT falls short in recommending hashtags that align with users' individual interests and linguistic styles, highlighting a limitation in its ability to fully capture user-specific preferences. Wei \textit{et al.} \cite{wei2019personalized} constructed a heterogeneous graph where nodes represent users, hashtags, and micro-videos, and edges encode relationships such as video-hashtag interactions, user-video associations, and user-hashtag co-preferences. Their approach used GCNs to propagate and update node embeddings through message passing. Updated embeddings were then used to refine representations of user interests and hashtag semantics, enabling personalized hashtag recommendation for micro-videos.
High-order relationships and multiple node interactions have also been utilized to improve personalization and accuracy. PAC-MAN \cite{padungkiatwattana2022pac} explicitly modeled relationships across three node types: users, posts, and hashtags, as well as high-order relations such as hashtag-hashtag co-occurrence and user-user social links. Graph embeddings were computed using Multi-Relational GNNs, and resulting representations were fused with word embeddings for word-level personalization. Finally, embeddings were applied to recommend hashtags most aligned with the user’s preferences and the semantic relevance of the post content. Khalil \textit{et al.} \cite{khalil2023cross} proposed a keyword-guided graph that integrated textual embeddings from transformers with graph embeddings generated by a GCN, producing highly contextualized recommendations. The hybrid model \cite{bansal2024hybrid} demonstrated superior performance in recommending hashtags for micro-video platforms by balancing content-based filtering (video/post content) and collaborative filtering (user-to-user and user-hashtag links) via graph embeddings. The authors constructed a graph with nodes representing users, hashtags, and micro-videos, while edges modeled complex interactions such as user-to-hashtag, user-to-video, and video-to-hashtag associations. This hybrid system leveraged GNNs to encode diverse, multimodal entity relationships. 

Temporal aspects have also been integrated into graph-based systems to combat evolving social media trends. Kolyszko \textit{et al.} \cite{kolyszko2024dynamic} modeled dynamic graphs where nodes and edges adapted to new trends (e.g., emerging hashtags). Similarly, Mehta \textit{et al.} \cite{mehta2021open} constructed a temporal graph for trending hashtag prediction, where embeddings updated in real time were used to track and recommend popular hashtags.

These studies illustrate the increasing sophistication of graph-based hashtag recommendation systems, which now integrate heterogeneous and multimodal graphs \cite{,chen2021tagnet,khalil2024mrlkg,bansal2024hybrid}, dynamic temporal updates \cite{kolyszko2024dynamic,mehta2021open}, and high-order relational modeling \cite{padungkiatwattana2022pac}. These approaches capitalize on graph embeddings’ ability to encode rich, structural information, which is subsequently used in ranking or classification models to recommend contextually appropriate hashtags. This evolution reflects the diverse challenges of hashtag recommendation, such as cold-start issues \cite{bansal2024hybrid}, long-tail hashtags \cite{li2019long}, and the fusion of multimodal data \cite{chen2021tagnet,bansal2024hybrid}.

Despite these advancements, gaps remain in scaling to massive social graphs, addressing sparsity in hashtags (long-tail), and generalizing to unseen users. Emerging paradigms such as contrastive graph learning, meta-learning, and large pretrained GNNs (GraphGPT) represent promising avenues to improve robustness and scalability. This integration of multimodal data, graph embeddings, and real-time updates highlights the expanding horizons of hashtag recommendation research and its potential for improving user engagement on social media platforms.

\subsection{LLM-based Methods for Hashtag Recommendation}
LLMs, with their extensive pre-trained knowledge and advanced reasoning capabilities, are well-suited for capturing the dynamic and context-dependent nature of hashtags. Unlike traditional recommendation tasks that rely on sparse numerical features such as UserIDs or ItemIDs, hashtags contain rich semantic information, making them particularly amenable to LLM-based approaches. The ability of LLMs to infer meaning from limited data allows for enhanced contextual understanding, thereby improving the accuracy and relevance of hashtag recommendations. However, integrating LLMs into multimodal hashtag recommendation systems presents several critical challenges, including efficient modality alignment for processing diverse input types, sensitivity to input order when incorporating multiple historical user samples, and substantial computational overhead associated with fine-tuning, necessitating strategies to optimize the number of trainable parameters.

To address these challenges, LLM-Augmented Multimodal Hashtag Recommendation (LLM-HR) introduces key innovations that leverage LLMs effectively. The Order-Agnostic Embedding (OAE) module plays a crucial role in modality alignment and mitigates sensitivity to input order by generating soft prompts that facilitate seamless integration of multimodal content with LLM knowledge. This approach ensures that multiple microposts are processed efficiently without being influenced by historical sample order. Furthermore, a hybrid prompt learning strategy, combining soft prompts with manually constructed hard prompts, enables more efficient fine-tuning by reducing computational costs while maintaining robust performance. By employing an end-to-end design, LLM-HR eliminates complex preprocessing steps and streamlines deployment through the use of a single LLM, thereby enhancing scalability and reducing overall computational burden. Unlike prior approaches such as MACON \cite{zhang2019hashtag}, which lack strong inference capabilities and perform suboptimally in few-shot or zero-shot learning scenarios, LLM-HR overcomes these limitations by leveraging its OAE module and hybrid prompt learning framework.

Recent advancements in LLM-based hashtag recommendation further underscore the potential of these models in multimodal contexts. LLM-MHR \cite{tan2024llm} has demonstrated robust processing of multimodal data through LLM pre-training, while Kumar \textit{et al.} \cite{kumar2019fully} have explored embedding-based zero-shot learning for the prediction of previously unseen hashtags. Additionally, GPT-2-based models \cite{jafari2023popular} have been employed for aligning image-text inputs to facilitate trend-sensitive recommendations. The adoption of transformer-based architectures, including BERT-based models \cite{cantini2021learning,jain2024nlp} and generative frameworks based on GPT \cite{jafari2023popular}, has significantly advanced contextual understanding in both unimodal and multimodal hashtag recommendation systems. These innovations have enabled zero-shot and few-shot learning, allowing systems to recommend novel and previously unseen hashtags without requiring extensive task-specific training. Despite these advancements, challenges remain in ensuring effective cross-modality alignment, optimizing computational efficiency for fine-tuning, and improving the robustness of models against real-world noise. Addressing these issues is crucial for further enhancing the performance and applicability of LLM-based hashtag recommendation systems in dynamic social media environments.

In summary, the field of hashtag recommendation has progressed from traditional content analytics to transformer-powered multimodal systems, emphasizing user personalization, real-time trends, and data diversity. This evolution reflects a growing focus on leveraging advanced representation learning, graph-based relational modeling, and LLMs to address the unique challenges of dynamic social media environments. The reviewed works collectively provide a comprehensive foundation for understanding the landscape of hashtag recommendation methodologies and the remaining gaps in the field.
\subsection{External Knowledge-based Methods for Hashtag Recommendation}
These knowledge sources enhance the tagging process by resolving lexical ambiguities, inferring semantic relationships, and improving the interpretability of the target content \cite{vairavasundaram2015data}.
Hashtag recommendation, especially for short-text content such as microblogs, benefits significantly from incorporating external knowledge. This enriches contextual understanding and improves the accuracy and relevance of suggested hashtags. Various approaches have been explored leveraging complementary knowledge sources,
such as ontologies \cite{al2019multi}, Wikipedia \cite{hong2018semantic,kumar2021hashtag} and Linked Open Data \cite{jayaratne2017content} to enhance hashtag generation by providing semantic information and contextual clues. 
\begin{table}\footnotesize
\centering
  \caption{Categorization of Papers Employing External Knowledge Bases for Hashtag Recommendation}
\label{table:external_knnowledge}
  \begin{tabular}{cccc}
    \toprule
    \textbf{Paper} & \textbf{Problem Tackled} & \textbf{External Knowledge Source }\\
    \midrule
    Kumar \textit{et al.} \cite{kumar2021hashtag} & Data Sparsity, Semantic Gap Problem & Wikipedia\\
    Li \textit{et al.} \cite{li2019long} & Long Tail Hashtag Distribution & Wordnet, Synnets\\
    Won \textit{et al.} \cite{won2023extra} & Contextual Understanding and Relevance& Linked Open Data\\
    \bottomrule
  \end{tabular}
\end{table}
\begin{itemize}
\item Wikipedia: Wikipedia provides comprehensive articles offering rich contextual information about entities, concepts, and relationships. This helps understand the semantics of a post more deeply, going beyond just the words present in the text. Wikipedia can also resolve ambiguity in short texts by providing disambiguated entities and concepts. For example, if a post mentions ``apple", Wikipedia can help determine whether it refers to the fruit or the company.
\item Linked Open Data (LOD): LOD sources contain structured information about entities and their relationships, allowing for concept expansion and linking related concepts. This can help suggest more relevant hashtags that may not be explicitly mentioned in the text. For instance, if a post mentions a movie title, LOD sources such as DBpedia can provide information about the movie's genre, actors, directors, and related movies, leading to suggestions such as \#scifi, \#actionmovie, or hashtags related to the actors or director.
\end{itemize}
Won \textit{et al.} \cite{won2023extra} showcased the effectiveness of using the Open Directory Project (ODP) for multimodal hashtag recommendation. Their system, EXTRA, employs a BERT-based classifier trained on the ODP dataset to categorize web pages and extract relevant categories from both image captions (generated using an OFA model) and post text. This external knowledge is then integrated into a transformer-based architecture with a pre-trained multimodal alignment model (FLAVA) to process visual and textual information for generating hashtag suggestions. EXTRA \cite{won2023extra} employs a
pre-trained multimodal alignment model, FLAVA, to simultaneously process visual and textual information. Notably, EXTRA does not utilize UserIDs, resulting in suboptimal utilization of user historical information. Li \textit{et al.} \cite{li2019long} enhanced hashtag recommendation by incorporating external knowledge to construct a hashtag relation graph. This graph captures semantic relationships among hashtags, including compositional, super-subordinate, positive, and co-occurrence, which are encoded in a correlation matrix. The authors utilize this external knowledge to inform a propagation mechanism within a graph convolutional network. This mechanism facilitates knowledge transfer from frequent to less frequent, long-tail hashtags, thereby improving the model's ability to represent and recommend less common but relevant hashtags. The integration of external knowledge allows for a richer understanding of hashtag relationships beyond simple co-occurrence, addressing the long-tail phenomenon and improving the overall effectiveness of hashtag recommendation.
Kumar \textit{et al.} \cite{kumar2021hashtag} enhanced hashtag recommendation for short tweets by leveraging Wikipedia as an external knowledge source. The system employs Stanford CoreNLP to extract named entities, which are then linked to corresponding Wikipedia pages for contextual information. To address ambiguity, a relevance scoring mechanism selects the most appropriate Wikipedia page for each entity based on the tweet's content.  Furthermore, the system extracts semantically related entities and anchor texts from the chosen Wikipedia page to expand the contextual knowledge. To ensure conciseness and relevance, only the top three paragraphs of each Wikipedia page are utilized, filtering out noise and irrelevant details. By integrating this extracted knowledge, the system enriches the contextual representation of the tweet, bridging the semantic gap often present in short texts, leading to more informed and relevant hashtag suggestions. Tajbakhsh \textit{et al.} \cite{tajbakhsh2016microblogging} used semantic similarity (WordNet metrics) combined with TF-IDF for improved short-text hashtag recommendation. Ben \textit{et al.} \cite{ben2017extended} proposed a spreading activation technique using semantic networks (e.g., DBpedia) for hashtag recommendation. Dovogpol \textit{et al.} \cite{dovgopol2015twitter} used external sources (WordNet, Wikipedia) to mitigate sparsity and improve traditional recommendation methods. \autoref{table:external_knnowledge} depicts the categorization of papers employing external knowledge bases for hashtag recommendation and problems tackled.

By incorporating external knowledge from Wikipedia and LOD, hashtag recommendation systems can move beyond simple keyword matching and provide more accurate, relevant, and diverse suggestions, thereby enhancing user experience and content discoverability.
\section{Datasets for Hashtag Recommendation}
\label{sec:datasets}
There are numerous publicly available datasets for hashtag recommender systems, each containing different information, thus researchers have different focuses. 
In the context of real-world multimodal hashtag recommendation tasks, it is common to encounter numerous hashtags that
do not exist in the candidate set. Microposts are inherently
temporal, and a plethora of compound word hashtags and
meme hashtags emerge over time.
Chen \textit{et al.} \cite{chen2021tagnet} utilized a large-scale Instagram dataset and measures recall/precision improvements with TAGNet. Kolyszko \textit{et al.} \cite{kolyszko2024dynamic} extends HARRISON dataset by adding temporal dimensions for tracking trends.
Zhang \textit{et al.} \cite{zhang2019hashtag} developed X-based hashtag benchmark datasets, introducing co-attention modeling pipelines.
Zafar \textit{et al.} \cite{zafar2024novel} constructed InstaHash dataset with 12,345 posts categorized by multimodal features for hashtag insights.
Badami \textit{et al.} \cite{badami2018cross} constructed a graph-based dataset to investigate hashtag similarity and discover latent story threads. \autoref{table:datasets} extensively describes existing datasets constructed for hashtag recommendation in social media.
\begin{table}\scriptsize
\centering
\caption{Existing Datasets for Hashtag Recommendation}
\label{table:datasets}
\resizebox{\textwidth}{!}{
\begin{tabular}{@{\extracolsep{\fill}} lccccccccc}
\toprule
\textbf{Proposed by} & \textbf{Dataset Name}& \textbf{Year} &\textbf{Modality} & \textbf{SNS} & \textbf{Available at:} & \textbf{Samples} &\textbf{Hashtags} & \textbf{Personalized} & \textbf{Features}\\
\midrule
Park \textit{et al. }\cite{park2016harrison} & HARRISON &2016 & Image & Instagram & \href{https://github.com/minstone/HARRISON-Dataset}{Link}& 57,383& 997 & No& Images only\\
Chua \textit{et al.} \cite{chua2009nus}&NUS WIDE & 2009 &Image & Flickr & \href{https://www.kaggle.com/datasets/xinleili/nuswide}{Link}&2,45,603 & 5018 &No& Images only\\
Thomee \textit{et al.} \cite{thomee2016yfcc100m}& YFCC100M & 2016&Micro-videos & Flickr & \href{https://github.com/chi0tzp/YFCC100M-Downloader}{Link}& 1,34,992 & 23,054 & Yes&Multimodal\\
Li \textit{et al.} \cite{li2019long}& INSVIDEO &2019& Micro-videos & Instagram & \href{https://anon425.wixsite.com/v2ht}{Link}& 48,888 & 12,194 & Yes&Multimodal+Users\\
Zhang \textit{et al.} \cite{zhang2019hashtag}&-& 2019& Text+Image & Instagram & \href{https://github.com/SoftWiser-group/MaCon/tree/master/data}{Link}& 6,24,520 & 3896 & Yes&Multimodal+User\\
Wei \textit{et al.} \cite{wei2019personalized}&- & 2019&Micro-videos& Instagram & -& 48,888 & 12,194 & No&\\
Bansal \textit{et al.} \cite{bansal2024multilingual} & IndicHash&2024& Text& X & -& 81,944 & 37,151 & Yes & 7 languages+ English\\
Bansal \textit{et al.}\cite{bansal2022hybrid} & TINS& 2022 & Text & X & -& 23,868 & 9,780 & Yes & User history\\
Chen \textit{et al.} \cite{chen2016micro}& TMALL & 2016 & Micro-videos &Vine & \href{http://acmmm2016.wix.com/}{Link}&3,03,242 & 3000+ & Yes& \\
Al \textit{et al.} \cite{al2024wasm} & WASM & 2024 & Text & X& -& 1,01,099 & 87 & No & Arabic Language\\
Yang \textit{et al.} \cite{yang2020amnn} & MMINS & 2020 & Image+Text& Instagram & \href{https://github.com/w5688414/AMNN}{Link}& 56,861 & 6,67,227 & Yes & PostID, userid\\
Zhang \textit{et al.} \cite{zhang2022twhin} & 2022 &-& Text & X & \href{https://github.com/xinyangz/TwHIN-BERT}{Link}& 1,00,000 & 500 & No & 50 Languages\\
Chakrabarti \textit{et al.} \cite{chakrabarti2023hashtag} & - & 2023 & Text & X & -& 15,000 & 21,536 & Yes & User+ Popularity attributes\\
Jain \textit{et al.} \cite{jain2024nlp} & - &2024 & Text & X& -& 1,00,000 & 1,350 & No & Covid-related\\
Kang \textit{et al.} \cite{kang2020leveraging} & - & 2020 & Text& Instagram & -& 87,872 & 907 & No & Location, Time\\
Jeong \textit{et al.} \cite{jeong2022demohash} & -&2022 &Text+Image&Instagram&\href{https://github.com/dxlabskku/DemoHash/tree/main/Data}{Link}&3,840 & 80,871 & Yes & Race, gender, emotion, age\\
Chowdhury \textit{et al.} \cite{chowdhury2020identifying} & - & 2020 & Text & X & \href{https://github.com/JRC1995/Tweet-Disaster-Keyphrase}{Link}& 67,288 & - & No & 37 Disaster Events\\
Liu \textit{et al.} \cite{liu2020user}& - &2020 & Micr-videos & Musical.ly & -& 10,291 & 669 & Yes & Gender, age, country, user history\\
Kou \textit{et al.} \cite{kou2018hashtag}& - & 2018 & Text & Sina Weibo & -& 67,835 & 4,061 & No & -\\
Gong \textit{et al.} \cite{gong2017phrase} & - & 2017 & Text & Sina Weibo & -& 50,000 & 3,174 & No & Phrase hashtags\\
Mao \textit{et al.} \cite{ma2018temporal} & - & 2018 & Text & Sina Weibo & -& 1,69,250 & 2,000 & No & Chinese\\
Zhang \textit{et al.} \cite{zhang2021howyoutagtweets} & - & 2021 & Text & X & \href{https://github.com/polyusmart/Personalized-Hashtag-Preferences}{Link}& 33,881 & 22,320 & Yes & User history\\
Lei \textit{et al.} \cite{lei2020tag} & TPA & 2020 & Text & AMiner & -& 18,464 & 5 Categories & No & Academic articles\\
Lei \textit{et al.} \cite{lei2020tag} & AG & 2020 & Text & AG & -& 1276000 & 4 Categories & No & News articles\\
Denton \textit{et al.} \cite{denton2015user}& - &2015 & Images & Facebook & -& 20 Million & 4.6 Million & & Age, gender, home city, country\\
Mao \textit{et al.} \cite{mao2022attend} & THG & 2022 & Text &X & -& 2,24,097 & - &No &-\\
Mao \textit{et al.} \cite{mao2022attend} & WHG & 2022 & Text &Sina Weibo&\href{https://github.com/OpenSUM/HashtagGen}{Link}& 3,11,401 & - & No & -\\
Gomez \textit{et al.} \cite{gomez2018learning}&InstaNY100K & 2018 & Image+Text & Instagram & \href{https://gombru.github.io/2018/08/01/InstaCities1M/}{Link}& 1,00,000 & 1,64,243 & No &-\\
Zheng \textit{et al}. \cite{zheng2021news}& Tweets2018 & 2018 & Text & X & -& 35,966 & 19,635 & No & Entity Hashtags\\
Zheng \textit{et al.} \cite{zheng2021news}& Tweets2020 & 2018 & Text & X & -& 27,418 & 14,687 &No & Entity Hashtags\\
Manawathilake \textit{et al.} \cite{manawathilake2024optimizing} & - & 2024 & Videos & YouTube & -& 6,000 & - &No & Number of likes, views, comments\\
\bottomrule
\end{tabular}
}
\label{tab:datasets}
\end{table}
\section{Evaluation Methodologies for Hashtag Recommendation Systems}
\label{sec:evaluation}
Evaluating hashtag recommendation systems is essential to ensure their effectiveness, scalability, and relevance in real-world applications. This section provides a detailed examination of evaluation methodologies, encompassing both quantitative and qualitative analysis, as well as hybrid approaches. We begin with a quantitative analysis, focusing on objective metrics that measure system performance. Next, we explore qualitative analysis, which emphasizes user-centric evaluation through human assessment and contextual relevance. Additionally, we discuss hybrid evaluation techniques that combine the strengths of quantitative and qualitative methods to provide a more comprehensive assessment. We also compare the advantages and limitations of qualitative and quantitative analysis. Finally, we examine evaluation approaches, including offline and online evaluation strategies.
\subsection{Quantitative Analysis of Hashtag Recommendation}
Quantitative analysis employs numerical metrics to objectively measure the performance of hashtag recommendation systems. These metrics are essential for evaluating various aspects of system efficacy, including accuracy, ranking quality, semantic relevance, diversity, and alignment with user preferences. The evaluation metrics used in hashtag recommendation systems are deeply rooted in domains such as machine learning and information retrieval. These metrics serve as critical tools for assessing the effectiveness of recommendation algorithms. We broadly classify evaluation methods for recommendation systems into three primary categories: ranking-based, accuracy-based, and sequence-based metrics. Each category addresses specific dimensions of performance, such as the correctness of predictions, the precision of ranked outputs, and the semantic or textual similarity between recommended and ground-truth hashtags.
\paragraph{Ranking-based Metrics}
Ranking-based metrics evaluate whether recommended hashtags appear in the top positions when sorted by relevance.
\begin{enumerate}
\item Normalized Discounted Cumulative Gain (NDCG)
NDCG measures ranking quality by prioritizing the most relevant hashtags at the top. It is widely employed in ranking-based hashtag recommendations \cite{kowald2017temporal}.
Discounted Cumulative Gain (DCG) at position $(k)$ is given by:
\begin{equation}
DCG@k = \sum_{i=1}^{k} \frac{2^{rel(h_{g_i})} - 1}{\log_2(i + 1)}
\end{equation}
where, $( rel(h_{g_i}) )$ is the relevance score of the generated hashtag ($h_{g_i}$) at position $(i)$ (binary or scaled), $(i)$ is the ranked position in the predicted hashtag list. Ideal Discounted Cumulative Gain (IDCG) is the maximum possible DCG for the perfect ranking. NDCG is computed as:
\begin{equation}
NDCG@k = \frac{DCG@k}{IDCG@k} 
\end{equation}
where IDCG@k is the DCG obtained when hashtags are sorted in the ideal ranking order.
Studies such as \cite{kowald2017temporal} employ NDCG@5 and NDCG@10 to assess ranking effectiveness.

\textbf{Application}: Used when ranking hashtag relevance is essential \cite{cao2020hashtag}.

\textbf{Disadvantages:} Requires ground-truth relevance scores, making it infeasible for datasets without human-labeled relevance levels.
\item Leave-One-Out (LOO) Evaluation for Ranking: It determines whether a system can accurately recover an omitted hashtag when given the rest of a post’s content.
Used in studies on hashtag prediction datasets \cite{cao2020hashtag}.
\item Mean Reciprocal Rank (MRR): MRR evaluates ranking effectiveness by computing the position of the first relevant hashtag:
\begin{equation}
MRR = \frac{1}{|Q|} \sum_{i=1}^{|Q|} \frac{1}{rank(h^{(q)}_{first})}
\end{equation}
Here, $Q$ represents the entire set of posts (or queries) in the test set. Each post $q in Q$ is an instance for which the hashtag recommendation system generates a list of recommended hashtags, $h^{(q)}_{first}$ represents the first relevant hashtag recommended by the system for the query
$q$. The relevance of a hashtag is typically determined by ground-truth data. $rank(h^{(q)}_{first})$ is the rank position of the first relevant hashtag in the list of recommended hashtags for the query $q$. 

\textbf{Application}: Suitable for ranking multiple valid hashtags.

\textbf{Disadvantages}: Limited to first relevant match, suboptimal for recall-based tasks.

\end{enumerate}
\paragraph{Accuracy-based Metrics}
Accuracy-based metrics assess the overall correctness of hashtag predictions.
\begin{enumerate}
\item Accuracy / Hit Rate
\begin{equation}
\text{Hit Rate}@k = \frac{\sum_{i=1}^{N} \mathbf{1}(h_{g_i} \in H_{gen})}{N}
\end{equation}
where, $(N)$ is the total number of evaluated posts.
$( \mathbf{1}(h_{g_i} \in H_{gen}))$ is an indicator function returning 1 if a correct hashtag was predicted within top-k, else 0.

\textbf{Application}: Commonly used in classification-based hashtag models.

\textbf{Disadvantages}: Does not reflect ranking awareness.
\item Precision: It measures the proportion of correctly predicted hashtags relative to all returned hashtags.
\begin{equation}
\text{Precision@k} = \frac{| H_{gen} \cap H_{ref} |}{| H_{gen} |}
\end{equation}
where, $(H_{gen})$ is the set of predicted hashtags.
$(H_{ref})$ is the ground truth set of hashtags.
$(|\cdot|)$ represents the number of elements in a set.

\textbf{Example Usage}: Studies on embedding-based hashtag ranking models use Precision@5 and Precision@10.
\item Recall: It measures how many of the ground-truth hashtags were successfully retrieved.
\begin{equation}
\text{Recall@k} = \frac{| H_{gen} \cap H_{ref} |}{| H_{ref} |} 
\end{equation}
High recall implies better coverage of relevant hashtags. Trade-off exists between precision and recall.
Example Usage: Commonly reported for hashtag retrieval models in \cite{kowald2017temporal,alsini2020hit}.
\item F1-Score: It combines precision and recall into a single measure.
\begin{equation}
F1@k = 2 \times \frac{\text{Precision@k} \times \text{Recall@k}}{\text{Precision@k} + \text{Recall@k}} 
\end{equation}

\textbf{Application}: Used to report overall effectiveness in hashtag ranking tasks \cite{park2016harrison}.

\textbf{Disadvantages}: High precision may miss diverse hashtags, high recall can retrieve noisy hashtags.
\end{enumerate}
These metrics are used in numerous hashtag recommendation studies, especially large-scale evaluations of classification-based and ranking-based models.
\paragraph{Sequence-based Metrics}
Sequence-based metrics evaluate hashtag generation models that generate new hashtags rather than ranking existing ones. These are commonly used in transformer-based models, such as GPT or BERT-based language models \cite{fan2024right}.
These metrics assess the similarity between model-generated hashtags and reference (gold-standard) hashtags.
\begin{enumerate}
\item BERTScore: It leverages pre-trained contextual embeddings from BERT to assess the semantic similarity between generated and reference hashtag sequences. It computes a similarity score by comparing the contextualized representations of corresponding tokens in the two sequences.
\begin{equation}
    BERTScore(H_{\text{gen}}, H_{\text{ref}}) = \frac{1}{|H_{\text{gen}}|} \sum_{h_g \in H_{\text{gen}}} \max_{r \in H_{\text{ref}}} \cos(h_g, h_r)
\end{equation}
where, $H_{\text{gen}}$ represents the generated hashtag sequence.
$H_{\text{ref}}$ represents the reference hashtag sequence, $h_g$ and $h_r$ are the contextualized embeddings of individual hashtags in $H_{\text{gen}}$ and $H_{\text{ref}}$, respectively, $cos(h_g,h_r)$ denotes the cosine similarity between embeddings $h_g$ and $h_r$.

\textbf{Strengths:}
\begin{itemize}
\item Captures semantic similarity: Effectively assesses the meaning and relatedness of hashtags.
\item Robust to lexical variation: Rewards semantically similar hashtags even if they differ in wording.
\end{itemize}

\textbf{Weaknesses:}

Computationally expensive: Requires significant computational resources for embedding generation and comparison.
Potential bias from BERT: May inherit biases present in pre-trained BERT.
\item Metric for Evaluation of Translation with Explicit ORdering (METEOR): It considers synonymy, stemming, and word order alongside token-based overlap. More adaptable to hashtag variations than BLEU.
\begin{equation}
\text{METEOR} = F_{mean} \times (1 - Penalty)
\end{equation}
where, $(F_{mean})$ balances precision and recall. Penalty reduces the score for disordered outputs.

\textbf{Example Usage}: Rarely used in hashtag studies, but relevant for semantic analysis.
\item Recall-Oriented Understudy for Gisting Evaluation (ROUGE): It measures recall-based n-gram overlap.
ROUGE-1 and ROUGE-2 track unigram/bigram overlap, while ROUGE-L uses longest common subsequence matching.
\begin{equation}
\text{ROUGE-L} = \frac{\text{LCS-length}}{\text{Reference Length}}
\end{equation}
where, LCS-length is the longest common subsequence between generated and true hashtags.

\textbf{Example Usage}: Common for hashtag generation studies \cite{yu2023generating}.

\textbf{Strengths:}
\begin{itemize}
\item Simple and efficient to compute: Relatively easy to implement and requires minimal computational resources.
\item Suitable for variable-length sequences: Can handle variations in the number of generated hashtags.
\end{itemize}
\textbf{Weaknesses:}
\begin{itemize}
\item Limited semantic understanding: Primarily focuses on lexical overlap and may not fully capture semantic relationships between hashtags.
\item Sensitive to word order: May not adequately reward partially correct sequences with different hashtag order.
\end{itemize}
\item Bilingual Evaluation Understudy (BLEU): 
Measures n-gram overlap between generated and ground-truth hashtags.
\begin{equation}
\text{BLEU} = BP \times \exp\left(\sum_{n=1}^{N} (w_n \log p_n \right)
\end{equation}
where, $(BP)$ signifies brevity penalty that prevents favoring short outputs, $w_n$ are weights assigned to n-gram precisions, $(p_n)$ represents precision of n-grams of size $(n)$.

\textbf{Example Usage}: Common in transformer-based sequence generation models for hashtags.

\textbf{Strengths:}
\begin{itemize}
\item Widely used and understood: A standard metric in machine translation and readily interpretable.
\item Easy to compute: Relatively efficient to calculate.
\end{itemize}
\textbf{Weaknesses:}
\begin{itemize}
\item Limited semantic understanding: Primarily focuses on lexical matches and may not capture semantic relationships between hashtags.
\item Sensitive to word order: May not fully reward partially correct sequences with different hashtag order.
\end{itemize}
\end{enumerate}



Each metric offers a unique perspective on the quality of generated hashtag sequences. While BERTScore excels in capturing semantic similarity, ROUGE-L and BLEU provide insights into lexical overlap and n-gram precision. Distance-1 offers a basic measure of character-level similarity. Selecting the most appropriate metric depends on the specific requirements and priorities of the hashtag recommendation task. Combining multiple metrics can provide a more comprehensive evaluation and a better understanding of the strengths and weaknesses of the recommendation framework.

Sequence-based metrics are most relevant for NLP-driven hashtag recommendation systems \cite{wang2019microblog,fan2024right}.
\subsection{Qualitative Analysis of Hashtag Recommendation}
Unlike quantitative metrics, qualitative analysis involves human judgment to assess hashtag relevance, creativity, coherence, and engagement potential.
\subsubsection{Human Evaluation Metrics}
Since automated metrics may miss contextual or creative relevance, human evaluation is often combined with quantitative scores. These metrcis are used to assess semantic coherence where automated metrics fail.
Despite automated metric dominance, human evaluation remains vital for assessing qualitative aspects of hashtag generation models.
\begin{enumerate}
\item Relevance Judgments
\begin{itemize}
\item Experts manually rate hashtags for contextual relevance (e.g., on a scale from 1-5) based on semantic fit.
\item Complementary to BLEU and ROUGE in evaluating semantic correctness to ensure meaningfulness.
\end{itemize}
\item Coherence and Interpretability:
Judges classify hashtags into coherent vs. random to reflect model reliability.

\textbf{Example Usage:} Used for explainability research in hashtag AI.
\item Creativity and Engagement Potential:
Evaluators check if hashtags align with post intent and are engagement-friendly (i.e., not generic).

\textbf{Example Usage}: Rarely studied but significant for influencer marketing.
\end{enumerate}
\subsection{Hybrid Evaluation}
Most modern studies combine quantitative and qualitative methods to ensure robust evaluation.
\begin{itemize}
\item BLEU and Human Ratings (for Hashtag Generation):
Ensures hashtags are well-formed (BLEU) and contextually effective (human evaluation) \cite{fan2024right}.
\item NDCG and Engagement Metrics (for Ranking):
Hybrid evaluations compare algorithmic ranking with real-world hashtag engagement \cite{park2016harrison}.
\end{itemize}
\subsection{Comparison of Qualitative and Quantitative Analysis}
\begin{table}\footnotesize
\centering
\caption{Comparison of Quantitative and Qualitative Analysis}
\label{table:quality_vs_quantity}
\begin{tabular}{@{\extracolsep{\fill}} lccccccccc}
\toprule
\textbf{Feature} & \textbf{Quantitative Analysis} & \textbf{Qualitative Analysis} \\
\midrule
Key Focus & Numerical performance & Human judgment \\
Metrics Used & Precision, Recall, BLEU, NDCG & Relevance, Coherence, Creativity \\
Scalability & High & Low \\
Subjectivity & Low (objective) & High (context-dependent) \\
Content Sensitivity & Ignores deep semantics & Captures deeper meaning \\
Example Studies & \cite{fan2024right,wang2019microblog,alsini2020hit}&\cite{bansal2022hybrid,bansal2024multilingual,bansal2024hybrid}\\
\bottomrule
\end{tabular}
\end{table}
The comparison of qualitative and quantitative analysis has been highlighted in \autoref{table:quality_vs_quantity}.
\subsection{Evaluation Approaches}
\begin{figure}
\includegraphics[width=\textwidth,keepaspectratio=True]{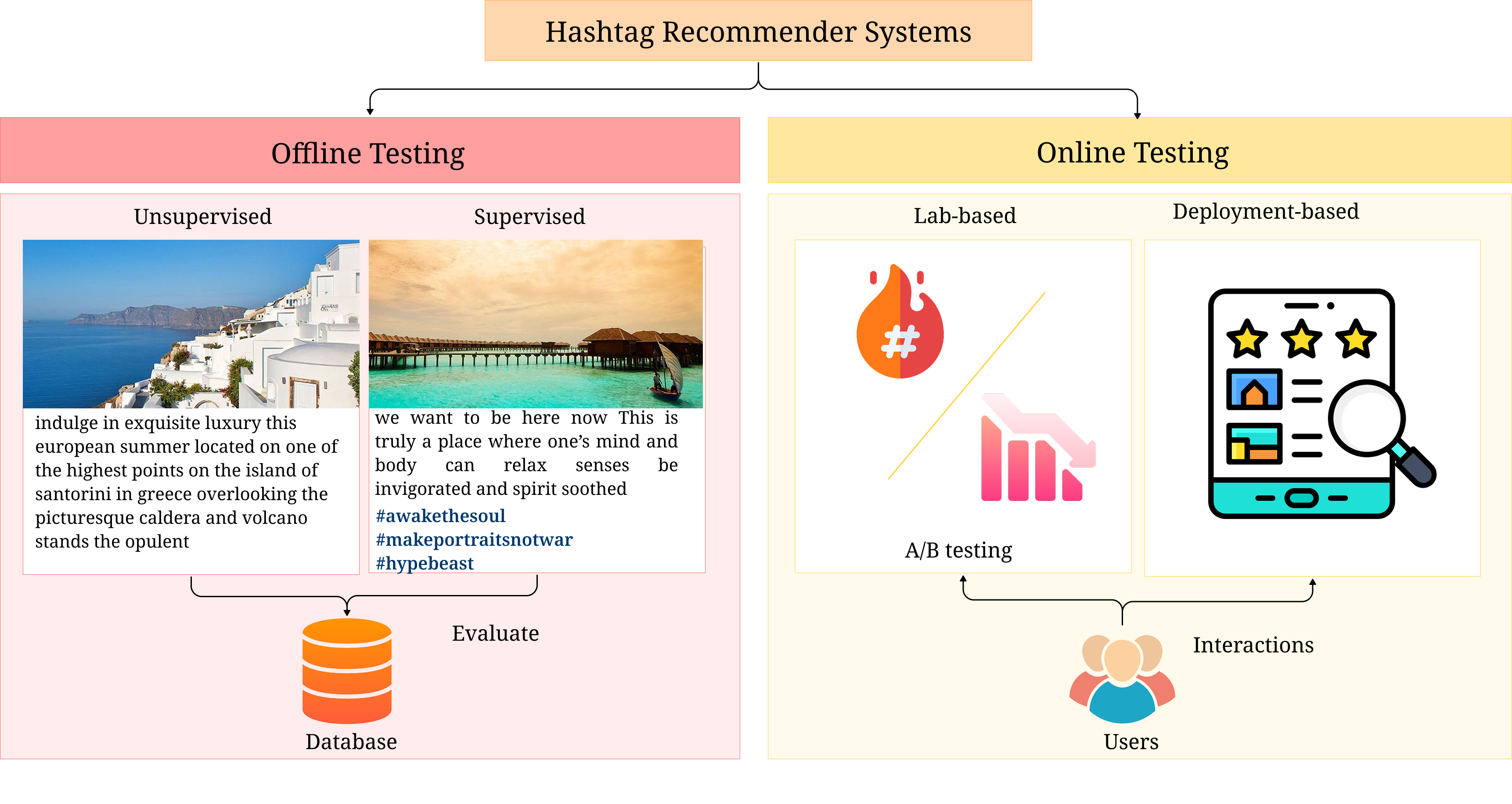}
\caption{Comparison of Offline and Online Testing for Hashtag Recommendation}
\label{fig:evaluation}
\end{figure}

Evaluation strategies for hashtag recommendation systems can be broadly grouped into offline and online evaluations. We discuss these in detail below followed by their comparative analysis as exhibited in \autoref{fig:evaluation}.
\subsubsection{Offline Evaluation}
Offline evaluation assesses models using benchmark datasets, applying predefined metrics without real-time user feedback. Most studies use this approach due to its reproducibility and controlled environment. It includes ranking-based, sequence-based, and accuracy-based measures.
\subsubsection{Online Evaluation}
Online evaluation involves real-time engagement measurements such as user click-through rates (CTR), interactions (likes, shares, comments), and A/B testing with live users. It is rarely covered in traditional literature due to platform restrictions on real-world deployment but provides the most accurate performance feedback. It is less common in academic literature due to platform constraints but crucial for practical applications.

Evaluating hashtag recommendation systems requires a multifaceted approach, integrating ranking-based, accuracy-based, and sequence-based metrics. While offline evaluation remains predominant, online methods incorporating user engagement are crucial for real-world assessments. Emerging evaluation techniques (e.g., semantic relevance, diversity-based scoring) are being explored to address gaps in automated scoring schemes. Bridging the gap between traditional ranking systems and neural generation models will require a combination of exact-match metrics (e.g., NDCG), semantic similarity measures (e.g., METEOR, ROUGE), and human evaluation for comprehensive assessment.
\section{Discussion}
\label{sec:discussion}
In this section, we examine challenges associated with hashtag recommendation in social networks, explore their practical implications, and discuss real-world applications. 
\subsection{Challenges in Hashtag Recommendation}
\label{sec:challenges}
\begin{figure}
\includegraphics[width=0.99\textwidth]{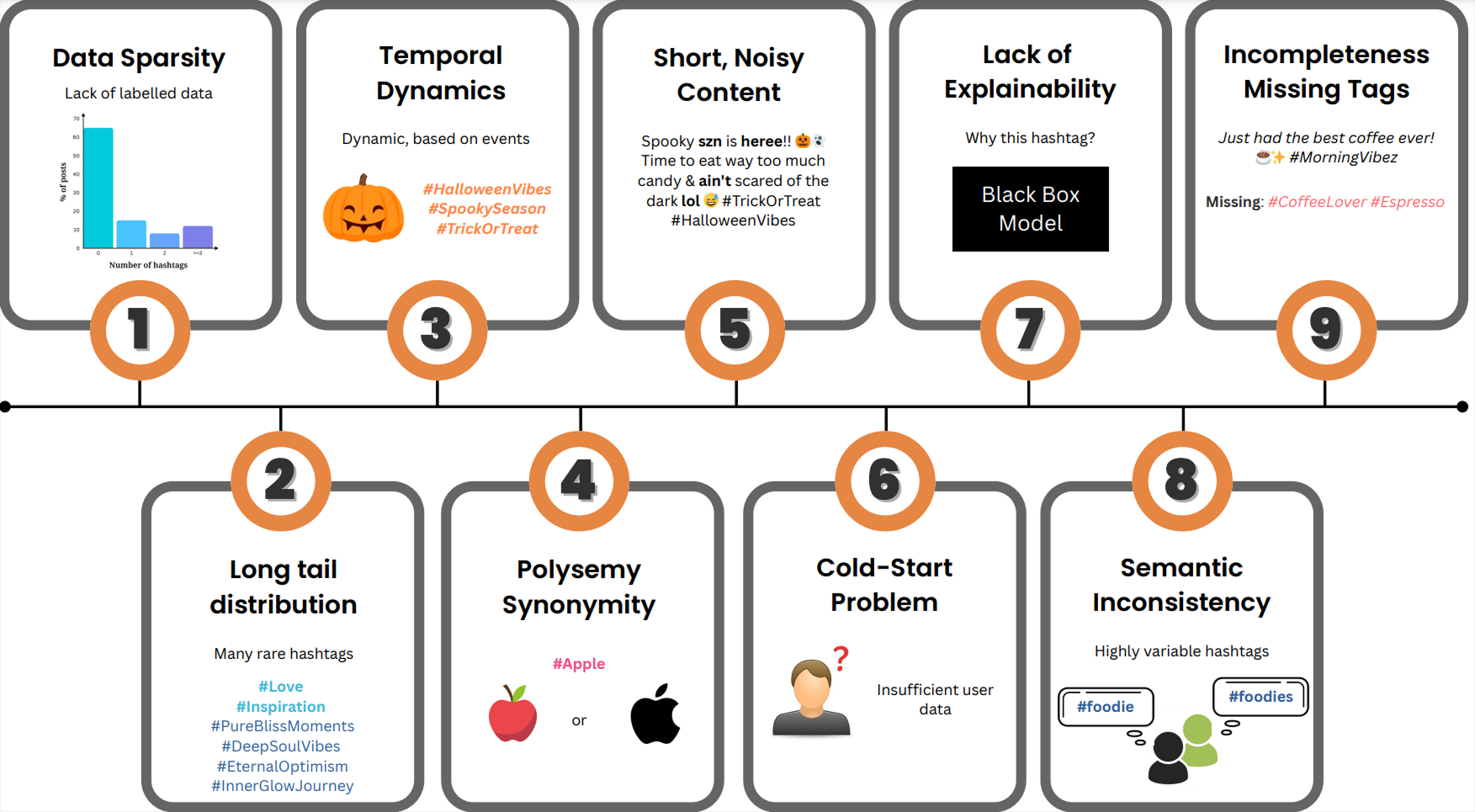}
\caption{Challenges in Hashtag Recommendation}
\label{fig:challenges}
\end{figure}

Despite its utility, the development of effective hashtag recommendation systems faces a range of complex challenges intrinsic to the domain of social media, which are exacerbated by the high-dimensional, dynamic, and noisy nature of social media data. These challenges include not only data sparsity, long-tail distributions, and dynamic contexts, but also core linguistic and systemic issues such as polysemy, cold-start problems, and the lack of explainable predictions depicted pictorially in \autoref{fig:challenges}.
\begin{enumerate}
\item Data Sparsity:
A fundamental challenge in hashtag recommendation is the lack of sufficient labeled data for modeling. Many posts on social media either lack hashtags or use them inconsistently. This creates a sparsity problem where hashtags associated with a large fraction of posts or users are insufficient to train robust models. This issue is exacerbated in cold-start scenarios where new users or hashtags are involved. Approaches to mitigate sparsity include graph-based and embedding-based representation methods. For example, GCNs leverage relationships between users, posts, and hashtags to propagate contextual information across sparse data points, enabling more robust predictions \cite{wei2019personalized,li2019long,bansal2024hybrid}. Zero-shot learning strategies, on the other hand, attempt to overcome the sparsity problem by predicting hashtags that the model has not seen during training. These methods map tweet semantics to hashtag embedding spaces and generalize to unseen data effectively \cite{kumar2019fully,tao2022personalized}.
\item Long-Tail Distribution of Hashtags:
Hashtag usage on social media platforms is highly skewed, following a long-tail distribution where a small number of hashtags dominate usage, while the majority of hashtags are rare and inconsistently used. This imbalance makes it difficult for models to accurately recommend less frequent or personalized hashtags. Graph-based methods explicitly address this challenge by propagating features from frequent hashtags to less frequent ones as part of a shared graph structure \cite{khalil2024mrlkg,li2019long,bansal2024hybrid}. Contrastive learning approaches refine hashtag representations by distinguishing semantically relevant hashtags from irrelevant ones within an embedding space, improving performance for rare hashtags \cite{kumar2019fully,tao2022personalized}. Hybrid solutions, such as those combining GCNs and multimodal transformers, further improve performance by incorporating both user-hashtag interactions and diverse content modalities \cite{bansal2022hybrid,wei2019personalized}.
\item Dynamic Context and Temporal Dynamics:
Hashtag usage trends and meanings are highly dynamic, influenced by temporal patterns, events, and cultural shifts. The meanings and usage trends of hashtags evolve rapidly over time, further complicating their recommendation. Hashtag drift refers to scenarios where the semantic meaning of a hashtag changes based on trends or contexts. Popular hashtags can emerge suddenly and decline just as rapidly, creating a need for models that adapt over time. Dynamic temporal models, such as LSTM networks with temporal attention mechanisms, capture time-sensitive patterns, enabling recommendations that align with real-time trends \cite{ma2018temporal}. Incremental learning approaches such as class-incremental GCN frameworks adapt to the introduction of new hashtags and evolving contexts without retraining entire models \cite{kolyszko2024dynamic}.
\item Polysemy and Synonymity: In natural language, hashtags often suffer from polysemy (a single hashtag having multiple meanings, e.g., \#Apple as a fruit or a company) and synonymity (different hashtags meaning the same thing, e.g., \#Car and \#Automobile). These semantic ambiguities complicate hashtag recommendations, leading to inconsistent or irrelevant predictions.

Researchers address polysemy using contextual embeddings (BERT) that disambiguate meaning based on surrounding text \cite{cantini2021learning,kaviani2020emhash}.
Methods such as semantic clustering or knowledge graphs help map related hashtags into a shared representation space, resolving issues of synonymity \cite{chen2021tagnet,khalil2024mrlkg}.
Embedding techniques, such as sentence-level attention and keyword-guided graphs, refine semantic understanding for hashtags within noisy contexts \cite{cantini2021learning,chen2021tagnet}.
\item Short and Noisy Content: Social media content, such as tweets or microblogs, is typically short, often poorly structured, and noisy (e.g., containing typos, informal slang, or abbreviations). This increases the difficulty of learning meaningful semantic representations.

Seq2seq models with bidirectional attention have been used to extract key information from noisy data and handle short text by modeling context more effectively \cite{wang2019microblog}. NTMs incorporate high-level thematic information to enhance predictions in noisy settings while mitigating sensitivity to outliers \cite{tangpong2023hashtag}.
\item Cold-Start Problem:
The cold-start problem occurs when recommendations must be made for new users or posts with no prior hashtag usage. This is particularly challenging in situations where historical behavior is sparse or nonexistent.
\begin{enumerate}
\item Collaborative Filtering (CF): While CF traditionally suffers from cold-start scenarios, hybrid models integrate CF with content-based approaches to alleviate this issue \cite{bansal2024hybrid}.
\item Zero-Shot and Meta-Learning: ZSL frameworks predict hashtags for new users or posts by leveraging semantic embeddings and unseen concepts, while meta-learning quickly adapts to new scenarios with minimal data 
\cite{kumar2019fully,tao2022personalized}.
\end{enumerate}
\item Lack of Explainability: Modern deep learning models used in hashtag recommendation often function as black boxes, offering little transparency into why a particular hashtag was recommended. This lack of explainability makes it difficult to evaluate trustworthiness and interpretability in practical applications.

Few studies tackle this issue explicitly. One approach involves incorporating attention mechanisms and visualizing attention weights on input features (e.g., images, words) to improve interpretability \cite{gong2016hashtag,yang2020amnn}.
Knowledge graph-based systems provide an interpretable structure by explicitly modeling semantic relationships among hashtags, users, and content \cite{chen2021tagnet}.
\item Semantic Inconsistency and Hashtag Explosion: Hashtags tend to be user-generated and highly variable, leading to problems of inconsistency and explosion in the potential hashtag vocabulary. For example, users may use hashtags with slight alterations (\#Foodie vs. \#Foodies) or irrelevant tags, making modeling difficult.

\begin{enumerate}
\item Sequence Generation Models: Seq2seq frameworks with attention mechanisms can mitigate explosion by generating hashtags relevant to both context and semantics \cite{wang2019microblog,yang2020amnn}.
\item Shared Embedding Spaces: Joint semantic spaces for text-hashtag embedding (e.g., BERT-based or graph-based representations) establish coherence across inconsistent hashtags \cite{cantini2021learning,chen2021tagnet}.
\end{enumerate}
\item Incompleteness and Missing Tags: Many social media posts lack sufficient context or detail, making it difficult to recommend meaningful hashtags. Incomplete labels occur when users fail to annotate posts with all relevant hashtags, leading to partial supervision for training datasets.

Weakly supervised attention mechanisms attempt to predict hashtags under noisy and incomplete conditions \cite{javari2020weakly}.
Contrastive representation learning methods enable robust predictions by learning to differentiate between relevant and irrelevant hashtags, even under incomplete data scenarios \cite{tao2022personalized}.
\end{enumerate}
\subsection{Practical Implications of Hashtag Recommendation}
The development and deployment of hashtag recommendation systems have significant practical implications for social media platforms, users, and content creators, which are enumerated below.
\begin{enumerate}
\item Improved User Experience: A hashtag recommendation system can help readers find articles that are relevant to their interests quickly and easily. This can improve
user engagement and satisfaction with the news website.
\item Increased Engagement: By suggesting relevant tags, the system can encourage readers to explore more articles on the same topic, leading to increased
engagement and time spent on the website.
\item Better Content Discovery: The system can help surface articles that might have been overlooked or buried in the website's archives, leading to better content
discovery for readers.
\item Improved Search Engine Optimization: Annotating UGC with relevant keywords can improve their search engine rankings and visibility, leading to increased traffic to the
website.
\item Personalization: A hashtag recommendation system can be personalized for each user based on their reading history and preferences, providing a unique
and tailored experience.
\item Cost-saving: With the help of hashtag recommendation systems, news organizations can save time and money by reducing the manual labor of
tagging articles.
\item Improving Advertisement Targeting: By understanding user interests and preferences through their search and reading habits, the tag recommendation system can help improve targeting of advertisements, making them more relevant to users.
\end{enumerate}
\subsection{Real-world Applications of Hashtag Recommendation}
\label{sec:applications}
In this section, we explore the real-world applications of hashtag recommendation across a multitude of sectors. Additionally, we discuss downstream tasks that leverage hashtag recommendation.
\subsubsection{Different Sectors}
\begin{figure}
\centering
\caption{Applications of Hashtag Recommendation in Different Sectors}
\includegraphics[width=0.82\textwidth,keepaspectratio=True]{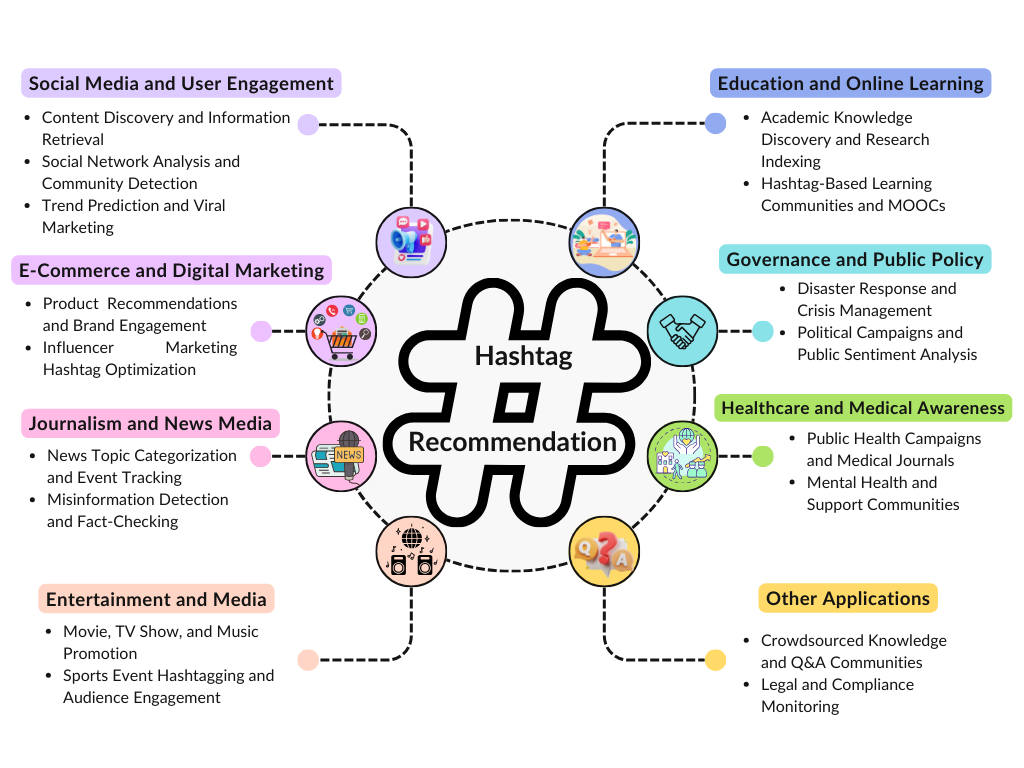}
\label{fig:applications}
\end{figure}

Hashtag recommendation is widely applied across various domains, leveraging its ability to categorize, enhance discoverability, and optimize user engagement across social media platforms as illustrated in \autoref{fig:applications}. Below are key sectors where hashtag recommendation is critical, along with their respective downstream applications.
\begin{enumerate}
\item Social Media and User Engagement
\begin{itemize}
\item Content Discovery and Information Retrieval:
These systems improve content visibility and information retrieval on platforms such as X, Instagram, TikTok, YouTube, and LinkedIn through learning-to-rank approaches and transformer-based ranking \cite{li2016hashtag}.
For example, personalized hashtag recommendations improve post visibility on X \cite{zhang2021howyoutagtweets}. Influencers and brands use trending hashtags for engagement \cite{jafari2023popular}.
\item Social Network Analysis and Community Detection:
Facebook Groups, Reddit, Discord, and Sina Weibo support the formation of online communities and subcultures. Hashtag-based topic modeling can identify these communities by analyzing the use of shared hashtags within the platform. A common approach involves using graph-based methods for community clustering \cite{khalil2024mrlkg,li2019long,padungkiatwattana2022pac}. Researchers have identified subcultures on Reddit by analyzing shared hashtags in discussion threads and detected like-minded user communities through hashtag graphs  \cite{javari2020weakly}.
\item Trend Prediction and Viral Marketing:
Platforms such as X, Instagram, and TikTok leverage trending hashtag recommendations to predict emerging conversations by utilizing temporal analysis and attention-based transformer models \cite{kolyszko2024dynamic, ma2019co}. For example, social media managers often rely on automated tools to suggest viral hashtags \cite{jafari2023popular}, while brands implement auto-suggestion systems to identify and capitalize on trending topics for greater engagement.
\end{itemize}
\item E-Commerce and Digital Marketing
\begin{itemize}
\item Product Recommendations and Brand Engagement:
Hashtags are often used in product listings on e-commerce sites including Facebook Marketplace, Instagram Ads, Shopify, and Amazon to improve cross-platform discoverability and searchability. It increases product visibility by facilitating product discovery through search queries and algorithm-driven suggestions. Multi-modal transformer encoders and visual hashtag embedding models are employed by several works to produce relevant hashtags \cite{bansal2024hybrid,yu2023generating}. 
For instance, websites such as Amazon use automated systems to create relevant hashtags, which aid merchants in improving the visibility and reach of their product listings \cite{he2021tagpick}.
\item Influencer Marketing Hashtag Optimization:
Influencer Marketing Hashtag Optimization is widely used on platforms such as Instagram, and X. They use personalized, user-history-aware models that recommend optimal hashtags to maximize visibility and engagement \cite{zhang2021howyoutagtweets,tao2022personalized}. These systems analyze user preferences and posting history to improve reach. Automated tools also optimize hashtags for sponsored content to enhance exposure.
\end{itemize}
\item Journalism and News Media
\begin{itemize}
\item News Topic Categorization and Event Tracking:
X, Google News, and Reddit use auto-tagging to categorize news articles, enhancing searchability and content discovery. These systems often employ learning-to-rank algorithms and topic-enhanced embeddings to generate relevant hashtags \cite{li2016hashtag,shi2016learning}. Automated hashtag suggestions help streamline the tagging process for news articles \cite{madisetty2019identification}. Additionally, hashtags contribute to more effective topic modeling, especially for breaking news stories \cite{zheng2021news}.
\item Misinformation Detection and Fact-Checking:
Facebook, X, and WhatsApp Fact-Check Services uses hashtags to identify misinformation clusters and viral hoaxes.
They employ graph-based hashtag communities to track misinformation propagation \cite{kolyszko2024dynamic}.
For example, coordinated hashtag campaigns in disinformation networks  to amplify false narratives and manipulate public opinion\cite{kolyszko2024dynamic,dadgar2022novel}. Such fake news dissemination can be prevented by filtering misleading tags \cite{dadgar2022novel}.
\end{itemize}
\item Entertainment and Media
\begin{itemize}
\item Movie, TV Show, and Music Promotion:
Official accounts on platforms such as X, Instagram, YouTube, and Spotify use optimized hashtags to boost reach. Sequence-to-sequence hashtag ranking models can be employed for promotional content \cite{yang2020sentiment, mehta2021open}. Movie trailer posts can utilize auto-suggested hashtags based on previous engagement trends \cite{guarascio2024movie}, while AI systems suggest personalized music hashtags for new releases.
\item Sports Event Hashtagging and Audience Engagement:
Sports leagues use auto-generated hashtags to enhance fan engagement across platforms such as ESPN, X, and Instagram Stories. Temporal and popularity-driven hashtag recommendation systems are commonly employed in these contexts \cite{mireshghallah2023simple, chakrabarti2023hashtag}. For instance, automatically generating event-based hashtags for real-time sports results, as well as AI-based hashtag generation for NBA and FIFA discussions, can significantly increase audience engagement.
\end{itemize}
\item Education and Online Learning
\begin{itemize}
\item Academic Knowledge Discovery and Research Indexing:
Platforms such as ResearchGate, Publons, and Google Scholar use hashtags to assist in tagging and indexing academic content. Hashtags help categorize research papers, making them easier to discover and track. These systems often leverage semantic embeddings and external knowledge graph-based tagging \cite{ghenname2015hashtags, jafari2023unsupervised}. For example, AI-based indexing tools can automatically generate hashtags that summarize the key points of a paper's abstract. Additionally, hashtags support academic knowledge graphs, aiding in citation tracking and enhancing research visibility.
\item Hashtag-based Learning Communities and MOOCs:
Platforms such as Coursera, Udemy, and X use hashtags to enhance the discovery of learning materials. Personalized recommendations based on user history help educators and students find relevant content \cite{babinec2017education}. Tweets about online courses often include autogenerated educational hashtags for greater visibility. AI-generated hashtags also connect learners with resources in discussion forums.
\end{itemize}
\item Governance and Public Policy
\begin{itemize}
\item Disaster Response and Crisis Management:
X, GovAlert, and Crisis Mapping Networks uuseag recommendation to organize emergency responses. The embedding of the community graph and the high-order relations helps to track the hashtags of the crisis \cite{chowdhury2020identifying}. Disaster response hashtags are recommended for relief efforts based on emerging keyword trends, while government agencies automatically generate informative tags related to ongoing crises.
\item Political Campaigns and Public Sentiment Analysis:
 Facebook, X, and YouTube use automated hashtags to support politicians in sentiment analysis and campaign tracking. Sentiment-aware hashtag generation and supervised attention models enable real-time recommendations during events such as election debates \cite{yang2020sentiment, javari2020weakly}. These systems also monitor viral political hashtags to track public opinion trends.
\end{itemize}
\item Healthcare and Medical Awareness
\begin{itemize}
\item Public Health Campaigns and Medical Journals:
Organizations such as WHO Twitter, CDC Health Alerts, and PubMed Social Sharing use hashtags to promote health awareness campaigns. Transformer-based contextual hashtag generation enhances the relevance and reach of these initiatives \cite{jain2024nlp,wang2023doctor}. Examples include hashtag suggestions for COVID-19 vaccine awareness and clinically-relevant tags for academic health articles \cite{wang2023doctor}.
\item Mental Health and Support Communities:
Facebook Support Groups and Twitter Mental Health Chats utilizes hashtags to help users connect with supportive communities. GNN-based community detection enhances engagement for those seeking mental health support \cite{alsini2020utilizing}. Hashtag recommendations prevent content isolation in niche mental health discussions and helps detect sensitive topics and suggest appropriate tags.
\end{itemize}
\item Other Applications
\begin{itemize}
\item Crowdsourced Knowledge and Q\&A Communities:
On platforms such as Stack Overflow, Quora, and Reddit, hashtags play a crucial role in organizing knowledge bases and categorizing user questions. Automatic tagging of posts improve searchability, helping users find similar questions \cite{he2022ptm4tag,he2025ptm4tag+}. Many approaches rely on unsupervised methods and keyword-based graph embeddings to enhance topic classification and content discovery \cite{wang2022tkgat}. Hashtags also support question similarity detection.
\item Legal and Compliance Monitoring:
Lawyers and compliance officers use hashtags to track legal trends on LinkedIn\footnote{https://in.linkedin.com/}, Law Twitter, or Compliance Journals. Knowledge graphs and semantic embeddings are employed to generate relevant hashtags \cite{wang2022tkgat, mezni2021temporal}. For example, Legal AI tools suggest trending, case-specific tags for compliance professionals. Additionally, hashtags assist in the automatic extraction of topics from regulatory documents \cite{wu2022judgment}.
\end{itemize}
\end{enumerate}
Hashtag recommendation methods have widespread applications, spanning across social media engagement, online education, digital marketing, crisis response, and research knowledge dissemination. Recent advances in transformer-based ranking, graph-based community detection, multimodal fusion, and retrieval-augmented generation enable more efficient and context-aware hashtag recommendations across various domains.
\subsubsection{Downstream Tasks}
Hashtag recommendation extends beyond simple tag suggestion, serving as a foundational task with significant downstream applications. It directly impacts content popularity forecasting, sentiment-based classification, hate speech detection, and automated text generation. Research shows hashtags influence engagement metrics, making them crucial signals for social media monitoring, misinformation management, and multimodal content optimization. Hashtag recommendations play a critical role in the dissemination and interpretation of content across social media platforms. In recent years, hashtags have become integral to social networks, serving as a means to provide timely and relevant information about user-generated content. Their utility extends to a wide range of applications, including event detection \cite{atefeh2015survey}, information diffusion \cite{chang2010new}, sentiment analysis \cite{davidov2010enhanced}, information retrieval \cite{efron2010hashtag}, text classification \cite{wang2011topic}, and event analysis \cite{xing2016hashtag}. For instance, Konjengbam \textit{et al.} \cite{konjengbam2020unsupervised} introduced a Tagging Product Review (TPR) system designed to generate informative and readable tags for popular products, summarizing reviews that highlight various product aspects. The authors utilized transfer learning to address the challenge of generating tags for less popular or ``cold" products. Furthermore, hashtags are employed in diverse scenarios, such as popularity prediction \cite{samanta2017lmpp,yamasaki2017folkpopularityrank}, immersive search \cite{gao2017hashtag}, and enterprise applications \cite{mahajan2016hashtag}, underscoring their versatility and significance in both academic and practical contexts.
\begin{enumerate}
\item Tweet Classification: Generated hashtags to significantly improve tweet classification tasks, including emoji prediction, emotion classification, hate speech detection, irony detection, offensive content detection, sentiment analysis, and stance detection \cite{diao2023hashtag}. HashTation generates hashtags for low-resource tweet classification using tweet and enrtity attention modules, which can indirectly help better categorization of tweets, boosting discoverability.
\item Sentiment Analysis: Many hashtags carry implicit sentiment cues (e.g., \#Happy, \#Sad) that models can leverage for weakly supervised sentiment classification. Automatically generated hashtags provide additional context to disambiguate short or vague tweets \cite{yang2020sentiment,diao2023hashtag}. 
\item Popularity Prediction and Social Media Engagement: Hashtags boost post visibility on algorithm-driven feeds via content categorization. Personalized hashtag selection improves user engagement by aligning with audience preferences \cite{,zhang2019hashtag,bansal2022hybrid}.
Yang \textit{et al.} \cite{yang2020sentiment} developed sentiment-augmented hashtag learning, showing that people engage more with emotion-aligned hashtags.
Chakrabarti \textit{et al.} \cite{chakrabarti2023hashtag} optimized hashtag recommendations for post visibility and engagement metrics, introducing a popularity prediction model tailored for real-world tweets. 
\item Caption Generation: Hashtags provide implicit summarization of image content, which can serve as context seeds for caption generation models.
This is particularly valuable for image-based social media platforms (Instagram, TikTok). Gaur \textit{et al}. \cite{gaur2019generation} explicitly explored hashtag-enhanced caption generation, where generated hashtags are inputs to a character-level RNN or transformer model to create storylike, engaging captions. Al \textit{et al.} \cite{al2022image} used CNN-RNN-based architectures to jointly generate image captions and hashtags, improving caption relevance compared to traditional captioning models.
\item Misinformation Detection: Hashtags tend to cluster misinformation campaigns (e.g., conspiracy-related hashtags).
Graph-based methods \cite{kolyszko2024dynamic} identified abnormal hashtag spread patterns, aiding misinformation detection.
\end{enumerate}

\section{Conclusion and Future Research Directions}
\label{sec: conclusion_and_fw}
Hashtag recommendation systems have evolved significantly over the past decade, driven by advances in deep learning, natural language processing, and multimodal data analysis. This survey synthesizes research from 2015 to 2024. By introducing a hierarchical taxonomy, we have categorized a vast body of research based on modality, problem formulation, filtering strategies, methods, evaluation metrics, challenges, and applications. This systematic approach has allowed us to trace the evolution of the field, highlighting the transition from rudimentary frequency-based approaches to sophisticated deep learning models capable of capturing semantic relationships and cross-modal dependencies. Our analysis reveals several key trends. Transformer-based models such as BERT and GPT-4 have emerged as dominant architectures, achieving state-of-the-art performance in hashtag recommendation. Multimodal systems that fuse information from various data sources, such as text, images, and videos, are gaining traction. Multimodal approaches outperform unimodal approaches, demonstrating the importance of leveraging diverse data sources. Furthermore, retrieval-augmented methods that leverage external knowledge bases are showing promise in enhancing the diversity and relevance of recommended hashtags. Data sparsity, long-tail hashtag distributions, cold-start problems for new users, and semantic ambiguity in short-text UGC continue to pose hurdles. 
Beyond technical advancements, this survey has emphasized real-world applications of hashtag recommendation systems. By enhancing content discoverability, fostering user engagement, and facilitating trend analysis, these systems play a crucial role in shaping social media ecosystems. Moreover, hashtag recommendation systems are proving valuable in downstream tasks such as sentiment analysis, misinformation detection, and viral content prediction.
\begin{figure}
\includegraphics[width=\textwidth]{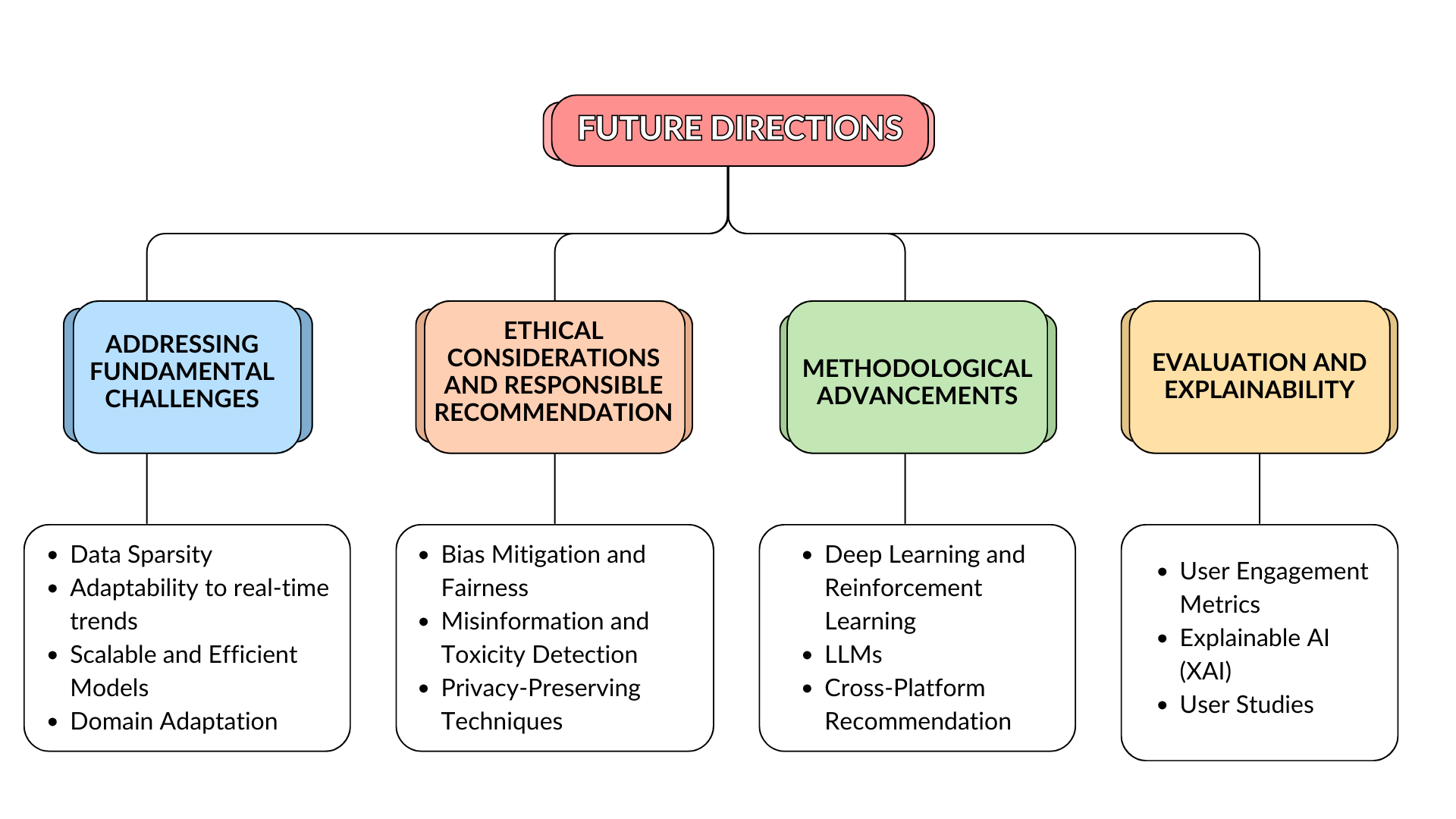}
\caption{Future Research Directions in the Domain of Hashtag Recommendation}
\label{fig:future_works}
\end{figure}
However, several critical challenges necessitate further investigation to fully realize the potential of these systems. \autoref{fig:future_works} outlines key future research directions in the domain of hashtag recommendation.
\begin{enumerate}
\item \textbf{Addressing Fundamental Challenges}: Current hashtag recommendation models face limitations related to data sparsity, adaptability to emerging hashtags, contextual interpretation, and semantic ambiguity. These challenges are further compounded by the real-time processing demands of large-scale, high-velocity social media data \cite{kolyszko2024dynamic, bansal2022hybrid}. Future research should focus on developing scalable and efficient models capable of handling these complexities. Exploring self-supervised learning techniques to leverage unlabeled social media data is crucial for mitigating data sparsity. To improve real-time adaptability, models should be designed to dynamically recognize and adjust to new hashtags and trending topics. Incorporating real-time trend analytics, similar to open-domain zero-shot trend models \cite{mehta2021open}, can enhance a system’s ability to capture emerging discussions.
\item \textbf{Ethical Considerations and Responsible Recommendation}:
Ethical concerns surrounding bias, fairness, and transparency are critical in hashtag recommendation.
\begin{itemize}
\item Bias Mitigation and Fairness
Research should focus on developing fairness-aware algorithms to ensure equitable hashtag distribution and mitigate the perpetuation of harmful stereotypes \cite{banbhrani2021sc, djenouri2022toward}, such as underrepresenting minority voices in activism-related hashtags (e.g., \#StopAAPIHate).
\item Misinformation and Toxicity Detection:
Hashtag recommendation systems should be designed to detect and reduce the spread of misinformation and hate speech. Robust strategies are needed to prevent the recommendation of hashtags linked to harmful content.
\item Privacy-Preserving Techniques:
Research should explore privacy-enhancing methods, such as differential privacy and federated learning, to protect user data while maintaining recommendation effectiveness.
\end{itemize}
\item \textbf{Methodological Advancements}:
\begin{itemize}
\item Deep Learning and Reinforcement Learning:
Advancing deep learning architectures, including transformer-based models, is essential for improving recommendation performance. Additionally, reinforcement learning can optimize hashtag recommendations for long-term user engagement and adapt to evolving user preferences.
\item LLMs: Integrating LLMs and advanced multimodal approaches holds significant promise for improving recommendation accuracy and contextual understanding.
\item Cross-Platform Recommendation:
Developing unified cross-platform recommendation frameworks \cite{bansal2022hybrid,won2023extra} (e.g., adapting to TikTok's video-centric UGC versus X’s text-heavy posts) will improve model transferability and robustness.
\end{itemize}
\item \textbf{Evaluation and Explainability}: Future evaluations should incorporate diversity-focused analysis and human-rated assessments to improve personalization metrics.
\begin{itemize}
\item User Engagement Metrics:
Online engagement indicators, such as click-through rates and real-time user interactions, should be explored despite data limitations.
\item Explainable AI (XAI):
Developing explainable AI techniques is crucial for enhancing transparency and trust by providing insights into recommendation rationales. Transparent models are particularly important in sensitive applications, such as political campaigns (\#Vote2024).
\item User Studies:
Conducting user studies will offer valuable insights into how individuals perceive and interact with hashtag recommendation systems.
\end{itemize}
By addressing these challenges and advancing these research directions, hashtag recommendation systems can contribute to a more engaging, ethical, and responsible social media experience. Ongoing research is essential for developing robust, fair, and efficient recommendation models that align with the evolving needs of users and platforms. As social media continues to evolve, these systems will play a crucial role in facilitating communication, fostering communities, reducing information overload, and shaping digital interactions.
\end{enumerate}

\bibliographystyle{elsarticle-num} 

\bibliography{crefs}
\end{document}